\documentclass[aps,pre,twocolumn,superscriptaddress,10pt]{revtex4-2}
\usepackage[T1]{fontenc}
\usepackage[utf8]{inputenc}
\usepackage{silence}
\usepackage{amsmath, amssymb}
\usepackage{graphicx}
\usepackage{bm}
\usepackage{xcolor}
\usepackage{csquotes}
\usepackage{accents}
\usepackage[version=4]{mhchem}
\usepackage{siunitx}
\usepackage{enumitem}

\definecolor{tab_red}{HTML}{D62728}

\let\orighat\hat
\def\hat#1{\accentset{\wedge}{#1}}

\def\dd{\text{d}}
\def\ii{\text{i}}

\usepackage[colorlinks=true,allcolors=teal]{hyperref}
\usepackage{bookmark}
\begin{document}

\title{Hall conductance of dilute electrolytes from odd stochastic density functional theory}

\author{Yael Avni}
\affiliation{University of Chicago, James Franck Institute, 929 E 57th Street, Chicago, IL 60637}
\affiliation{Department of Physics of Complex Systems, Weizmann Institute of Science, 76100 Rehovot, Israel}

\author{Michel Fruchart}
\affiliation{Gulliver, ESPCI Paris, Université PSL, CNRS, 75005 Paris, France}

\author{David Martin}
\affiliation{LPTMC, CNRS UMR 7600, Université Pierre et Marie Curie, 75252 Paris, France}

\author{Tali Khain}
\affiliation{John A. Paulson School of Engineering and Applied Sciences, Harvard University, Cambridge, MA 02138, USA}

\author{Vincenzo Vitelli}
\affiliation{University of Chicago, James Franck Institute, 929 E 57th Street, Chicago, IL 60637}
\affiliation{Leinweber Institute for Theoretical Physics, University of Chicago, Chicago, IL 60637, USA}

\begin{abstract}
Electrolytes under an external magnetic field exhibit a Hall conductivity, much like metals and semiconductors. 
However, existing theories fail to quantitatively account for their measured Hall conductivities. 
To make progress toward such a theory, we develop a stochastic density functional framework for electrolytes in a magnetic field. 
This framework, which we dub odd SDFT, extends the celebrated stochastic density functional theory (SDFT) to incorporate transverse mobilities and chiral time-correlated noise. Using this framework, we calculate the Hall conductivity in the absence of hydrodynamic solvent effects through three approaches: (i) direct response, (ii) Green-Kubo relation, and (iii) a hybrid fluctuation-dissipation relation. 
We further show that odd SDFT applies more broadly to fluids with broken time-reversal and mirror symmetry, including odd-diffusive fluids and chiral active matter.
Our results generalize the relaxation correction of Debye-Hückel-Onsager theory to magnetized electrolytes and provide a foundation for incorporating additional effects mediated by the solvent that could quantitatively reconcile theory and experiments.
\end{abstract}

\maketitle

\section{Introduction}

The theory of electrolyte conductivity was pioneered by Debye, H\"uckel, and Onsager, who derived the Debye-H\"uckel-Onsager (DHO) limiting law~\cite{onsager1927report,onsager2002irreversible,bernard1992conductance,chandra1999ion,altenberger1983theory} for the conductivity $\kappa$ at low ionic concentration ${c}$, whereby $\kappa = \mathcal{A}{c}+\mathcal{B} {c}^{3/2}$ where the coefficients $\mathcal{A}$ and $\mathcal{B}$ are explicit functions of the solvent and ion parameters. 
The linear term corresponds to the infinite-dilution conductivity, while the ${c}^{3/2}$ term gives the leading correction due to ionic interactions. 
This result provides an accurate description of electrolyte conductivity at sufficiently low concentrations~\cite{robinson2002electrolyte,bockris1998modern}, but it does not apply to situations where a magnetic field is present.

When the electrolyte is subject to an external magnetic field, it also exhibits a Hall conductivity $\kappa_{\text{H}}$ similar to that observed in metals and semiconductors~\cite{wolynes1980dynamics}, leading to a Hall current perpendicular to both the applied electric and magnetic fields.
Theoretical predictions for the Hall conductivity, including an analogue of the DHO limiting law, have been derived using liquid-theory techniques \cite{Friedman1968}. 
However, experimental measurements differ substantially from the theoretical prediction even at lowest order in concentration \cite{meton1976hall}, likely because the Hall effect acts as a probe of ion-solvent interactions which involve both the dielectric and hydrodynamic responses of the solvent \cite{sung1987microscopic,hubbard1981electrohydrodynamic}.
To date, none of the theoretical predictions capture experimental data \cite{gerard1990hall}.
The goal of this work is to make progress in this direction by developing a stochastic field theoretical approach that can both capture the behavior of electrolytes in a magnetic field and later on be combined with modern approaches to describe the structured dielectric response of solvents \cite{Kornyshev1986,Belaya1986,Levy2012,Berthoumieux2019,Berthoumieux2021,Illien2024,Blossey2022,Blossey2022b,Berthoumieux2024,dean2026dielectric,adar2018dielectric}.

To do so, we build on a recent reformulation of the DHO theory within the modern framework of stochastic density functional theory (SDFT), commonly refereed to as the Dean-Kawasaki equation~\cite{dean1996langevin,kawasaki1994stochastic,demery2016conductivity,peraud2017fluctuation,avni2022conductivity,avni2022conductance}, which has allowed for extensions to a broad range of regimes and other transport phenomena in electrolytes~\cite{robin2024correlation,dean2026dielectric,kruger2018stresses,benois2023enhanced,Berthoumieux2024}. 
This framework is well-suited to incorporate a principled effective description of the solvent, but it does not take into account the presence of a magnetic field. 
We fill this gap by developing a generalization of the Dean-Kawasaki equation that handles the Gaussian non-white noise induced by magnetic Lorentz forces.
This theory, that we dub \enquote{odd SDFT}, is summarized from a broader perspective in the companion letter \cite{avni2026odd}, including its relation to chiral active matter~\cite{caprini2023chiral, abdoli2026dynamical,goerlich2026particleresolvedrheologicalstudychirality,kuroda2023microscopic}, while here we focus on electrolytes under magnetic field.

\section{Overview and outline}

The theory of electrolyte conductance is typically formulated in the overdamped regime (i.e. where inertial effects are neglected), and it is formally derived in the low mass limit of the full underdamped dynamics. This is also the regime underlying the Dean-Kawasaki equation~\cite{dean1996langevin}, which is essentially a field description of a system of coupled overdamped Langevin equations driven by white noise. 
It was recently realized by Chun et al.~\cite{chun2018emergence} that in the presence of a magnetic field, the low mass limit of the Langevin equation is singular: it involves a Gaussian but nonwhite noise, for which the standard Dean-Kawasaki formalism is not applicable.

In Sec.~\ref{SecModel}, we derive the analogue of the Dean-Kawasaki equation for charged particles under Lorentz force, hereafter referred to as the odd Dean-Kawasaki equation.
The derivation, presented in Sec.~\ref{SecModel}, is done by first obtaining a field theory description of the underdamped dynamics, and then taking the low mass limit from which the nonwhite noise naturally emerges. We give a pedagogical example of how to calculate a transport coefficient from odd SDFT using both direct linear response and Green-Kubo relation. The latter highlights the importance of the nonwhite noise in odd SDFT, as it is necessary for the Green-Kubo relation to correctly capture the antisymmetric part of transport coefficients.

In Sec.~\ref{SecHall}, we formulate odd SDFT for multiple ionic species. We then evaluate the conductivity of a symmetric $q\!:\!q$ electrolyte. To obtain the first correction beyond the infinite dilution limit (the $c^{3/2}$ term), we go beyond the mean-field level by performing the equivalent of a one-loop expansion in field theory. We calculate the conductivity using (i) direct linear response, (ii) Green-Kubo relation, and (iii) a hybrid fluctuation-dissipation relation derived from the perspective of first-order dynamics, and show that all three approaches yield the same result. We obtain explicit expressions for Hall conductance as a function of the magnetic field. It takes the form 
\begin{equation}
\kappa_{\rm H} = \mathcal{A_{\rm H}}c+\mathcal{B_{\rm H}} c^{3/2}    
\end{equation}
similar to the DHO theory, with the $c^{3/2}$ term opposing the linear term. When truncated at linear order in the magnetic field $B$, our results agree exactly with a cluster-expansion calculation done in 1965 by Friedman~\cite{friedman1965calculation}, which we review in Appendix~\ref{FridmanComparison2}. While this represents an important step toward calculating electrolyte conductivity within the Dean-Kawasaki formalism, our treatment neglects both fluid advection and the interplay between the dielectric response and the magnetic field. Incorporating these effects would require supplementing the odd Dean-Kawasaki equation with additional hydrodynamic equations, as discussed in Sec.~\ref{Conclusion}.

Finally, beyond its focus on electrolytes, this paper provides detailed derivations of several results summarized in a companion Letter~\cite{avni2026odd}. Readers interested specifically in the derivation of odd SDFT are referred to Sec.~\ref{SecModel}, while those interested in the connection between odd SDFT and chiral active matter are referred to Appendices~\ref{chiral_active}--\ref{mobility_active}.

\section{Odd stochastic density functional theory}
\label{SecModel}

In the following, we derive a stochastic field-theoretical description of a single species of charged particles in a magnetic field. We generalize the formulation to multiple ionic species in the next section \ref{SecHall}. We begin with the single-species case both for simplicity and to elucidate the general structure of odd SDFT, which can also be applied to systems beyond electrolytes, including odd diffusive fluids and chiral active matter. A high-level schematic overview of the derivation and its relation to the existing literature is shown in Fig.~\ref{SchematicEquation}.

\subsection{Inertial interacting charged particles in a magnetic field} \label{fullLangevin}
We consider a system of $N$ particles of charge $q$ and mass $m$, immersed in a fluid at temperature $T$, and subjected to an external fixed magnetic field $\boldsymbol{B}=B\hat{z}$. We describe the motion of the particles by the Langevin equation
\begin{subequations}\label{Langevin1}
\begin{align} \label{Langevin1a}
\dot{\boldsymbol{x}}_{i}&=\boldsymbol{v}_{i}\\
 \label{Langevin1b}
m\dot{\boldsymbol{v}}_{i} & =-\frac{\partial U}{\partial\boldsymbol{x}_{i}}+q\boldsymbol{v}_{i}\times{\bf B}-\gamma\boldsymbol{v}_{i}+\bm{\xi}_{i}(t)
\end{align}
\end{subequations}
where $\boldsymbol{x}_i$ and $\boldsymbol{v}_i$ are the position and velocity, respectively, of the $i$'th particle ($i=1,...,N$), $\gamma$ is a friction coefficient, and $\bm{\xi}_{i}$ is a Gaussian white noise vector with zero mean, $\langle \boldsymbol{\xi}_{i}(t)\rangle =0$, and correlations
\begin{equation} \label{Langevin2}
\langle \xi_{i}^{\mu}(t)\xi_{j}^{\nu}(t')\rangle=2\gamma k_{B}T\delta_{ij}\delta_{\mu\nu}\delta(t-t').
\end{equation}
The potential $U$ includes both pairwise interactions between the particles and external potentials, and is given by
$U(\boldsymbol{x}_{1},...,\boldsymbol{x}_{N})=\frac{1}{2}\sum_{i\neq j=1}^{N}V\left(|\boldsymbol{x}_{i}-\boldsymbol{x}_{j}|\right)+\sum_{i}V_{{\rm ext}}\left(\boldsymbol{x}_{i}\right)
.$
Collecting terms that depend on $\bm v_i$, Eq.~\eqref{Langevin1b} can be written as
\begin{equation}
\label{single_species_eom_v}
m\dot{\boldsymbol{v}}_{i}=-\frac{\partial U}{\partial\boldsymbol{x}_{i}}-\mathsf{M}^{-1}\boldsymbol{v}_{i}+\boldsymbol{\xi}_{i}\left(t\right)
\end{equation} where $\mathsf{M}$ is the $3\times3$ matrix
\begin{equation} \label{G_matrix}
\mathsf{M}\equiv \frac{1}{\gamma} \left(\begin{array}{ccc}
1 & -\frac{qB}{\gamma} & 0\\
\frac{qB}{\gamma} & 1 & 0\\
0 & 0 & 1
\end{array}\right)^{-1}.
\end{equation}

The dimensionless parameter $\omega_c \tau=qB/\gamma$ appearing in $\mathsf{M}$ compares two time scales involving the mass of the particle: the velocity relaxation time scale $\tau$ and the inverse cyclotron frequency $\omega_c^{-1}$, respectively given by
\begin{equation} \label{timescales}
\tau \equiv \frac{m}{\gamma}
\qquad
\text{and}
\qquad
\omega_c^{-1} \equiv \frac{m}{qB}.
\end{equation}
This parameter measures how much the particle velocity is rotated by the magnetic field during one relaxation time (the same parameter controls the weak- and strong-field regimes of the Hall effect in the Drude model~\cite{ashcroft1976solid}). When $\omega_c \tau\ll 1$, friction relaxes the velocity before the magnetic field can substantially deflect it. This corresponds to the low $B$ regime where
\begin{equation} \label{MlinearB}
\mathsf{M}= \frac{1}{\gamma} \left(\begin{array}{ccc}
1 & \omega_c \tau & 0\\
-\omega_c \tau & 1 & 0\\
0 & 0 & 1
\end{array}\right) + \mathcal{O}(B^2).
\end{equation}
For $B=0$, this reduces to $\gamma^{-1}\mathsf{I}$.
When $\omega_c \tau\gg 1$, the particle undergoes many cyclotron rotations before its velocity is significantly relaxed. This corresponds to the high $B$ regime where
\begin{equation} \label{MlimithighB}
\mathsf{M}= \frac{1}{\gamma} \left(\begin{array}{ccc}
(\omega_c \tau)^{-2} & (\omega_c \tau)^{-1} & 0\\
-(\omega_c \tau)^{-1} & (\omega_c \tau)^{-2} & 0\\
0 & 0 & 1
\end{array}\right) + \mathcal{O}(B^{-3}),
\end{equation}
reducing to $M_{\mu\nu}=\gamma^{-1}\delta_{\mu 3}\delta_{\nu 3}$ in the $B\to \infty$ limit. Hence, a strong magnetic field suppresses the motion perpendicular to $\bm B$, while leaving the parallel motion unaffected.

\textit{Conventions---} Throughout the text, Greek indices $\mu,\nu,\rho,\sigma$ label spatial components, e.g. $\mu=1,2,3$, whereas Greek indices $\alpha,\beta,\gamma,\delta$ label particle species, e.g. positively and negatively charged ions (relevant in Sec.~\ref{SecHall}). Both types of indices may appear as either subscripts or superscripts, with no distinction implied by their position. Repeated spatial indices are summed over according to the Einstein summation convention. Species indices, in contrast, are not summed over unless the summation is written explicitly. Vectors are denoted by bold symbols and matrices by sans-serif symbols. Their components are written in ordinary italic font, e.g. $\bm{x}$ becomes $x_{\mu}$ and $\mathsf{M}$ becomes $M_{\mu\nu}$. The identity matrix is denoted by $\mathsf{I}$, with components $I_{\mu\nu}=\delta_{\mu\nu}$, and the superscript ${}^{\mathsf{T}}$ denotes matrix transposition.

\subsection{The low mass limit of the Langevin equation}

In view of the two time scales in Eq.~\eqref{timescales}, taking $m\to0$ is equivalent to saying that both $\tau$ and $\omega_c^{-1}$ are small compared to any other relevant time scale (e.g. given by $U$), yet $\omega_c\tau$ is held fixed. To assess whether this assumption is fulfilled, consider as an example standard salts such as $\rm NaCl$ in water at room temperature. 
In this case, $\tau\sim 10^{-2}\,\mathrm{ps}$ and $\omega_c^{-1} \sim 10^5\,\mathrm{ps}/B[\mathrm{T}]$~\cite{haynes2016crc}. 
The Debye time $\tau_{D}=\lambda_D^2 \gamma/k_B T$ (where $\lambda_D$ is the screening length) gives an order of magnitude of the timescales associated to electrostatic interactions in the solution~\cite{bazant2004diffuse}.
Since $\tau_{D}\sim 100 \,\mathrm{ps} / {c}[\mathrm{M}]$, $\tau\ll \tau_{D}$ is easily satisfied in most cases, yet $\omega_c^{-1}\ll \tau_{D}$ is restricted to very dilute solutions or strong magnetic fields. 
When the system is not in the low mass limit, one should retain both Eqs.~\eqref{Langevin1a} and \eqref{Langevin1b}. 
In contrast, in the low mass limit, we can eliminate $\bm{v_i}$ to get an equation on $\bm x_i$ only, simplifying the subsequent calculations.

\begin{figure}
\centering
{\includegraphics[width=0.48\textwidth,draft=false]{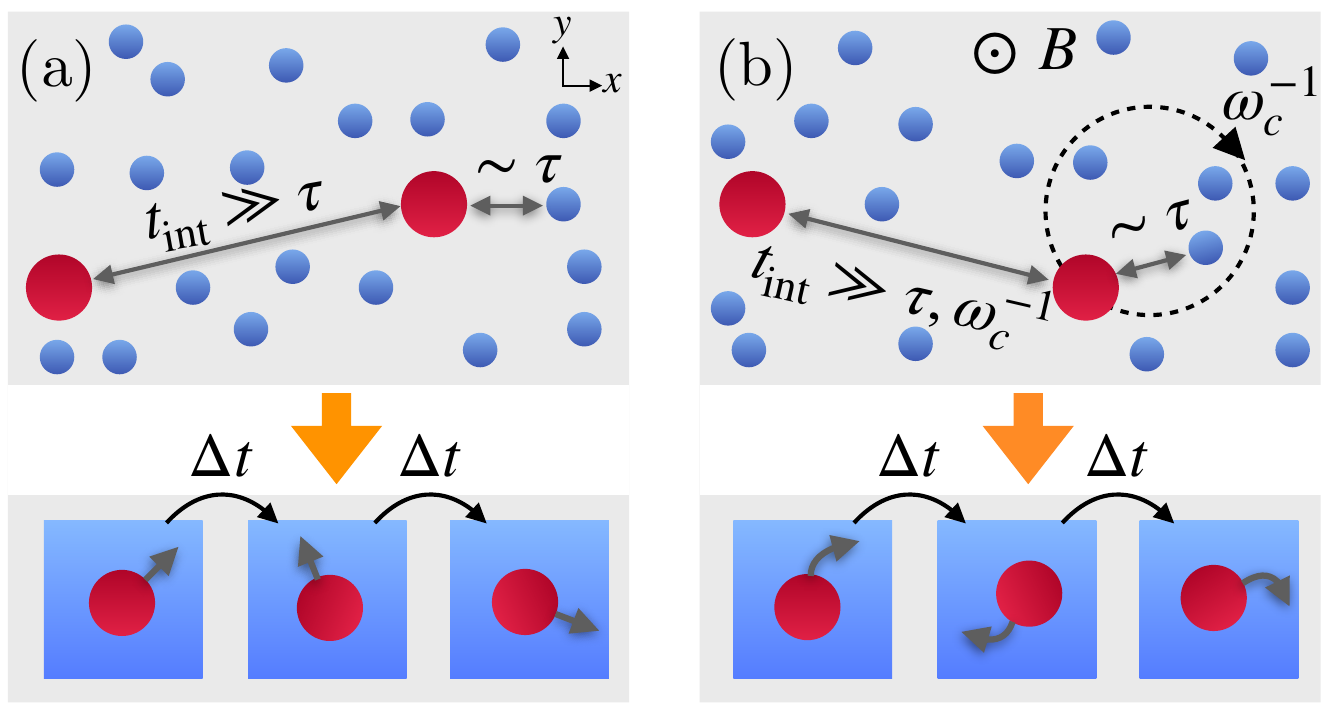}}
\caption{\textbf{Schematic drawing of the low mass limit with explicit and implicit solvent.} (a) A particle interacting with solvent molecules (top) with characteristic relaxation time $\tau$ much smaller than any other relevant time scale can be described by an effective bath (bottom) with random kicks represented by white noise. (b) A magnetic field introduces an additional time scale: the period of the cyclotron orbit, $~\sim \omega_c^{-1}$. If $\tau$ and $\omega_c^{-1}$ are both much smaller than any other relevant time scale, but their ratio is finite, the particle's bath has nonwhite noise with traces of chirality (bottom). \label{SchematicSystem}}
\end{figure}

Intuitively, the low mass limit means that while the time over which a single particle loses its momentum is very small, so is the time it takes this particle to complete a full cyclotron orbit without the solvent (Fig.~\ref{SchematicSystem}). Hence, on average, the particle performs a finite fraction of the circular motion before resetting (the exact fraction depends on the ratio $\tau/\tau_{c}$). Finally, note that taking the high $\gamma$ limit, rather than the low mass limit, would yield $\omega_c\tau\to0$ and $\mathsf{M}\to\gamma^{-1}\mathsf{I}$, eliminating any signature of the magnetic field. Since we are interested in the effects of the magnetic field, the low mass limit is therefore the appropriate one here.

\begin{figure*}
\centering
{\includegraphics[width=1\textwidth,draft=false]{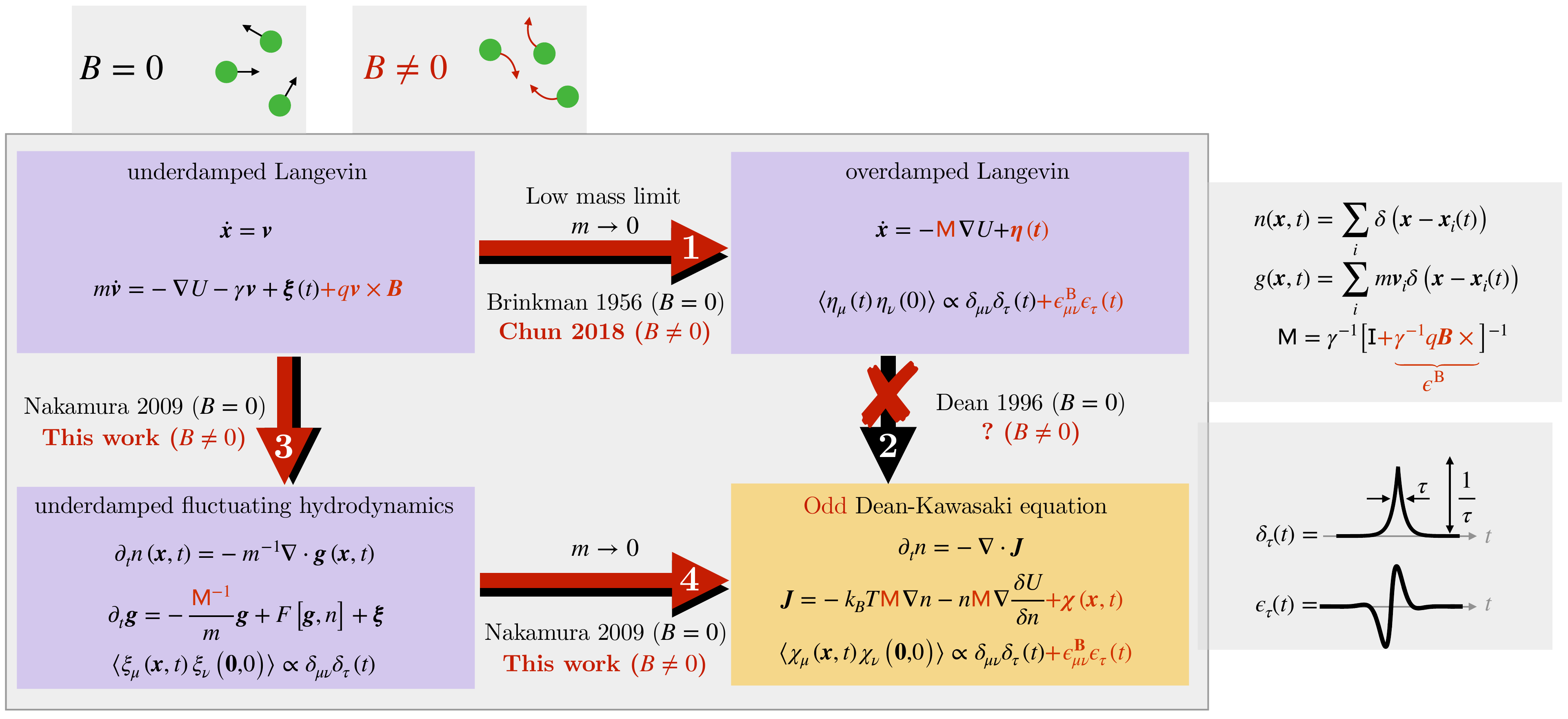}}
\caption{\textbf{Schematic summary of the derivation of the standard and odd Dean-Kawasaki equations and their relation to the existing literature.} Starting from inertial Langevin dynamics, one may take the low-mass limit (arrow 1) to obtain first-order Langevin dynamics~\cite{brinkman1956brownian}. For a system of multiple particles, this first-order dynamics can then be expressed in terms of a single density field (arrow 2), yielding the standard Dean-Kawasaki equation~\cite{dean1996langevin}. Alternatively, one may start from the second-order dynamics of multiple particles and derive coupled equations for the number-density and momentum-density fields (arrow 3), before taking the low-mass limit (arrow 4) to recover the same Dean-Kawasaki equation~\cite{nakamura2009derivation}. In the presence of a magnetic field (red arrows and text), arrow 1 instead leads to a Langevin equation with chiral, temporally correlated noise~\cite{chun2018emergence}, making the direct extension of arrow 2 nontrivial. In this work, we therefore generalize the route corresponding to arrows 3 and 4 to the case of a magnetic field and obtain the odd Dean-Kawasaki equation with chiral noise.}
\label{SchematicEquation}
\end{figure*}

At first sight, one might expect the low-mass limit of Eq.~\eqref{Langevin1} to be obtained simply by neglecting the inertial term. However, as shown in Ref.~\cite{chun2018emergence}, this limit is more subtle (see Fig.~\ref{SchematicEquation}). In the low-mass limit, Eq.~\eqref{Langevin1} reduces to
\begin{equation}
\label{Chun_result1}\dot{\boldsymbol{x}}_{i}=-\mathsf{M}\frac{\partial U}{\partial\boldsymbol{x}_{i}}+\boldsymbol{\eta}_{i}\left(t\right)
\end{equation}
with $\bm \eta_i$ being a nonwhite Gaussian vector with zero mean and correlations
\begin{subequations}
\label{Chun_result2}
\begin{align}\langle\boldsymbol{\eta}_{i}(t)\boldsymbol{\eta}_{j}^{\mathsf{T}}(t')\rangle & =\delta_{ij}\mathsf{C}(t-t')\\
\intertext{with}\label{Climit}\!\!\mathsf{C}(u)\equiv\lim_{m\to0}\frac{k_{B}T}{m}e^{-\frac{\gamma|u|}{m}} & \begin{pmatrix}\cos(\frac{qB}{m}u) & \sin(\frac{qB}{m}u) & 0\\[0.1cm]
-\sin(\frac{qB}{m}u) & \cos(\frac{qB}{m}u) & 0\\[0.1cm]
0 & 0 & 1
\end{pmatrix}\\
=\lim_{\substack{\tau\to0\\
\omega_{c}\tau=\mathrm{const}
}
}\frac{k_{B}T}{\gamma\tau}e^{-\frac{|u|}{\tau}} & \begin{pmatrix}\cos(\omega_{c}u) & \sin(\omega_{c}u) & 0\\[0.1cm]
-\sin(\omega_{c}u) & \cos(\omega_{c}u) & 0\\[0.1cm]
0 & 0 & 1
\end{pmatrix}
\end{align}\end{subequations}
The limit in Eq.~\eqref{Climit} can be expressed as~\cite{chun2018emergence}
\begin{equation} \label{Chun_result3}
\mathsf{C}(u)=k_B T\left[\mathsf{M}\delta_{+}\left(u\right)+\mathsf{M}^{\mathsf{T}}\delta_{-}\left(u\right)\right]
\end{equation}
where $\delta_{+}(u)$ and $\delta_{-}(u)$ are variants of the Dirac delta function restricted to the nonnegative and nonpositive real lines, respectively, so they vanish when $u\neq 0$ and satisfy $\int_{0}^{\infty}\delta_{+}(u){\rm d}u=\int_{-\infty}^{0}\delta_{-}(u){\rm d}u=1
$ and $\int_{0}^{\infty}\delta_{-}(u){\rm d}u=\int_{-\infty}^{0}\delta_{+}(u){\rm d}u=0
$.

Our goal in the next two subsections is to find a density-field description of Eqs.~(\ref{Chun_result1}--\ref{Chun_result2}), analogous to the Dean-Kawasaki equation, which is a density-field description of the standard overdamped Langevin equation with white noise. 

While we focus on charged particles under magnetic field, Eq.~\eqref{Langevin1} has been used to model odd-diffusive fluids~\cite{abdoli2026dynamical,kalz2022collisions,kalz2026reversal}, and therefore the following derivation applies to such systems as well in the low mass limit.
Moreover, Eq.~\eqref{Chun_result1} describes chiral active particles~\cite{caprini2023chiral} under certain limits and mappings specified in Appendix~\ref{chiral_active}. Hence, the odd Dean-Kawasaki equation we derive can also be extended to describe such models (see Appendix~\ref{chiral_active} for details).

\subsection{Derivation of a fluctuating hydrodynamic equation from the underdamped Langevin equation}

The standard derivation of the Dean-Kawasaki equation starts from the overdamped Langevin equation~\cite{dean1996langevin}. However, it relies on Ito calculus, applicable for white noise only. Since in our case the the low-mass limit has a nonwhite noise (Eq.~\eqref{Chun_result2}), it is not obvious how to  follow the same path of the original derivation.
Instead, we derive a fluctuating hydrodynamic equation for the number density field and momentum density field, using a formalism introduced in Ref.~\cite{nakamura2009derivation}, and take the low-mass limit at the density field level (see Fig.~\ref{SchematicEquation}).

The number density field is defined as
\begin{equation} \label{density_def}
n(\boldsymbol{x},t)\equiv\sum_{i}\delta(\boldsymbol{x}-\boldsymbol{x}_{i}(t))
\end{equation}
and the momentum density field as
\begin{equation}\label{momentum_density_def}
\boldsymbol{g}(\boldsymbol{x},t)\equiv \sum_{i}m\boldsymbol{v}_{i}(t)\delta(\boldsymbol{x}-\boldsymbol{x}_{i}(t)).
\end{equation}
Taking the time derivative of Eq.~\eqref{density_def} and using Eq.~\eqref{Langevin1}, we obtain
\begin{align} \label{n_equation}
\partial_{t}n(\boldsymbol{x},t) & =-\sum_{i}\boldsymbol{v}_{i}(t)\cdot\boldsymbol{\nabla}\delta(\boldsymbol{x}-\boldsymbol{x}_{i}(t))\nonumber \\
 & =-\frac{1}{m}\boldsymbol{\nabla}\cdot\boldsymbol{g}(\boldsymbol{x},t)
\end{align}
Similarly, taking the time derivative of Eq.~\eqref{momentum_density_def} we obtain
\begin{align}\partial_{t}\boldsymbol{g}(\boldsymbol{x},t) & =m\sum_{i}\dot{\boldsymbol{v}}_{i}\delta(\boldsymbol{x}-\boldsymbol{x}_{i}(t))\\
 & -m\sum_{i}\boldsymbol{v}_{i}(\dot{\boldsymbol{x}}_{i}(t)\cdot\boldsymbol{\nabla}\delta(\boldsymbol{x}-\boldsymbol{x}_{i}(t)))\nonumber\\
 & =-\frac{1}{m}\mathsf{M}^{-1}\boldsymbol{g}-n\nabla\frac{\delta U}{\delta n}-\boldsymbol{\nabla}\cdot\mathsf{G}+\boldsymbol{\xi}\nonumber
\end{align}
where, in the last step, we used the relation
${\sum_{i}\frac{\partial U}{\partial\boldsymbol{x}_{i}}\delta(\boldsymbol{x}-\boldsymbol{x}_{i}(t))=n(\boldsymbol{x},t)\nabla\frac{\delta U}{\delta n(\boldsymbol{x},t)}}$
that arise from the expression
\begin{equation*}
    {U=\frac{1}{2}\int{\rm d}\boldsymbol{x}{\rm d}\boldsymbol{x}'n(\boldsymbol{x})V(\boldsymbol{x}-\boldsymbol{x}')n(\boldsymbol{x}')+\int n(\bm{x})V_{{\rm ext}}(\bm{x})} \dd\bm{x}
\end{equation*}
of the potential, and defined
\begin{subequations}
\begin{align}
\boldsymbol{\xi}(\boldsymbol{x},t)\equiv & \sum_{i}\boldsymbol{\xi}_{i}(t)\delta(\boldsymbol{x}-\boldsymbol{x}_{i}(t))\\
\mathsf{G}(\boldsymbol{x},t)\equiv\, & m\sum_{i}\boldsymbol{v}_{i}(t)\boldsymbol{v}^{\mathsf T}_{i}(t)\delta(\boldsymbol{x}-\boldsymbol{x}_{i}(t)).
\end{align}
\end{subequations}
Using Eq.~\eqref{Langevin2} and following the same argument as the original derivation by Dean~\cite{dean1996langevin}, namely that $\delta(\bm{x}-\bm{x}_{i})\delta(\bm{x}'-\bm{x}_{i})=\delta(\bm{x}-\bm{x}')\delta(\bm{x}-\bm{x}_{i})
$, we find the new noise correlations to be
\begin{align}\label{noise2_corr}
\!\!\!\!\langle\boldsymbol{\xi}(\boldsymbol{x},t)\boldsymbol{\xi}^{\mathsf{T}}(\boldsymbol{x}',t')\rangle & =2\gamma k_{B}Tn(\boldsymbol{x},t)\delta(\boldsymbol{x}-\boldsymbol{x}')\delta(t-t')\mathsf{I}\end{align}
Following Ref.~\cite{nakamura2009derivation}, we assume that
\begin{equation} \label{no_overlap_identity}
\delta(\boldsymbol{x}_{i}(t)-\boldsymbol{x}_{j}(t))=\delta(\boldsymbol{x}_{i}(t)-\boldsymbol{x}_{j}(t))\delta_{ij}
\end{equation}
which is valid if the particles do not overlap (this is justified if the potential has a repulsive core but also in the dilute limit or simply by considering point-like particles in three dimensions). 
We then insert a representation of unity in terms of delta functions, which yields
\begin{align}G^{\mu\nu} & =m\sum_{i}v_{i}^{\mu}(t)v_{i}^{\nu}(t)\delta(\boldsymbol{x}-\boldsymbol{x}_{i}(t))\frac{\sum\limits _{j}\delta(\boldsymbol{x}-\boldsymbol{x}_{j}(t))}{\sum\limits _{k}\delta(\boldsymbol{x}-\boldsymbol{x}_{k}(t))}\nonumber\\
 & =\frac{m}{n}\sum_{i}v_{i}^{\mu}(t)\delta(\boldsymbol{x}-\boldsymbol{x}_{i}(t))\sum_{j}v_{j}^{\nu}(t)\delta(\boldsymbol{x}-\boldsymbol{x}_{j}(t))\nonumber\\
 & =\frac{g_{\nu}(\boldsymbol{x},t)g_{\nu}(\boldsymbol{x},t)}{m\,n(\boldsymbol{x},t)}.
\end{align}
In going from the first line to the second, we used the definition of $n$, Eq.~\eqref{density_def}, together with the relation~\eqref{no_overlap_identity}; from the second line to the third, we used the definition of $g$, Eq.~\eqref{momentum_density_def}.
Finally, we find that the evolution of the momentum density field is given by
\begin{equation} 
\label{g_equation}
\partial_{t}g_{\mu}(\boldsymbol{x},t)=-\frac{(\mathsf{M}^{-1})_{\mu\nu}g_{\nu}}{m}-n\partial_{\mu}\frac{\delta U}{\delta n}-\partial_{\nu}\frac{g_{\mu}g_{\nu}}{m\,n}+\xi_{\mu}.
\end{equation}

\subsection{Derivation of the odd Dean-Kawasaki equation from the low mass limit} \label{ODK_derivation}

Next, we eliminate $\bm g$ by taking the low mass limit. In the following derivation it will be useful to work with the dimensionless matrix
\begin{equation}
\mathsf{\Gamma}\equiv\gamma^{-1}\mathsf{M}^{-1}=\left(\begin{array}{ccc}
1 & -\frac{qB}{\gamma} & 0\\
\frac{qB}{\gamma} & 1 & 0\\
0 & 0 & 1
\end{array}\right).
\end{equation}
The use of $\mathsf{\Gamma}$ instead of $\mathsf{M}$ is restricted to this  subsection.
Equation~\eqref{g_equation} can be formally solved as
\begin{equation} \label{g_decomposed}
\frac{1}{\tau}\boldsymbol{g}(\boldsymbol{x},t)=\boldsymbol{\Upsilon}+\boldsymbol{\Xi}+\boldsymbol{\Pi}
\end{equation}
where
\begin{subequations}
\begin{align}\label{Upsilon_{d}ef}\boldsymbol{\Upsilon}(\boldsymbol{x},t) & \equiv-\frac{1}{\tau}\int\limits _{-\infty}^{t}{\rm d}se^{-\frac{\mathsf{\Gamma}(t-s)}{\tau}}n(\boldsymbol{x},s)\boldsymbol{\nabla}\frac{\delta U}{\delta n(\boldsymbol{x},s)}\\
\label{Xi_def}\boldsymbol{\Xi}(\boldsymbol{x},t) & \equiv\frac{1}{\tau}\int\limits _{-\infty}^{t}{\rm d}se^{-\frac{\mathsf{\Gamma}(t-s)}{\tau}}\boldsymbol{\xi}(\boldsymbol{x},s)\\
\label{Pi_def}\boldsymbol{\Pi}(\boldsymbol{x},t) & \equiv-\frac{1}{\tau^{2}\gamma}\int\limits _{-\infty}^{t}{\rm d}se^{-\frac{\mathsf{\Gamma}(t-s)}{\tau}}\boldsymbol{\nabla}\cdot\frac{\boldsymbol{g}(\boldsymbol{x},s)\boldsymbol{g}^{\mathsf{T}}(\boldsymbol{x},s)}{n(\boldsymbol{x},s)}
\end{align}
\end{subequations}
and the matrix elements of $e^{-\mathsf{\Gamma}u/\tau}$ are
\begin{equation} \label{ElementsGamma}
e^{-\mathsf{\Gamma}u/\tau}=e^{-u/\tau}\left(\begin{array}{ccc}
\cos\left(\omega_c u\right) & \sin\left(\omega_c u\right) & 0\\
-\sin\left(\omega_c u\right) & \cos\left(\omega_c u\right) & 0\\
0 & 0 & 1
\end{array}\right).
\end{equation}
This is an implicit solution because $\bm{\Pi}$ depends on $\bm{g}$.

The new noise, $\bm \Xi$, is a Gaussian noise with zero mean, and substituting Eq.~\eqref{noise2_corr} its correlations are,
\begin{align}\langle\bm{\Xi}(\boldsymbol{x},t)\boldsymbol{\Xi}^{\mathsf{T}}(\boldsymbol{x}',t')\rangle & =\frac{2\gamma k_{B}T\delta(\boldsymbol{x}-\boldsymbol{x}')}{\tau^{2}}\\
 & \times\int\limits _{-\infty}^{t_<}{\rm d}s\,e^{-\mathsf{\Gamma}\frac{(t-s)}{\tau}}n(\boldsymbol{x},s)e^{-\mathsf{\Gamma}^{\mathsf{T}}\frac{(t'-s)}{\tau}}\nonumber
\end{align}
where $t_<\equiv {\min(t,t')}$.
Using trigonometric identities and the relation ${t+t'-2s=\left|t-t'\right|+2(t_<-s)
}$ we obtain
\begin{align}\label{correlationfinitem}\langle\bm{\Xi}(\boldsymbol{x},t)\boldsymbol{\Xi}^{\mathsf{T}}(\boldsymbol{x}',t')\rangle & =\frac{2\gamma^{2}}{\tau}\delta(\boldsymbol{x}-\boldsymbol{x}')\mathsf{C}_{m}(t-t')\nonumber\\
 & \times\int\limits _{-\infty}^{t_{<}}{\rm d}s\,e^{-\frac{2(t_{<}-s)}{\tau}}n(\boldsymbol{x},s)
\end{align}
where 
\begin{equation} \label{C_def2}
\mathsf{C}_m(u)\equiv\frac{k_{B}T}{\gamma\tau}e^{-\left|u\right|/\tau}\left(\begin{array}{ccc}
\cos\left(\omega_c u\right) & \sin\left(\omega_c u\right) & 0\\
-\sin\left(\omega_c u\right) & \cos\left(\omega_c u\right) & 0\\
0 & 0 & 1
\end{array}\right).
\end{equation}
We now use the fact that $n(\boldsymbol{x},t)$ varies only weakly over the fast time scales $\tau$ and $\omega_c^{-1}$
(unlike $\boldsymbol{g}(\boldsymbol{x},t)$). Hence, in the low mass limit, $\frac{1}{\tau}\int\limits _{-\infty}^{t_{<}}{\rm d}s\,e^{-\frac{2(t_{<}-s)}{\tau}}n(\boldsymbol{x},s)\to\frac{1}{2}n(\boldsymbol{x},t_{<})
$. Moreover, as was proven in~Ref.~\cite{chun2018emergence}, in the low mass limit $\lim_{m\to0}\mathsf{C}_{m}(u)=\mathsf{C}(u)
$ (with $\mathsf C$ given by Eq.~\eqref{Chun_result3}), so we obtain
\begin{equation} 
\label{XiXilimit}
\!\!\lim_{m\to0}\langle\boldsymbol{\Xi}(\boldsymbol{x},t)\boldsymbol{\Xi}^{\mathsf{T}}(\boldsymbol{x}',t')\rangle=\gamma^{2}n(\boldsymbol{x},t)\delta(\boldsymbol{x}-\boldsymbol{x}')\mathsf{C}(t-t').
\end{equation}
Note that we have assumed Ito prescription. For a different prescription, one has to integrate Eq.~\eqref{n_equation} over a time $\Delta t\gg \tau, \omega_c^{-1}$ and evaluate the multiplicative part of the noise that depends on $n$ at $t^{*}$
where $t<t^{*}<t+\Delta t$ and $\Delta t$
(see Ref.~\cite{kuroiwa2013brownian} for details).

To evaluate $\bm \Upsilon$, we notice that for $u>0$, $e^{-\mathsf{\Gamma}u/\tau}=\frac{\gamma \tau}{k_B T}\mathsf{C}_m(u)$ (compare Eq.~\eqref{ElementsGamma} and~\eqref{C_def2}). Hence, for a function $h(t)$ that varies on time scale much larger than both $\tau$ and $\omega_c^{-1}$,
\begin{align}\nonumber\lim_{m\to0}\frac{1}{\tau}\int_{-\infty}^{t}\!\!\!{\rm d}s\,e^{-\frac{\mathsf{\Gamma}(t-s)}{\tau}}h(s) & =\frac{\gamma}{k_{B}T}\int_{-\infty}^{t}\!\!\!{\rm d}s\,\mathsf{C}(t-s)h(s)\\
 & =h(t)\mathsf{\Gamma}^{-1},
\end{align}
hence we obtain
\begin{equation} \label{final_Upsilon}
\lim_{m\to0}\boldsymbol{\Upsilon}(\boldsymbol{x},t)=-n(\boldsymbol{x},t)\mathsf{\Gamma}^{-1}\boldsymbol{\nabla}\frac{\delta U}{\delta n(\boldsymbol{x},t)}.
\end{equation}

We are left with evaluating Eq.~\eqref{Pi_def} in the low-mass limit. As is pointed out in Ref.~\cite{nakamura2009derivation}, the lowest order contribution to $\boldsymbol{g}$ in powers of $\tau$ (or $m$), as given in Eq.~\eqref{g_decomposed}, is
\begin{equation}
\boldsymbol{g}(\boldsymbol{x},s)\approx\tau\boldsymbol{\Xi}.
\end{equation}
Indeed, $\boldsymbol{\Xi}(\boldsymbol{x},t)\sim \tau^{-1/2}$, as follows from $\langle\boldsymbol{\Xi}(\boldsymbol{x},t)\boldsymbol{\Xi}^{\mathsf T}(\boldsymbol{x},t)\rangle\sim \tau^{-1}$ (see Eq.~\eqref{C_def2}), whereas $\boldsymbol{\Upsilon}(\boldsymbol{x},t)\sim \tau^{0}$. As we show below, the remaining contribution is also self-consistently of order $\boldsymbol{\Pi}(\boldsymbol{x},t)\sim\tau^0$. Hence, we can approximate $\boldsymbol{\Pi}$ to leading order in $\tau$ as
\begin{equation} \label{PiApprox}
\boldsymbol{\Pi}(\boldsymbol{x},t)\approx-\frac{1}{\gamma}\int\limits _{-\infty}^{t}{\rm d}se^{-\frac{\mathsf{\Gamma}(t-s)}{\tau}}\boldsymbol{\nabla}\cdot\frac{\boldsymbol{\Xi}(\boldsymbol{x},s)\boldsymbol{\Xi}^{\mathsf T}(\boldsymbol{x},s)}{n(\boldsymbol{x},s)}.
\end{equation}
Note that the right hand side of Eq.~\eqref{PiApprox} is stochastic. However, it suffices to retain only its average (deterministic) contribution, since the fluctuating part is subdominant to $\bm \Xi$ (when integrated over a time interval $\Delta t$ it scales as $\Delta t$, while $\Xi$ scales with $\Delta t^{1/2}$). 
Following Eq.~\eqref{correlationfinitem}, the average of $\bm{\Xi}(\boldsymbol{x},s)\boldsymbol{\Xi}^{\mathsf{T}}(\boldsymbol{x},s)$ is
\begin{align}\langle\bm{\Xi}(\boldsymbol{x},s)\boldsymbol{\Xi}^{\mathsf{T}}(\boldsymbol{x},s)\rangle & =\frac{2\gamma k_{B}T}{\tau^{2}}\mathsf{I}\delta(\boldsymbol{x}-\boldsymbol{x})\nonumber\\
 & \times\int\limits _{-\infty}^{s}{\rm d}s'\,e^{-\frac{2(s-s')}{\tau}}n(\boldsymbol{x},s')
\end{align}
where we used $\mathsf{C}_m(0)=k_B T/(\gamma \tau)$ (see Eq.~\eqref{C_def2}).
While this expression appears problematic because of the $\delta(\boldsymbol{x}-\boldsymbol{x})$, we follow Ref.~\cite{nakamura2009derivation} and use the relation $n(\boldsymbol{x},t)^{2}=\delta(\boldsymbol{x}-\boldsymbol{x})n(\boldsymbol{x},t)
$ derived from substituting the definition in Eq.~\eqref{density_def} and using Eq.~\eqref{no_overlap_identity} (for more rigorous justification see Ref.~\cite{nakamura2009derivation}).
We obtain, to leading order in $\tau$,
\begin{align}
\boldsymbol{\Pi}(\boldsymbol{x},t) & =-\frac{2k_{B}T}{\tau^{2}}\boldsymbol{\nabla}\int\limits _{-\infty}^{t}{\rm d}se^{-\frac{\mathsf{\Gamma}(t-s)}{\tau}}\nonumber\\[-5pt]
 & \times\frac{1}{n(\boldsymbol{x},s)}\int\limits _{-\infty}^{s}{\rm d}s'\,e^{-\frac{2(s-s')}{\tau}}n(\boldsymbol{x},s')^{2}
\end{align}
which, in the $m\to0$ limit, becomes
\begin{equation} \label{final_Pi}
\lim_{m\to0}\boldsymbol{\Pi}(\boldsymbol{x},t)=-k_{B}T\mathsf{\Gamma}^{-1}\boldsymbol{\nabla}n(\boldsymbol{x},t).
\end{equation}
Inserting the results of Eqs.~\eqref{XiXilimit}, \eqref{final_Upsilon} and~\eqref{final_Pi} into Eq.~\eqref{g_decomposed} we obtain
\begin{subequations}\label{EOM}
\begin{align}\label{EOM1}\partial_{t}n(\boldsymbol{x},t) & =-\boldsymbol{\nabla}\cdot\boldsymbol{J}\\
\label{EOM2}\boldsymbol{J}  =-k_{B}&T\mathsf{M}\boldsymbol{\nabla}n-n\mathsf{M}\boldsymbol{\nabla}\frac{\delta U}{\delta n}+\boldsymbol{\chi}(\boldsymbol{x},t)\\ \label{correlationsChi}
\langle\boldsymbol{\chi}(\boldsymbol{x},t)\boldsymbol{\chi}&(\boldsymbol{x}',t')^{\mathsf{T}}\rangle  =  n(\boldsymbol{x},t)\delta(\boldsymbol{x}-\boldsymbol{x}')\mathsf{C}(t-t')
\end{align}
\end{subequations}
where $\bm J(\bm x,t)=\bm g(\bm x,t)/m$ is the density current and we defined the rescaled Gaussian noise $\boldsymbol{\chi}\equiv \boldsymbol{\Xi}/\gamma$ with $\langle\bm \chi\rangle=0$.  Equation~\eqref{EOM} constitutes the odd Dean-Kawasaki equation for particles under a Lorentz force. A modified version for chiral active particles is presented in Appendix~\ref{chiral_active}.

\subsection{Mobility}
 \label{GK_nonint}
We conclude this section by demonstrating how the odd Dean-Kawasaki equation can be used to calculate a transport coefficient in a simple setup. 
Readers interested in the Hall conductivity can directly skip to Sec.~\ref{SecHall}.
Consider a collection of particles subject to an external force that is fixed and uniform, $\bm F=F_0\hat{n}$.
The system's response can be characterized by the
collective mobility tensor $\mathsf{L}$, which relates the average density current to the external force through
\begin{equation} \label{LF}
\langle\boldsymbol{J}\rangle=\mathsf{L}\,\boldsymbol{F}.
\end{equation}
We can derive $\mathsf{L}$ from the odd Dean-Kawasaki equation using either direct response or Green-Kubo relation, as described below. For simplicity, we first consider the case of noninteracting particles ($V=0$); the more complicated case with $V\neq0$, together with a generalization of Eq.~\eqref{LF} to a spatiotemporal force, is treated in Appendix~\ref{CollectiveMobilitySec}. Additionally, the collective mobility of chiral active particles is derived from a modified odd Dean-Kawasaki equation in Appendix~\ref{mobility_active}. We show that, in this case, the Green-Kubo relation is no longer valid.

\subsubsection{Odd mobility from direct response}
In the direct response approach, we substitute $-\nabla(\delta U/\delta n)=\boldsymbol{F}$ into Eq.~\eqref{EOM2} and then take the ensemble average. Since the system is homogeneous, $\langle n(\boldsymbol{x},t)\rangle= {c}$, where $c$ denotes the mean concentration, and as the particles are noninteracting, we obtain
\begin{equation} \label{M_linear}
\mathsf{L}={c}\mathsf{M}=\frac{{c}}{\gamma\left(1+\frac{q^{2}B^{2}}{\gamma^{2}}\right)}\left(\begin{array}{ccc}
1 & \frac{qB}{\gamma} & 0\\
-\frac{qB}{\gamma} & 1 & 0\\
0 & 0 & 1+\frac{q^{2}B^{2}}{\gamma^{2}}
\end{array}\right).
\end{equation}
The derivation of Eq.~\eqref{M_linear} via the direct-response approach is not affected by the non-white noise term in the odd Dean-Kawasaki equation; the same remains true in the interacting case, as shown in Appendix~\ref{CollectiveMobilitySec}. By contrast, the Green-Kubo derivation of $\mathsf{L}$ is sensitive to this noise term, as shown below.

\subsubsection{Odd mobility from Green-Kubo relation} \label{mobilityGKNoInt}
Alternatively to the direct response approach, response functions can be calculated from correlations of current fluctuations in the unperturbed system from fluctuation-dissipation relations~\cite{risken1989fokker}. In the present context, these manifest as the Green-Kubo relation
\begin{equation} \label{GK_M}
    \mathsf{L}=\frac{1}{\mathcal{V}k_{B}T}
    \int\limits_{0}^{\infty}\dd t
    \int\dd\bm{x}
    \int\dd\bm{x'}
    \langle\bm{J}(\bm{x},t)\bm{J}(\bm{x'},0)^{\mathsf{T}}\rangle_{0}
\end{equation}
where $\mathcal{V}$ is the system's volume and $\langle\cdot\rangle_0$ denotes ensemble average over the unperturbed system, i.e., at $F_0=0$.  
It is not entirely obvious that Eq.~\eqref{GK_M} holds for systems under Lorentz force because the external magnetic field breaks time-reversal symmetry.
Indeed, Green-Kubo relations can be violated in nonequilibrium settings, see Refs.~\cite{chun2021nonequilibrium,han2021fluctuating} and references therein for discussions.
It turns out that Eq.~\eqref{GK_M} does hold in the present context, as we show in Appendix~\ref{GK_proof}, see also Refs.~\cite{Martin1967,Corkum1974,Pavliotis2010,Gaspard2013,Shalchi2011,faedi2026velocityforceautocorrelationsbrownian} for discussions on Green-Kubo relations in magnetic systems.

In the unperturbed system, assuming no interactions, the density current is simply
\begin{equation} \label{current_no_force}
\boldsymbol{J}(\boldsymbol{x},t)=-k_{B}T\mathsf{M}\boldsymbol{\nabla}n(\boldsymbol{x},t)+\boldsymbol{\chi}(\boldsymbol{x},t).
\end{equation}

The gradient term in Eq.~\eqref{current_no_force} does not contribute to the integral in Eq.~\eqref{GK_M} other than a boundary term taken to zero. Hence, Eq.~\eqref{GK_M} simplifies into
\begin{align}\label{M_GK}\nonumber\mathsf{L} & =\frac{1}{\mathcal{V}k_{B}T}\int\limits _{0}^{\infty}{\rm d}t\int{\rm d}\boldsymbol{x}\int{\rm d}\boldsymbol{x}'\langle\boldsymbol{\chi}(\boldsymbol{x},t)\boldsymbol{\chi}(\boldsymbol{x}',0)^{\mathsf{T}}\rangle\\
 & =\frac{1}{\mathcal{V}k_B T}\int\limits _{0}^{\infty}{\rm d}t\int{\rm d}\boldsymbol{x}\int{\rm d}\boldsymbol{x}'n(\boldsymbol{x},t)\delta(\boldsymbol{x}-\boldsymbol{x}')\mathsf{C}(t)
\end{align}
where we inserted Eq.~\eqref{correlationsChi} in the second equality.
Using the spatial average $\frac{1}{\mathcal{V}}\int{\rm d}\boldsymbol{x}\,n(\boldsymbol{x},t)={c}$, and the relation
\begin{equation}
    \int\limits _{0}^{\infty}{\rm d}t\,\mathsf{C}(t-t')= k_B T\,\mathsf{M}
\end{equation}
that follows Eq.~\eqref{Chun_result3}, we obtain $\mathsf{L}={c}\mathsf{M}$, which is the same as the direct response result, Eq.~\eqref{M_linear}. 
Moreover, defining $H_{\mu\nu}(t)\equiv\int{\rm d}\boldsymbol{x}\int{\rm d}\boldsymbol{x}'\langle\chi_{\mu}(\boldsymbol{x},t)\chi_{\nu}(\boldsymbol{x}',0)\rangle$ and invoking time-translational invariance at steady state, the collective mobility calculated in Eq.~\eqref{M_GK} is shown to satisfy
\begin{equation} \label{Msymmetric}
\mathsf{L}-\mathsf{L}^{\mathsf{T}}=\frac{1}{\mathcal{V}k_{B}T}\int\limits _{0}^{\infty}{\rm d}t [\mathsf{H}(t)-\mathsf{H}(-t)].
\end{equation}
Hence, $\mathsf L$ is not symmetric only if ${\mathsf H}(t)\neq{\mathsf H}(-t)$, which
highlights the necessity of having temporal asymmetric correlations in the noise in order to obtain asymmetric response functions via fluctuation-dissipation relations.
\bigskip

We finally note that the equivalence between the direct-response and Green-Kubo approaches is limited to the linear response regime. Since we have assumed noninteracting particles, the response is necessarily linear, whereas for interacting systems the equations of motion in the direct response calculation must be linearized with respect to the perturbation, as is shown in Appendix~\ref{CollectiveMobilitySec}.

\section{Hall conductivity in electrolytes} \label{SecHall}
We now extend the odd Dean-Kawasaki framework to ionic solutions under a magnetic field. 
These are described by $N_{\rm s}$ ionic species ($N_{\rm s}\geq 2$ for electroneutrality to hold), with particles of species $\alpha$ ($\alpha=1,...,N_{\rm s}$) having charge $q_{\alpha}$, mass $m_{\alpha}$, and friction coefficient $\gamma_{\alpha}$.
We assume that the particles follow the dynamics
\begin{equation}
m_{\alpha}\dot{\boldsymbol{v}}_{i}^\alpha=-\frac{\partial U}{\partial\boldsymbol{x}_{i}^\alpha}-\mathsf{M}_\alpha^{-1}\boldsymbol{v}_{i}^\alpha+\boldsymbol{\xi}_{i}^\alpha\left(t\right)
\label{eom_v_ions}
\end{equation} 
in which $\bm{x}_i^\alpha$ is the position of the $i$th particle of species $\alpha$, $\bm{v}_i^\alpha \equiv \dot{\bm{x}}_i^\alpha$ is its velocity, and
\begin{equation}\mathsf{M_{\alpha}}  \equiv\frac{1}{\gamma_{\alpha}}\left(\begin{array}{ccc}
1 & -\frac{q_{\alpha}B}{\gamma_{\alpha}} & 0\\
\frac{q_{\alpha}B}{\gamma_{\alpha}} & 1 & 0\\
0 & 0 & 1
\end{array}\right)^{-1}.
\end{equation}
We reiterate that species indices do not follow the Einstein summation convention, unlike spatial indices.
Equation \eqref{eom_v_ions} generalizes Eq.~\eqref{single_species_eom_v} to multiple species.
It models the solvent implicitly (see Fig.~\ref{SchematicSystem}) in the sense that its effect is captured by the friction force and the noise (rather by an explicit description of the molecules constituting the solvent), see Refs.~\cite{Molina2009,Lesnicki2021,Illien2024,Nair2025,Demery2026} for recent discussions on this approximation and Ref.~\cite{sung1987microscopic} for a discussion in the context of the Hall effect.
Ions interact via the Coulomb potential and may also be subject to an external electric field. The corresponding potential energy is
    \begin{align}
        U  = \;&\frac{1}{2}\sum_{\alpha,\beta}\int{\rm d}\boldsymbol{x}{\rm d}\boldsymbol{x}'n_{\alpha}(\boldsymbol{x})V_{\alpha\beta}(\boldsymbol{x}-\boldsymbol{x}')n_{\beta}(\boldsymbol{x}')\\
&+\sum_{\alpha}\int{\rm d}\bm{x}V_{{\rm ext}}(\bm{x},t)n_{\alpha}(\boldsymbol{x})\nonumber
  \end{align}
where
\begin{equation}
V_{\alpha\beta}(\boldsymbol{x})=q_{\alpha}q_{\beta}/(4\pi\varepsilon\left|\bm{x}\right|)
\quad
\text{and}
\quad
q_{\alpha}\boldsymbol{E}=-\nabla V_{\alpha}^{{\rm ext}}.
\end{equation}
Here, $\varepsilon=\varepsilon_{0}\varepsilon_r$ is the permittivity of the solvent, where $\varepsilon_{0}$ is the vacuum permittivity and $\varepsilon_r$ is the solvent's relative permittivity.

The odd Dean-Kawasaki Eq.~\eqref{EOM} generalizes to
\begin{subequations} \label{ODKelectrolyte}\begin{align}\partial_{t}n_{\alpha}= & -\boldsymbol{\nabla}\cdot\boldsymbol{J}_{\alpha}
\end{align}
by defining a number density field $n_\alpha(\bm{x},t)\equiv\sum_{i=1}^{N_\alpha}\delta(\bm{x}-\bm{x}_{i}^\alpha(t))$ for each species and carrying identical steps as in Sec.~\ref{SecModel}. 
We find that the density current of species $\alpha$ is
\begin{align}
\label{Jalpha}\boldsymbol{J}_{\alpha}(\boldsymbol{x},t)= & -k_{B}T\mathsf{M}_{\alpha}\boldsymbol{\nabla}n_{\alpha}+n_{\alpha}\mathsf{M}_{\alpha}q_{\alpha}\boldsymbol{E}+\boldsymbol{\chi}_{\alpha}\\
 & -n_{\alpha}\mathsf{M}_{\alpha}\sum_{\beta}\int\boldsymbol{\nabla}V_{\alpha\beta}(\bm{x}-\bm{x'})n_{\beta}(\bm{x}',t){\rm d}\bm{x}'\nonumber
\end{align}
and the corresponding Gaussian noise fields satisfy
\begin{align}
\langle\boldsymbol{\chi}_{\alpha}(\boldsymbol{x},t)\boldsymbol{\chi}_{\beta}^{\mathsf{T}}(\bm x',t')\rangle=n_{\alpha}(\boldsymbol{x},t)\delta_{\alpha\beta}\delta(\bm x-\bm x')\mathsf{C}_{\alpha}(t-t')
\end{align}
\end{subequations}
where $\mathsf{C}_{\alpha}(t)  =k_B T\left[\mathsf{M}_{\alpha}\delta_{+}(t)+\mathsf{M}_{\alpha}^{\mathsf T}\delta_{-}(t)\right]$, as well as $\langle\boldsymbol{\chi}_{\alpha}(\boldsymbol{x},t)\rangle = 0$.

The conductivity tensor $\mathsf{K}$ connects the total electric current density $\boldsymbol{J}_{c}$ defined as
\begin{equation} \label{Jc}
\boldsymbol{J}_{c}\equiv\sum_{\alpha}q_{\alpha}\boldsymbol{J}_{\alpha},
\end{equation}
to the external electric field $\bm E=E_0\hat{n}$ (assumed here to be fixed in time and space) through the relation
\begin{equation} \label{K_def}
\langle\boldsymbol{J}_{c}\rangle=\mathsf{K}\boldsymbol{E}.
\end{equation}
in which, because of rotational invariance around the magnetic field axis, $\mathsf{K}$ must be of the form
\begin{equation}
\mathsf{K}=\begin{pmatrix}\kappa & \kappa^{\text{H}} & 0\\
-\kappa^{\text{H}} & \kappa & 0\\
0 & 0 & \kappa^{\parallel}
\end{pmatrix}.
\end{equation}

Our calculation of the conductivity from odd SDFT is setup in Sec.~\ref{IntegralExpressions} in three different ways. 
We perform a perturbative calculation of the relevant averages in Sec.~\ref{CorrelationsPert} and put everything together in Sec.~\ref{resultsHall}.

For simplicity, from here on we will restrict ourselves to two symmetric ionic species with charges $+q$ and $-q$, where $q=ze$ and $e$ is the elementary charge. We denote the positively and negatively charged species by $\alpha=+$ and $\alpha=-$, respectively, and denote their average concentration, which is the same for both species by electroneutrality, by ${c}$. The friction coefficients $\gamma_+$ and $\gamma_-$ are generally different. The calculation can be straightforwardly generalized to a larger number of species or to asymmetric valencies.

\subsection{Three ways to the Hall conductivity}
\label{IntegralExpressions}
In the following, we derive three different formulas for the conductivity, based on (i) direct response, (ii) the Green-Kubo relation, and (iii) a hybrid fluctuation-dissipation relation. 

\subsubsection{Direct response}
\label{method1} 

Taking the ensemble average of Eq.~\eqref{Jalpha} and substituting it in Eq.~\eqref{Jc}, we obtain the average electric current density
\begin{equation}
\langle\boldsymbol{J}_{c}\rangle={c}q^{2}(\mathsf{M}_{+}+\mathsf{M}_{-})\boldsymbol{E}-\sum_{\alpha,\beta}q_{\alpha}\mathsf{M}_{\alpha}\langle n_{\alpha}\boldsymbol{\nabla}V_{\alpha\beta}*n_{\beta}\rangle,
\end{equation}
where $*$ denotes convolution in space, $(f*g)(\boldsymbol{x})\equiv\int \mathrm{d}\boldsymbol{x}'\,f(\boldsymbol{x}-\boldsymbol{x}')g(\boldsymbol{x}')$, and we used the fact that $\langle n_{+}(\boldsymbol{x},t)\rangle=\langle n_{-}(\boldsymbol{x},t)\rangle={c} $.
Hence, we can decompose $\mathsf{K}$ into
\begin{equation}
\mathsf{K} = \mathsf{K}_0 + \delta\mathsf{K}
\end{equation}
where $\mathsf{K}_0$ is the conductivity without interactions
\begin{equation} \label{K0}
\mathsf{K}_{0}={c}q^{2}(\mathsf{M}_{+}+\mathsf{M}_{-}),
\end{equation}
while its interaction-induced correction $\delta\mathsf{K}$, within linear response, is given by
\begin{equation} \label{deltaK_def}
\delta K_{\mu\nu}=-\sum_{\alpha\beta}q_{\alpha}M_{\alpha}^{\mu\rho}\frac{\partial}{\partial E_{\nu}}\langle n_{\alpha}\partial_{\rho}V_{\alpha\beta}*n_{\beta}\rangle\Big|_{\bm{E}=0}.
\end{equation}
This correction is computed explicitly in Sec.~\ref{CorrelationsPert}.
The noninteracting conductivity can be expressed as
\begin{subequations} \label{K0_approx}
\begin{align}
&\quad\quad\quad\mathsf{K}_{0}=\begin{pmatrix}\kappa_{0} & \kappa_{0}^{\text{H}} & 0\\
-\kappa_{0}^{\text{H}} & \kappa_{0} & 0\\
0 & 0 & \kappa^{\parallel}_{0}
\end{pmatrix}
\\
\intertext{where}
\label{kappa0}
        \kappa_0 & = q^{2}{c}
        \frac{(\gamma_{+}+\gamma_{-})(\gamma_+ \gamma_- +q^2 B^2)}{(\gamma_+^2+q^2 B^2)(\gamma_-^2+q^2 B^2)}
        \\
        \kappa^{\parallel}_0 & = q^{2}{c}
        (\gamma_{+}^{-1} + \gamma_{-}^{-1})
        \\[0.2cm]
\kappa^{\text{H}}_0 & = q^{3}{c}B
\big(
[\gamma_{+}^{2}+q^{2}B^{2}]^{-1}
-
[\gamma_{-}^{2}+q^{2}B^{2}]^{-1}
\big).
\end{align}
\end{subequations}
These quantities are plotted in Fig.~\ref{HallCond}(a) as a function of the normalized magnetic field ${b\equiv2q B/(\gamma_++\gamma_-)}$ with fixed $2\gamma_+/(\gamma_++\gamma_-)=3/2$, where it is shown how $\kappa_0$ monotonically decays with $b$ while $|\kappa_0^{\rm H}|$ peaks at finite $b$ and $\kappa_0^{\parallel}$ remains constant.
As expected, $\kappa_0$ and $\kappa_0^{\parallel}$ are even functions of $B$, while $\kappa_0^{\rm H}$ is an odd function of $B$. Moreover, a nonzero $\kappa_0^{\rm H}$ requires the positive and negative ions to have different friction coefficients, $\gamma_+\neq \gamma_-$. This is because the two species drift in opposite directions along the applied electric field but are deflected by the Lorentz force in the same transverse direction. Their Hall-current contributions therefore cancel for $\gamma_+=\gamma_-$.
For small $B$, 
\begin{subequations}
\begin{align}
 \kappa_0&=\kappa^{\parallel}_0+\mathcal{O}(B^2)   
 \\
 \kappa_{0}^{\text{H}}&=q^{3}{c}B(\gamma_{+}^{-2}-\gamma_{-}^{-2})+\mathcal{O}(B^{3})
\end{align}
\end{subequations}
while both $\kappa_0$ and $\kappa_0^{\rm H}$ approach zero in the $B\to\infty$ limit, consistent with the suppression of motion perpendicular to $\bm B$ in the limit where the cyclotron period is much shorter than the the relaxation time (as discussed in Sec.~\ref{fullLangevin}). 

\subsubsection{Green-Kubo relation} \label{method2}
In the linear response regime, the conductivity can be derived from the Green-Kubo relation (see Appendix~\ref{GK_proof} for derivation with magnetic field),
\begin{equation} \label{GKconductivity}
\mathsf{K}=\frac{1}{\mathcal{V}k_{B}T}\int\limits _{0}^{\infty}{\rm d}t\int{\rm d}\boldsymbol{x}\int{\rm d}\boldsymbol{x}'\langle\boldsymbol{J}_{c}(\boldsymbol{x},t)\boldsymbol{J}^{\mathsf T}_{c}(\boldsymbol{x}',0)\rangle_0
\end{equation}
where $\langle\cdot\rangle_0$ denotes ensemble average at $E_0=0$.

As the density current Eq.~\eqref{Jalpha} is composed of three terms (in the absence of $\bm E$), multiple correlators can, in principle, contribute to Eq.~\eqref{GKconductivity}. To simplify that, we write the electric current as the sum 
\begin{equation}
\boldsymbol{J}_{c}=\boldsymbol{J}_{{\rm d}}+\boldsymbol{J}_{{\rm i}}+\boldsymbol{J}_{{\rm s}}
\end{equation}
of the diffusion, interaction, and stochastic contributions
\begin{subequations}
\begin{align}\boldsymbol{J}_{{\rm d}} & \equiv-\sum_{\alpha}q_{\alpha}k_{B}T\mathsf{M}_{\alpha}\boldsymbol{\nabla}n_{\alpha}\\
\label{Ji}\boldsymbol{J}_{{\rm i}} & \equiv-\sum_{\alpha,\beta}q_{\alpha}n_{\alpha}\mathsf{M}_{\alpha}\boldsymbol{\nabla}V_{\alpha\beta}*n_{\beta}\\
\boldsymbol{J}_{{\rm s}} & \equiv\sum_{\alpha}q_{\alpha}\bm{\chi}_{\alpha}.
\end{align}
\end{subequations}
We notice that $\boldsymbol{J}_{{\rm d}}$ does not contribute to the Green-Kubo integral because the spatial integral $\int{\rm d}\boldsymbol{x}\boldsymbol{J}_{{\rm d}}=-\sum_{\alpha}q_{\alpha}k_{B}T\mathsf{M}_{\alpha}\int{\rm d}\boldsymbol{x}\nabla n_{\alpha}$ vanishes apart from a boundary term which we set to zero.
We also note that substituting in Eq.~\eqref{GKconductivity} the stochastic currents alone gives the conductivity without interactions $\mathsf{K}_0$,
\begin{align}\nonumber&\frac{1}{\mathcal{V}k_{B}T}\int\limits _{0}^{\infty}{\rm d}t\int{\rm d}\boldsymbol{x}\int{\rm d}\boldsymbol{x}'\langle\boldsymbol{J}_{{\rm s}}(\boldsymbol{x},t)\boldsymbol{J}_{{\rm s}}^{\mathsf{T}}(\boldsymbol{x}',0)\rangle_0\nonumber\\
&=\frac{q^{2}}{\mathcal{V} k_B T}\sum_{\alpha}\int\limits _{0}^{\infty}{\rm d}t\int{\rm d}\boldsymbol{x}n_{\alpha}(\boldsymbol{x},t)\mathsf{C}_{\alpha}(t-t')\nonumber\\
&={c}q^{2}(\mathsf{M}_{+}+\mathsf{M}_{-})
\end{align}
which is the same as Eq.~\eqref{K0}. Therefore, the correction to the conductivity due to the electrostatic interactions has only two contributions given by
\begin{subequations} \label{deltaKGK}
\begin{align}
&\quad\quad\quad\delta\mathsf{K}=\delta\mathsf{K}^{\rm ii}+\delta\mathsf{K}^{\rm is},\label{deltaKGK0}\\
\label{K1}\delta K_{\mu\nu}^{{\rm ii}}= & \sum_{\alpha,\beta,\gamma,\delta}\frac{q_{\alpha}q_{\gamma}M_{\alpha}^{\mu\rho}M_{\gamma}^{\nu\sigma}}{\mathcal{V}k_{B}T}\int\limits _{0}^{\infty}{\rm d}t\int{\rm d}\boldsymbol{x}\int{\rm d}\boldsymbol{x}'\\
 & \times\langle\mathcal{F}_{\alpha\beta}^{\rho}(\boldsymbol{x},t)\mathcal{F}_{\gamma\delta}^{\sigma}(\boldsymbol{x}',0)\rangle_0\nonumber\\
\label{K2}\delta K_{\mu\nu}^{{\rm is}}= & -\sum_{\alpha,\beta,\gamma}\frac{q_{\alpha}q_{\gamma}M_{\alpha}^{\mu\rho}}{\mathcal{V}k_{B}T}\int\limits _{0}^{\infty}{\rm d}t\int{\rm d}\boldsymbol{x}\int{\rm d}\boldsymbol{x}'\\
 & \times\langle\mathcal{F}_{\alpha\beta}^{\rho}(\boldsymbol{x},t)\chi_{\gamma}^{\nu}(\boldsymbol{x}',0)\rangle_0,\nonumber
\end{align}
\end{subequations}
where $\mathcal{F}_{\alpha\beta}^{\mu}\equiv n_{\alpha}\partial_{\mu}V_{\alpha\beta}*n_{\beta}$. These terms come from $\boldsymbol{J}_{{\rm i}}$ and $\boldsymbol{J}_{{\rm i}}$ correlations (for $\delta\mathsf{K}^{\rm ii}$), and $\boldsymbol{J}_{{\rm i}}$ and $\boldsymbol{J}_{{\rm s}}$ cross correlations (for $\delta\mathsf{K}^{\rm is}$). The other cross term, $\delta \mathsf{K}^{\rm si}$, vanishes due to causality. 
The relevant correlations are computed explicitly in Sec.~\ref{CorrelationsPert}.

\subsubsection{Hybrid fluctuation-dissipation relation} \label{method3}
In the linear response regime, the conductivity can be derived by a different fluctuation-dissipation relation, derived in Appendix~\ref{hybrid_proof}, which is given by
\begin{align}\mathsf{K}= & \; {c}q^{2}(\mathsf{M}_{+}+\mathsf{M}_{-})+\delta\mathsf{K} 
\\
\intertext{with}
\label{hybrid}
\delta\mathsf{K}=-\frac{1}{\mathcal{V}k_{B}T} & 
\int_{0}^{\infty}\!\!\!{\rm d}t\!
\int\!\!{\rm d}\boldsymbol{x}\!\!
\int\!\!{\rm d}\boldsymbol{x}'
\langle\boldsymbol{J}_{{\rm i}}(\boldsymbol{x},t)\overleftarrow{\boldsymbol{J}}_{{\rm i}}^{\mathsf{T}}(\boldsymbol{x}',0)\rangle_0.
\end{align}
We refer to Eq.~\eqref{hybrid} as a hybrid fluctuation-dissipation relation because only the interaction-induced correction $\delta\mathsf{K}$ is expressed in terms of current fluctuations in the unperturbed systems, whereas the noninteracting contribution appears explicitly. Here, $\overleftarrow{\boldsymbol{J}}_{{\rm i}}$ is the time-reversed version of the current in Eq.~\eqref{Ji}, which is given by
\begin{equation} \label{Jtilde}
\overleftarrow{\boldsymbol{J}}_{{\rm i}}=-\sum_{\alpha,\beta}q_{\alpha}n_{\alpha}\mathsf{M}_{\alpha}^{\mathsf{T}}\sum_{\beta}\boldsymbol{\nabla}V_{\alpha\beta}*n_{\beta}
\end{equation}
and can be obtained by reversing the magnetic field ${\overleftarrow{\boldsymbol{J}}_{{\rm i}}(\bm B)=\boldsymbol{J}_{{\rm i}}(-\bm B)}$. Hence, Eq.~\eqref{hybrid} involves correlations between forward and backward currents.

The hybrid Green-Kubo relation \eqref{hybrid} is derived from the overdamped version of the Fokker-Planck equation~\cite{Felderhof1983,Felderhof1987,Jardat1999,hoang2023frequency}, unlike the relation \eqref{GKconductivity} of Sec.~\ref{method2} which is derived from the full underdamped Fokker-Planck equation~\cite{risken1989fokker}. 
It explicitly depends on microscopic interactions, but can be useful for simulations~\cite{Jardat1999,hoang2023frequency}.

\subsection{Perturbative calculation of correlations} \label{CorrelationsPert}
To compute the three expressions for $\delta \mathsf{K}$ in sections~\ref{method1}--\ref{method3}, all of which involve ensemble averages of products of fields, one would need to solve the odd SDFT equations \eqref{ODKelectrolyte}. 
This can be done perturbatively, in the same way as standard SDFT. To keep track of the orders in the expansion, we multiply the noise by a constant $\bm \chi_{\alpha}(\bm x,t)\to\epsilon \bm \chi_{\alpha}(\bm x,t)$ (to be set to $1$ later on), and expand
\begin{subequations} \label{expansion}\begin{align}\label{expansion_n}n_{\alpha}(\bm{x},t) & \simeq{c}+\epsilon\delta n_{\alpha}(\bm{x},t)+\epsilon^{2}\delta w_{\alpha}(\bm{x},t)+\mathcal{O}(\epsilon^{3})\\
\label{expansion_noise}\epsilon\boldsymbol{\chi}_{\alpha}(\bm{x},t) & \simeq\epsilon\boldsymbol{\chi}_{\alpha}^{0}(\bm{x},t)+\frac{\epsilon^{2}}{2{c}}\delta n_{\alpha}(\bm{x},t)\boldsymbol{\chi}_{\alpha}^{0}(\bm{x},t)+\mathcal{O}(\epsilon^{3})
\end{align}
\end{subequations}
where $\boldsymbol{\chi}^0_{\alpha}$ is a Gaussian noise vector field satisfying
\begin{equation} 
\label{noise0_corr}
\langle\boldsymbol{\chi}_{\alpha}^{0}(\boldsymbol{x},t)\boldsymbol{\chi}_{\beta}^{0^{\mathsf{T}}}(\boldsymbol{x}',t')\rangle={c}\delta_{\alpha\beta}\delta(\boldsymbol{x}-\boldsymbol{x}')\mathsf{C}_{\alpha}(t-t').
\end{equation}
To get Eq.~\eqref{expansion_noise}, we used the relation $\bm \chi_{\alpha}=\sqrt{\frac{n_{\alpha}}{c}}\bm \chi_{\alpha}^{0}$ and the expansion $\sqrt{\frac{n_{\alpha}}{c}}\approx1+\epsilon\frac{\delta n_{\alpha}}{2c}+\mathcal{O}\left(\epsilon^{2}\right)
$.

We substitute Eq.~\eqref{expansion} into Eq.~\eqref{ODKelectrolyte} and equate terms to linear order in $\epsilon$. Then, defining 
for any function $f(\bm x,t)$ its Fourier transform $\tilde{f}_{\bm k}(t) = \tilde{f}^{\bm k}(t) \equiv \int f(\boldsymbol{x},t)e^{-i\boldsymbol{k}\cdot\boldsymbol{x}}{\rm d}\bm{x}$, we obtain
\begin{equation} \label{Lin_algebra}
\partial_{t}\delta\tilde{n}_{\alpha}^{\boldsymbol{k}}=\sum_{\beta}D_{\alpha\beta}^{\boldsymbol{k}}\delta\tilde{n}_{\beta}^{\boldsymbol{k}}-i{\bm k}\cdot \tilde{\bm \chi}_{\alpha}^{0,\bm k}(t)
\end{equation}
with the dynamical matrix $D_{\alpha\beta}^{\boldsymbol{k}}$ defined as
\begin{align} 
D_{\alpha\beta}^{\boldsymbol{k}}  \equiv&-k_{B}Tk_{\mu}k_{\nu}M_{\alpha}^{\mu\nu}\delta_{\alpha\beta}-\frac{{c}q_{\alpha}q_{\beta}k_{\mu}k_{\nu}M_{\alpha}^{\mu\nu}}{\varepsilon k^{2}}\nonumber\\ \label{L_def}
 & -iq_{\alpha}k^{\mu}M_{\alpha}^{\mu\nu}E_{\nu}\delta_{\alpha\beta},
\end{align}
where we have inserted the Coulomb potential in Fourier space $\tilde{V}_{\alpha\beta}^{\bm k}=q_{\alpha}q_{\beta}/(\varepsilon k^{2})$.
The second term on the right hand side of Eq.~\eqref{Lin_algebra} is a Gaussian white noise scalar field with zero mean and correlations $\langle(-i\boldsymbol{k}\cdot\tilde{\boldsymbol{\chi}}_{\alpha}^{0,\boldsymbol{k}}(t))(-i\boldsymbol{k}'\cdot\tilde{\boldsymbol{\chi}}_{\beta}^{0,\boldsymbol{k}'}(t'))\rangle=W_{\alpha\beta}^{\bm k}(2\pi)^{3}\delta(\boldsymbol{k}+\boldsymbol{k}')\delta(t-t')
$, with 
\begin{equation}
W_{\alpha\beta}^{\bm k}=2k_{B}T{c}\delta_{\alpha\beta}k_{\mu}k_{\nu}M_{\alpha}^{\mu\nu}.
\end{equation}
Notably, since  $k_{\mu}k_{\nu}$ is symmetric under $\mu\leftrightarrow\nu$, $W_{\alpha\beta}$ contains only the symmetric part of $M_{\alpha}$.

The solution to Eq.~\eqref{Lin_algebra} in steady state is 
\begin{equation}\label{X_sol}
\delta\tilde{n}_{\alpha}^{\boldsymbol{k}}(t)=-i\sum_{\beta}\int\limits _{-\infty}^{t}{\rm d}s\,(e^{\mathsf{D}^{\boldsymbol{k}}(t-s)})_{\alpha\beta}\boldsymbol{k}\cdot\tilde{\boldsymbol{\chi}}_{\beta}^{0,\boldsymbol{k}}(s)
\end{equation}
where $\mathsf{D}^{\bm k}$ is a matrix in species space and
$e^{\mathsf{D}^{\bm k}}$ is its matrix exponential.

The steady-state correlations of the system of stochastic differential equations in Eq.~\eqref{Lin_algebra} are given by~\cite{gardiner2009stochastic}
\begin{equation} \label{XX_cor2}
\langle\delta\tilde{n}_{\alpha}^{\boldsymbol{k}}(t)\delta\tilde{n}_{\beta}^{\boldsymbol{k}'}(t')\rangle=(2\pi)^{3}\delta(\boldsymbol{k}+\boldsymbol{k}')A_{\alpha\beta}^{nn,\boldsymbol{k}}(t-t')
\end{equation}
with
\begin{equation}
A_{\alpha\beta}^{nn,\boldsymbol{k}}(t)\equiv\begin{cases}
\sum_{\gamma}(e^{{\mathsf D}^{\boldsymbol{k}}t})_{\alpha\gamma}A_{\gamma\beta}^{0,\boldsymbol{k}} & t>0\\
\sum_{\gamma}(e^{-{\mathsf D}^{-\boldsymbol{k}}t})_{\beta\gamma}A_{\alpha\gamma}^{0,\boldsymbol{k}} & t<0
\end{cases}
\end{equation}
where $A_{\alpha\beta}^{0,\boldsymbol{k}}$ is the equal-time correlator given by the following Lyapunov equation, written in matrix notation in species space:
\begin{equation}
\mathsf{D}_{\bm{k}}\mathsf{A}_{\bm{k}}^{0}+\mathsf{A}_{\bm{k}}^{0}(\mathsf{D}_{-\bm{k}})^{\mathsf{T}}=-\mathsf{W}_{\bm{k}}.
\end{equation}
This yields a closed-form expression for $A_{\alpha\beta}^{nn,\bm k}(t)$, which is evaluated using a computer algebra system (see Supplemental Material).

In addition to density-density correlations, density-noise correlations are also needed to evaluate the conductivity (see  Eq.~\eqref{K2}). Those can be calculated using Novikov's theorem~\cite{novikov1965functionals}, but here we compute them directly using Eq.~\eqref{noise0_corr}
and Eq.~\eqref{X_sol},obtaining
\begin{equation} \label{nchi_cor2}
\langle\delta\tilde{n}_{\alpha}^{\boldsymbol{k}}(t)\tilde{\chi}_{\beta,\mu}^{0,\boldsymbol{k}'}(t')\rangle=(2\pi)^{3}\delta(\boldsymbol{k}+\boldsymbol{k}')A_{\alpha\beta,\mu}^{n\chi,\boldsymbol{k}}(t-t')\end{equation}
with 
\begin{equation} \label{A_nnoise}
A_{\alpha\beta,\mu}^{n\chi,\boldsymbol{k}}(t)\equiv -i k_{B}T{c}(e^{\mathsf{D}_{\boldsymbol{k}}t})_{\alpha\beta}k_{\nu}(M^{\nu\mu}_{\beta}+M^{\mu\nu}_{\beta})\Theta(t).
\end{equation}

We have obtained all the relevant correlations of order $\epsilon^2$ in perturbation theory. As we will see in the next subsection, calculating the first nontrivial contribution to the conductivity in the Green-Kubo approach also requires $\delta w_{\alpha}$, the next order in the $n_{\alpha}$ expansion of Eq.~\eqref{expansion_n}. Equating terms of order $\epsilon^2$ in Eq.~\eqref{ODKelectrolyte} in Fourier space leads to
\begin{equation}
\partial_{t}\delta\tilde{w}_{\alpha}^{\boldsymbol{k}}=\sum_{\beta}D_{\alpha\beta}^{\boldsymbol{k}}\delta\tilde{w}_{\beta}^{\boldsymbol{k}}+f_{\alpha}^{\boldsymbol{k}}(t)
\end{equation}
with $D^{\bm k}_{\alpha\beta}$ as defined in Eq.~\eqref{L_def} and with $f_{\alpha}^{\boldsymbol{k}}(t)$ given by  the lower order terms $\delta n$ and $\boldsymbol{\chi}^0$,
\begin{align}\label{fk}\nonumber f_{\alpha}^{\boldsymbol{k}}(t) & \equiv-\frac{1}{(2\pi)^{3}}k_{\mu}M_{\alpha}^{\mu\nu}k'_{\nu}\sum_{\beta}\int\delta\tilde{n}_{\alpha}^{\boldsymbol{k}-\boldsymbol{k}'}\tilde{V}_{\alpha\beta}^{\boldsymbol{k}'}\delta\tilde{n}_{\beta}^{\boldsymbol{k}'}{\rm d}^{3}\boldsymbol{k}'\\
 & -\frac{1}{(2\pi)^{3}}\frac{ik_{\mu}}{2{c}}\int\delta\tilde{n}_{\alpha}^{\boldsymbol{k}'}\tilde{\chi}_{\alpha,\mu}^{0,\boldsymbol{k}-\boldsymbol{k}'}{\rm d}^{3}\boldsymbol{k}'.
\end{align}
Finally, the steady-state solution of $\delta \tilde{w}^{\bm k}_{\alpha}(t)$ is given by
\begin{equation} \label{wsolution}
\delta\tilde{w}_{\alpha}^{\boldsymbol{k}}(t)=\sum_{\beta}\int\limits _{-\infty}^{t}{\rm d}s\,(e^{\mathsf{D}^{\boldsymbol{k}}(t-s)})_{\alpha\beta}f_{\beta}^{\boldsymbol{k}}(s)
\end{equation}
where we see that $\delta w$ is non-Gaussian, as it is a quadratic function of the correlated Gaussian fields $\delta n$ and $\bm \chi^0$.

\subsection{Results} \label{resultsHall}
We now substitute the correlations derived in Sec.~\ref{CorrelationsPert} into the three relations in Sec.~\ref{IntegralExpressions}.\subsubsection{Direct linear response results}
In the direct response approach, $\delta\mathsf{K}$ is given by Eq.~\eqref{deltaK_def}. Substituting the Coulomb potential and going to Fourier space we obtain
\begin{align}\label{directresultintegral}\nonumber\delta K_{\mu\nu} & =-\sum_{\alpha,\beta}\frac{iq^{2}q_{\beta}M_{\alpha}^{\mu\rho}}{(2\pi)^{6}\varepsilon}\int{\rm d}\boldsymbol{k}\int{\rm d}\boldsymbol{k}'\,\frac{k_{\rho}'e^{i(\boldsymbol{k}+\boldsymbol{k}')\cdot\boldsymbol{x}}}{k'^{2}}\\
 & \times\frac{\partial}{\partial E_{\nu}}\langle\tilde{n}_{\alpha}^{\boldsymbol{k}}(t)\tilde{n}_{\beta}^{\boldsymbol{k}'}(t)\rangle\Big|_{\bm{E}=0}.
\end{align}
The lowest-order nonvanishing contribution to Eq.~\eqref{directresultintegral} is due to the term $\langle\delta\tilde{n}_{\alpha}^{\boldsymbol{k}}(t)\delta\tilde{n}_{\beta}^{\boldsymbol{k}'}(t)\rangle$ arising from expanding $\langle\tilde{n}_{\alpha}^{\boldsymbol{k}}(t)\tilde{n}_{\beta}^{\boldsymbol{k}'}(t)\rangle$ according to Eq.~\eqref{expansion}. 
At this level,
\begin{align}\label{deltaKE}\delta K_{\mu\nu} & =\sum_{\alpha,\beta}\frac{iq^{2}q_{\beta}M_{\alpha}^{\mu\rho}}{(2\pi)^{3}\varepsilon}\,\int{\rm d}\boldsymbol{k}\frac{k_{\rho}}{k^{2}}\frac{\partial}{\partial E_{\nu}}A_{\alpha\beta}^{0,nn,\boldsymbol{k}}\Big|_{\bm{E}=0}
\end{align}
where we used Eq.~\eqref{XX_cor2} and the fact that $A_{\alpha\beta}^{nn,\boldsymbol{k}}(0)=A_{\alpha\beta}^{0,\boldsymbol{k}}$.

    Performing the integral (see Supplemental Material) yields the result
\begin{subequations}
\label{Dkappamatrix}
\begin{equation}
    \delta\mathsf{K} = \begin{pmatrix}
\Delta\kappa & \Delta\kappa^{{\rm H}} & 0\\
-\Delta\kappa^{{\rm H}} & \Delta\kappa & 0\\
0 & 0 & \Delta\kappa^{\parallel}
\end{pmatrix}\\
\end{equation}
where
    \begin{align}
\frac{\Delta\kappa}{\Delta\kappa_{{\rm DHO}}} & =-\frac{3(\sqrt{2}-1)}{2-2\sqrt{2}}\frac{\alpha^{2}\beta^{2}\left(\left(\alpha\beta+b^{2}\right)^{2}-b^{2}(\alpha-\beta)^{2}\right)}{b^{3}f^{3}g\left(\alpha^{2}+b^{2}\right)\left(\beta^{2}+b^{2}\right)}\nonumber\\
 & \times\left(\left(\alpha^{2}+b^{2}\right)\left(\beta^{2}+b^{2}\right)\tan^{-1}\left(\frac{bf}{g}\right)-bfg\right)\\
\frac{\Delta\kappa^{{\rm H}}}{\Delta\kappa_{{\rm DHO}}} & =-\frac{3\alpha\beta g(\alpha-\beta)}{b^{2}f^{3}\left(\alpha^{2}+b^{2}\right)\left(\beta^{2}+b^{2}\right)}\\
 & \times\left(\left(\alpha^{2}+b^{2}\right)\left(\beta^{2}+b^{2}\right)\tan^{-1}\left(\frac{bf}{g}\right)-bfg\right)\nonumber\\
\frac{\Delta\kappa^{\parallel}}{\Delta\kappa_{{\rm DHO}}} & =\frac{3\left(\alpha^{2}+b^{2}\right)\left(\beta^{2}+b^{2}\right)\left(bf-g\tan^{-1}\left(\frac{bf}{g}\right)\right)}{b^{3}f^{3}}
\end{align}
\end{subequations}
with
    \begin{align} \label{balphadef}
    \nonumber
    b & \equiv\frac{qB}{\overline{\gamma}},\,\,\,\overline{\gamma}\equiv\frac{\gamma_{+}+\gamma_{-}}{2}\\
\alpha & \equiv\frac{\gamma_{+}}{\overline{\gamma}},\,\,\,\beta\equiv\frac{\gamma_{-}}{\overline{\gamma}}=2-\alpha
\end{align}
and with the functions
\begin{align}
\nonumber f\left(b,\alpha,\beta\right) & \equiv \sqrt{b^{2}+\alpha^{2}-\alpha\beta+\beta^{2}}\\
g\left(b,\alpha,\beta\right) & \equiv\sqrt{\alpha\beta\left(b^{2}+\alpha\beta\right)}.
\end{align}
The normalizing factor, $\Delta \kappa_{\rm DHO}$, is defined as
\begin{equation}
    \Delta \kappa_{\rm DHO}\equiv-\frac{\sqrt{2\pi}}{3}(2-\sqrt{2})\kappa_{0}(B=0)|z|^3\ell_{B}^{3/2}\sqrt{{c}}
\end{equation}
where $\kappa_0$ is given by Eq.~\eqref{kappa0}, $\ell_{B}\equiv e^{2}/(4\pi\varepsilon k_{B}T)$ is the Bjerrum length, and $z=q/e$ is the ionic valency. 
At linear order in $B$, the conductivity corrections take the simple form
\begin{subequations} \label{deltaK_linres}
    \begin{align}
 \label{DHO}
\Delta\kappa &= \Delta\kappa^{\parallel}=\Delta \kappa_{\rm DHO}+\mathcal{O}(B^{2})\\
\Delta\kappa^{\text{H}} &=2qB(\gamma_{+}^{-1}-\gamma_{-}^{-1})\Delta\kappa_{\rm DHO}+\mathcal{O}(B^{3}).
\end{align}
\end{subequations}
Without Lorentz force ($B=0$), we obtain $\delta\mathsf{K}=\Delta \kappa_{\rm DHO} \mathsf{I}$, which is the known result of the relaxation effect computed originally by Debye, H{\"u}ckel and Onsager~\cite{onsager1927report,onsager2002irreversible}. Moreover, our results truncated at first order in $B$ match exactly with the results of a cluster expansion calculation done in 1965 by Friedman~\cite{friedman1965calculation}, see Appendix~\ref{FridmanComparison2} for details.

\begin{figure}
\centering
{\includegraphics[width=0.4\textwidth,draft=false]{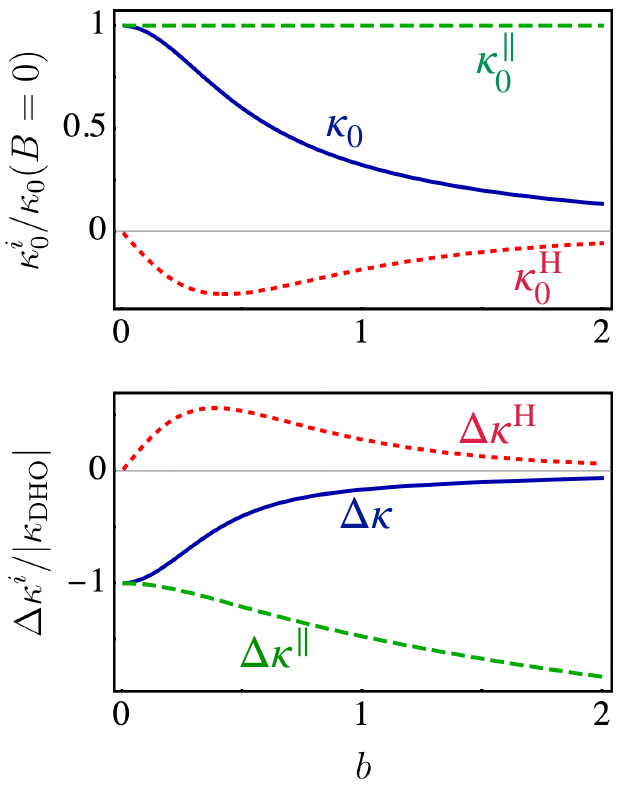}}
\caption{\textbf{Hall conductance of dilute electrolytes from odd SDFT.} The three independent components of the conductivity tensor, $\mathsf{K}_{xx}=\kappa$, $\mathsf{K}_{xy}=\kappa^{\rm H}$, and $\mathsf{K}_{zz}=\kappa^{\parallel}$, are shown as a function of $b=2q B/(\gamma_++\gamma_-)$ and with a fixed $\alpha=2\gamma_+/(\gamma_++\gamma_-)=3/2$.
Top: the mean-field conductivity valid in the infinite dilution limit given by Eq.~\eqref{K0_approx} and normalized by $\kappa_0(B=0)=q^2 {c} (\gamma_+^{-1}+\gamma_-^{-1})$. Bottom: lowest order correction at higher concentrations given by Eq.~\eqref{Dkappamatrix}, normalized by $|\Delta \kappa_{\rm DHO}|=\frac{\sqrt{2\pi}}{3}(2-\sqrt{2})\kappa_{0}(B=0)|z|^3\ell_{B}^{3/2}\sqrt{{c}}$.
\label{HallCond}}
\end{figure}

The conductivity corrections $\Delta \kappa$, $\Delta \kappa^{\parallel}$ and $\Delta \kappa^{\rm H}$, when normalized by $\Delta\kappa_{\rm DHO}$, depend only on two independent parameters: $b$ and $\alpha$, see Eq.~\eqref{balphadef} (for simple aqueous ionic solutions such as table salt in water, $b$ is of order $10^{-7}$ at $B=\SI{1}{\tesla}$, while $\alpha$ is of order one). 
In Fig.~\ref{HallCond}, we show them as a function of $b$ for $\alpha=3/2$, alongside the mean-field values $\kappa_0$, $\kappa_0^{\rm H}$, and $\kappa_0^{\parallel}$, normalized by $\kappa_0(B=0)$.
Note that at low concentrations, $\kappa_{\rm DHO}\ll \kappa_0(B=0)$ such that the corrections are small compared to the mean-field value. 
It is shown that all three corrections reduce the conductivity, i.e., they have the opposite sign to the mean-field contribution. Similar to the DHO theory, this reduction can be understood as a consequence of electrostatic correlations between oppositely charged ions. These correlations effectively associate positive and negative ions, thereby reducing the amount of free charge available to contribute to electrical conduction.

Similar to their respective mean-field values, $\Delta \kappa$ and $\Delta \kappa^{\rm H}$ vanish as $b\to \infty$, consistent with the suppression of particle motion in directions perpendicular to $\bm B$ in this regime (see Sec.~\ref{fullLangevin}). On the other hand, $\Delta \kappa^{\parallel}$ grows with $b$ and saturates at  $\lim_{b\to\infty}\Delta\kappa^{\parallel}/\Delta\kappa_{\rm DHO}= 3$ (this limit applies for any $\alpha$).  The saturation occurs outside the range of the plot, at around $b\approx 50$. This behavior contrasts with the mean-field longitudinal conductivity $\kappa^{\parallel}_0$ which is independent on $b$. The difference indicates that, although the Lorentz force does not directly affect the single-particle motion parallel to $\bm B$, electrostatic interactions couple the longitudinal response to fluctuations in the transverse degrees of freedom.

\subsubsection{Green-Kubo results}
In the Green-Kubo approach, $\delta\mathsf{K}$ is given by Eq.~\eqref{deltaKGK}. The lowest order nonzero contributions in the perturbative expansion are of order $\epsilon^4$ (see Eq.~\eqref{expansion}), and are given by
\begin{subequations} \label{twodeltaK}
\begin{align}\label{K1k}\delta K_{\mu\nu}^{{\rm ii}} & =-\sum_{\alpha,\beta,\gamma,\delta}\frac{q^{4}q_{\beta}q_{\delta}M_{\alpha}^{\mu\rho}M_{\gamma}^{\nu\sigma}}{(2\pi)^{6}\mathcal{V}k_{B}T\varepsilon^{2}}\int\limits _{0}^{\infty}{\rm d}t\int{\rm d}\boldsymbol{k}\int{\rm d}\boldsymbol{k}'\\
 & \quad\times\frac{k_{\rho}k_{\sigma}'}{k^{2}k'^{2}}\langle\delta\tilde{n}_{\alpha}^{-\boldsymbol{k}}(t)\delta\tilde{n}_{\beta}^{\boldsymbol{k}}(t)\delta\tilde{n}_{\gamma}^{-\boldsymbol{k}'}(0)\delta\tilde{n}_{\delta}^{\boldsymbol{k}'}(0)\rangle_0\nonumber\\
\label{K2k}\delta K_{\mu\nu}^{{\rm is}} & =-\sum_{\alpha,\beta,\gamma}\frac{iq^{2}q_{\gamma}q_{\beta}M_{\alpha}^{\mu\rho}}{(2\pi)^{6}2\mathcal{V}k_{B}T{c}\varepsilon}\int\limits _{0}^{\infty}{\rm d}t\int{\rm d}\boldsymbol{k}\frac{k_{\rho}}{k^{2}}\\
 & \quad\times\bigg[\int{\rm d}\boldsymbol{k}'\langle\delta\tilde{n}_{\alpha}^{-\boldsymbol{k}}(t)\delta\tilde{n}_{\beta}^{\boldsymbol{k}}(t)\delta\tilde{n}_{\gamma}^{-\boldsymbol{k}'}(0)\tilde{\chi}_{\gamma,\nu}^{0,\boldsymbol{k}'}(0)\rangle_0\nonumber\\
 & \quad-2(2\pi)^{3}{c}\bigg(\langle\delta\tilde{n}_{\alpha}^{\boldsymbol{k}}(t)\delta\tilde{w}_{\beta}^{-\boldsymbol{k}}(t)\tilde{\chi}_{\gamma,\nu}^{0,\boldsymbol{k}=0}(0)\rangle_0\nonumber\\
 & \quad-\langle\delta\tilde{n}_{\beta}^{\boldsymbol{k}}(t)\delta\tilde{w}_{\alpha}^{-\boldsymbol{k}}(t)\tilde{\chi}_{\gamma,\nu}^{0,\boldsymbol{k}=0}(0)\rangle_0\bigg)\bigg].\nonumber
\end{align}
\end{subequations}
Lower-order terms ($\sim\epsilon^n$ with $n<4$) vanish either because they yield integrals over full space derivatives, or because they contain an odd number of Gaussian fields, and hence vanish due to Wick's theorem (note also that $\delta n_{\alpha}$ is Gaussian but $\delta w_\alpha$ is a linear combination of products of two Gaussian fields).

Using Wick's theorem for the 4-point correlators in Eq.~\eqref{twodeltaK} and substituting the correlations calculated in Sec.~\ref{CorrelationsPert}, we obtain,
    \begin{align} & \langle\delta\tilde{n}_{\alpha}^{-\boldsymbol{k}}(t)\delta\tilde{n}_{\beta}^{\boldsymbol{k}}(t)\delta\tilde{n}_{\gamma}^{-\boldsymbol{k}'}(0)\delta\tilde{n}_{\delta}^{\boldsymbol{k}'}(0)\rangle_0\nonumber\\
 & \quad\quad\quad\quad=(2\pi)^{6}(\delta(\boldsymbol{k}+\boldsymbol{k}'))^{2}A_{\alpha\gamma}^{nn,-\boldsymbol{k}}(t)A_{\beta\delta}^{nn,\boldsymbol{k}}(t)\big|_{E_0=0}\nonumber\\
 & \quad\quad\quad\quad\quad+(2\pi)^{6}(\delta(\boldsymbol{k}-\boldsymbol{k}'))^{2}A_{\alpha\delta}^{nn,-\boldsymbol{k}}(t)A_{\beta\gamma}^{nn,\boldsymbol{k}}(t)\big|_{E_0=0}\label{Wick1}\\
 & \nonumber\langle\delta\tilde{n}_{\alpha}^{-\boldsymbol{k}}(t)\delta\tilde{n}_{\beta}^{\boldsymbol{k}}(t)\delta\tilde{n}_{\gamma}^{-\boldsymbol{k}'}(0)\tilde{\chi}_{\gamma,\nu}^{0,\boldsymbol{k}'}(0)\rangle_0\\
 & \quad\quad\quad\quad=(2\pi)^{6}(\delta(\boldsymbol{k}+\boldsymbol{k}'))^{2}A_{\alpha\gamma}^{nn,-\boldsymbol{k}}(t)A_{\beta\gamma,\nu}^{n\chi,\boldsymbol{k}}(t)\big|_{E_0=0}\nonumber\\
 & \quad\quad\quad\quad\quad+(2\pi)^{6}(\delta(\boldsymbol{k}-\boldsymbol{k}'))^{2}A_{\beta\gamma}^{nn,\boldsymbol{k}}(t)A_{\alpha\gamma,\nu}^{n\chi,-\boldsymbol{k}}(t)\label{Wick2}\big|_{E_0=0}.
\end{align}
Here, we have omitted Wick contractions that do not connect fields at different times, since they do not contribute to the Green-Kubo integral.
In addition, substituting the solution of $\delta \tilde{w}_{\alpha}^{\bm k}(t)$ in Eq.~\eqref{wsolution} and the definition of $f_{\alpha}^{\bm k}(t)$ in Eq.~\eqref{fk} we get
 \begin{align} & \label{Wick3}\nonumber\langle\delta\tilde{n}_{\alpha}^{\boldsymbol{k}}(t)\delta\tilde{w}_{\beta}^{-\boldsymbol{k}}(t)\tilde{\chi}_{\gamma,\nu}^{0,\bm{k}=0}(0)\rangle_0
\\
\nonumber & =\frac{i\mathcal{V}}{2}\int\limits _{-\infty}^{t}{\rm d}s\,\big(e^{\mathsf{D}^{-\boldsymbol{k}}(t-s)}\big)_{\beta\gamma}k_{\sigma}A_{\alpha\gamma}^{nn,\boldsymbol{k}}(t-s)\mathsf{C}_{\gamma}^{\sigma\nu}(s)\big|_{E_0=0}\\
 & =\frac{ik_{B}T\mathcal{V}}{2}(M_{\gamma}^{\sigma\nu}+M_{\gamma}^{\nu\sigma})\,\big(e^{\mathsf{D}^{-\boldsymbol{k}}t}\big)_{\beta\gamma}k_{\sigma}A_{\alpha\gamma}^{nn,\boldsymbol{k}}(t)\big|_{E_0=0}
\end{align}
where we have used the fact that contractions between $\delta\tilde n_{\alpha}^{\boldsymbol{k}}$ and $\tilde{\boldsymbol{\chi}}^{0,\boldsymbol{k}=0}$ vanish (see Eq.~\eqref{A_nnoise}).
Substituting Eqs.~(\ref{Wick1}--\ref{Wick3}) into Eq.~\eqref{twodeltaK} and using the relation $(2\pi)^{3}\delta(\boldsymbol{k}=0)=V$ we obtain
\begin{subequations}
\begin{align}\label{Kii}\delta K_{\mu\nu}^{{\rm ii}} & =\sum_{\alpha,\beta,\gamma,\delta}\frac{q^{4}q_{\beta}q_{\delta}M_{\alpha}^{\mu\rho}M_{\gamma}^{\nu\sigma}}{(2\pi)^{3}k_{B}T\varepsilon^{2}}\int\limits _{0}^{\infty}{\rm d}t\int{\rm d}\boldsymbol{k}\frac{k_{\rho}k_{\sigma}}{k^{4}}\nonumber\\
 & \times
 \Big(A_{\alpha\gamma}^{nn,-\boldsymbol{k}}(t)A_{\beta\delta}^{nn,\boldsymbol{k}}(t)-A_{\alpha\delta}^{nn,-\boldsymbol{k}}(t)A_{\beta\gamma}^{nn,\boldsymbol{k}}(t) \Big)
 \Big|_{E_0=0}\\
\delta K_{\mu\nu}^{{\rm is}} & =-\sum_{\alpha,\beta,\gamma}\frac{iq^{2}q_{\gamma}q_{\beta}M_{\alpha}^{\mu\rho}}{(2\pi)^{3}2{c} k_{B}T\varepsilon}\int\limits _{0}^{\infty}{\rm d}t\int{\rm d}\boldsymbol{k}\frac{k_{\rho}}{k^{2}}\nonumber\\
 & \times\bigg[A_{\alpha\gamma}^{nn,-\boldsymbol{k}}(t)A_{\beta\gamma,\nu}^{n\chi,\boldsymbol{k}}(t)-A_{\beta\gamma}^{nn,-\boldsymbol{k}}(t)A_{\alpha\gamma,\nu}^{n\chi,\boldsymbol{k}}(t)\nonumber\\
 & -i{c}k_{B}Tk_{\sigma}(M_{\gamma}^{\nu\sigma}+M_{\gamma}^{\sigma\nu})\nonumber\\
 & \times
 \Big( \big(e^{\mathsf{D}^{-\boldsymbol{k}}t}\big)_{\beta\gamma}A_{\alpha\gamma}^{nn,\boldsymbol{k}}(t)-(e^{\mathsf{D}^{-\boldsymbol{k}}t})_{\alpha\gamma}A_{\beta\gamma}^{nn,\boldsymbol{k}}(t)\Big)
 \bigg]\Big|_{E_0=0}.
\end{align}
\end{subequations}
Finally, performing the  $\delta\mathsf{K}^{\rm ii}$ and $\delta\mathsf{K}^{\rm is}$ integrals (see Supplemental Material) and adding them together (see Eq.~\eqref{deltaKGK0}), yields the same direct linear response result, Eq.~\eqref{Dkappamatrix}.

\subsubsection{Hybrid fluctuation-dissipation results}
In the hybrid fluctuation-dissipation relation approach, $\delta\mathsf{K}$ is given by Eq.~\eqref{hybrid}. 
The expression is almost identical to $\delta\rm{K}^{\rm ii}$ (Eq.~\eqref{K1}) except that it has an overall minus sign and that $\mathsf{M}_{\gamma}$ is transposed. Hence, we can use the result of Eq.~\eqref{Kii} and introduce those two modifications, so that
\begin{align}\delta K_{\mu\nu} & =-\sum_{\alpha,\beta,\gamma,\delta}\frac{q^{4}q_{\beta}q_{\delta}M_{\alpha}^{\mu\rho}M_{\gamma}^{\sigma\nu}}{(2\pi)^{3}k_{B}T\varepsilon^{2}}\int\limits _{0}^{\infty}{\rm d}t\int{\rm d}\boldsymbol{k}\frac{k_{\rho}k_{\sigma}}{k^{4}}\\
 & \quad\times(A_{\alpha\gamma}^{nn,-\boldsymbol{k}}(t)A_{\beta\delta}^{nn,\boldsymbol{k}}(t)-A_{\alpha\delta}^{nn,-\boldsymbol{k}}(t)A_{\beta\gamma}^{nn,\boldsymbol{k}}(t))\big|_{E_0=0}.\nonumber
\end{align}
Computing the integral (see Supplemental Material) yields the same result as the two previous methods, Eq.~\eqref{Dkappamatrix}.

\section{Summary and Discussion}
\label{Conclusion}

To sum up, we have computed, using odd SDFT \cite{avni2026odd}, the conductivity tensor
\begin{equation}
\mathsf{K}=\begin{pmatrix}\kappa & \kappa^{\text{H}} & 0\\
-\kappa^{\text{H}} & \kappa & 0\\
0 & 0 & \kappa^{\parallel}
\end{pmatrix}
\end{equation}
 of a model of a dilute symmetric $q\!:\!q$ electrolyte where electrophoretic effects (i.e. advection by the solvent) are ignored. 
Up to linear order in $B$, the components read
\begin{align}
        \kappa=\kappa^{\parallel} &= \mathcal{A} {c} + \mathcal{B} {c}^{3/2} + \cdots \\
        \label{kappa_H_discussion}
        \kappa_{\text{H}} &= \mathcal{A}_\text{H} {c} + \mathcal{B}_\text{H} {c}^{3/2} + \cdots
\end{align}
where ${c}$ is the ion concentration and in which
\begin{equation*}
\mathcal{A} = q^{2} (\gamma_{+}^{-1} + \gamma_{-}^{-1})
\quad
\text{and}
\quad
\mathcal{B} = -\frac{\sqrt{2\pi}}{3}(2-\sqrt{2}) \mathcal{A} |z|^3\ell_{\text{B}}^{3/2}
\end{equation*}
while
\begin{equation*}
\mathcal{A}_\text{H} = q^{3}B \big(\gamma_{+}^{-2} - \gamma_{-}^{-2}\big) 
\quad
\text{and}
\quad
\mathcal{B}_\text{H} = 2 q B(\gamma_{+}^{-1}-\gamma_{-}^{-1}) \mathcal{B}
\end{equation*}
where $\ell_{\text{B}}=e^{2}/(4\pi\varepsilon k_{B}T)$ is the Bjerrum length and $z=q/e$ is the ionic valency.
When $B=0$, we recover the result of the relaxation effect computed originally by Debye, H{\"u}ckel and Onsager~\cite{onsager1927report,onsager2002irreversible}, see Ref.~\cite{avni2022conductivity} for a simplified computation. 
At finite magnetic field, our results match with the relaxation part of the cluster expansion calculation done in 1965 by Friedman~\cite{friedman1965calculation}, despite being setup in a completely different way. 
In particular, the calculation of Ref.~\cite{friedman1965calculation} involves thirteen diagrams, five of which contribute to the relaxation part, the sum of which turns out to precisely match our results truncated at linear order in $B$, as shown in  Appendix~\ref{FridmanComparison2}. 
We have also computed the ${c}$ and ${c}^{3/2}$ coefficients beyond linear order in $B$, see full expressions in Eqs.~\eqref{K0_approx} and~\eqref{Dkappamatrix}. Our results can be straightforwardly generalized to more species and asymmetric electrolytes.

We emphasize that our calculation ignores two important effects related to the solvent: fluid advection and dielectric response.
These ingredients have been ignored in order to focus on the effect of the magnetic field on the formalism, not because they can be neglected.
In the Debye-Hückel-Onsager theory of electrolytes without magnetic field, advection produces an additional correction to the conductivity, known as the electrophoretic effect~\cite{robinson2002electrolyte,bockris1998modern}, which, in standard electrolytes, is comparable in magnitude to the relaxation effect analogous to the correction calculated here. 
The calculation of Friedman~\cite{friedman1965calculation} strongly suggests that a similar effect is at play for the Hall conductivity.

In addition, experimental and theoretical results suggest that the interplay between dielectric response and fluid advection play a key role in the Hall response of electrolytes, even at the lowest order in concentration.
Indeed, the Hall effect in ordinary electrolyte solutions has been experimentally reported by several groups~\cite{Mergault1964,LaforgueKantzer1964,LaforgueKantzer1965,Wendhausen1968,Hanebeck1974,Gerard1990,Frank1972,Abbes1982,Bellissent1971,meton1976hall,Meton1976b,Abbes1980,Hanebeck1974}, see Ref.~\cite{Frank1972} for a discussion of the early history and Refs.~\cite{wolynes1980dynamics} and \cite{Fahidy1983} for reviews.
\footnote{Similar effects have been reported in molten salts \cite{Plummer1972,GuerinOuler1982,GuerinOuler1983} and solid electrolytes \cite{Knotek1977,Stuhrmann2002}, see also \cite{Gagliardi2016,Mouhat2013,Coretti2020} for simulations.}. 
\nocite{Plummer1972,GuerinOuler1982,GuerinOuler1983, Knotek1977,Stuhrmann2002,Gagliardi2016,Mouhat2013,Coretti2020} 
For instance, Ref.~\cite{meton1976hall} reported the measurement of the Hall response of 25 different electrolytes using AC electric and magnetic fields. 
These measurements also revealed a surprising phenomenon. 
In the infinite-dilution limit, the theory considered in our work (as well as in early works on the Hall effect in electrolytes) predicts the Hall conductivity (at linear order in $B$)
\begin{equation}
\label{InfiniteDiluteTheory}
    \kappa_0^{\rm H}
    =
    q^3 {c} B
    \left(
    \gamma_+^{-2}
    -
    \gamma_-^{-2}
    \right).
\end{equation}
Experimental data is instead consistent with
\begin{equation}
\label{InfiniteDiluteExp}
    \kappa_0^{\rm H}
    =
    q^3 {c} B
    \left(
    h_+\gamma_+^{-2}
    -
    h_-\gamma_-^{-2}
    \right)
\end{equation}
where the quantities $h_+$ and $h_-$, called ionic Hall numbers, are of order unity but generally differ from one. In particular, cations and anions of the same Pauling ionic radii are reported to exhibit different Hall numbers~\cite{meton1976hall}.
This observation suggests that the response of an ion to the Lorentz force is characterized by a different mobility from its response to an electric field~\cite{wolynes1980dynamics}, effectively replacing the Lorentz force $\bm{F} = q (\bm{E} + \bm{v} \times \bm{B})$ with a modified Lorentz force $\bm{F}_{\text{eff}} = q (\bm{E} + h \bm{v} \times \bm{B})$ where $\bm{E}$ and $\bm{B}$ are the applied electric and magnetic fields, while $\bm{v}$ is the drift velocity of the ion of charge $q$ and ionic Hall number $h$.
This phenomenological modification of the magnetic Lorentz force allows one to fit the experimentally observed data at infinite dilution~\cite{meton1976hall,hubbard1981electrohydrodynamic,wolynes1980dynamics} and can even be included in calculations of the Hall conductance at finite dilution, roughly matching experimental data \cite{Budkov2026}.
Theoretical studies have attributed this renormalization of the magnetic Lorentz force to the response of the viscous dielectric solvent to motion of ions in an external electromagnetic field~\cite{sung1987microscopic,Friedman1968,Harris1970,Harris1972,Khanh1972,hubbard1981electrohydrodynamic,Kroh1988,Chechetkin1989,kroh1990hall}. 
The idea is that a moving ion polarizes and sets the surrounding solvent into motion. 
The resulting motion of the solvent polarization couples to the magnetic field and, in turn, produces an additional drag on the ion. 
These effects have been qualitatively reproduced by theories that incorporate such solvent-mediated effects, but these have not yet quantitatively reproduced the experimentally inferred ionic Hall numbers~\cite{gerard1990hall}.
A combination of our approach to handle the magnetic field with continuum theories of liquid dielectrics \cite{Kornyshev1986,Belaya1986,Levy2012,Berthoumieux2019,Berthoumieux2021,Illien2024,Blossey2022,Blossey2022b,Berthoumieux2024,dean2026dielectric,adar2018dielectric} could shed light on these aspects.

From a broader perspective, our work opens avenues for the theoretical analysis of electrolytes and ionic liquids under magnetic field.
For instance, a detailed understanding of the Hall effect may help shed light on the collective ion dynamics within room-temperature ionic liquids, with potential applications to supercapacitors or batteries~\cite{Feng2019}. 
Beyond the Hall effect, it could provide insights on electrochemistry under magnetic fields~\cite{Gatard2020,Luo2022,Chen2024,Li2026}, with potential applications to nanoscale ion transport in systems from biological ion channels to nanoporous electrodes~\cite{Chun2015,Kavokine2021,Marbach2019}.

\bigskip\bigskip

\begin{acknowledgments}
Y.A. dedicates this work to the memory of Rudolf Podgornik, a generous mentor and a true gentleman. V.V. acknowledges partial support from the Army Research Office under grant W911NF-22-2-0109 and W911NF-23-1-
0212. M.F. and V.V acknowledge partial support from
the France Chicago center through a FACCTS grant.
This research was partly supported from the National
Science Foundation through the Center for Living Systems (grant no. 2317138), the Simons Foundation and
the Chan Zuckerberg Foundation. \end{acknowledgments}

\appendix
\section{Relation to chiral active matter} \label{chiral_active}
In this appendix, we show how the odd Dean-Kawasaki equation can be adapted to describe certain models of chiral active matter. 

\subsection{Models of chiral active matter}

Consider the paradigmatic model of active particles given by~\cite{te2026colloquium,fodor2018statistical,martin2021statistical}
\begin{equation} \label{relation2}
\dot{\boldsymbol{x}}_{i}=-\frac{1}{\gamma}\nabla_{\bm{x}_{i}}U(\boldsymbol{x}_{1},...,\boldsymbol{x}_{N} )+v_{0}\boldsymbol{n}_{i}(t),
\end{equation}
where $\bm{n}_i$ specifies the direction of self-propulsion of particle $i$, located at position $\bm{x}_i$, $v_0$ is the self-propulsion speed, $U$ is the interaction potential, and $\gamma$ is the particle friction coefficient.
For simplicity, we have set to zero the translational noise that is associated with $\gamma$.

Chiral active particles may be captured within this description by endowing the dynamics of the orientation $\bm{n}_i$ with a given chirality \cite{caprini2023chiral,hargus2025odd}.
The simplest examples are (i) chiral active Brownian particles and (ii) chiral active Ornstein-Uhlenbeck particles. Both are typically studied in two dimensions, as we shall do here, although generalizations to three dimensions are possible~\cite{Kummel2013,Sevilla2016,Kuroda2025}.

\subsubsection{Chiral active Brownian particles}

Chiral active Brownian particles follow the dynamics
\begin{subequations}
     \label{CAB}
    \begin{align}
\bm{n}_{i}(t)&=(\cos\theta_{i},\sin\theta_{i})\\
\dot{\theta}_{i}&  =-\omega_{\rm R}+\sqrt{\frac{2}{\tau_{\rm R}}}\xi^{\rm R}_{i} \label{ABP_theta}
    \end{align}
\end{subequations}
where $\xi^{\rm R}_{i}$ is a scalar Gaussian white noise with zero mean and $\langle \xi^{\rm R}_{i}(t)\xi^{\rm R}_{j}(t')\rangle=\delta_{ij}\delta(t-t')$. The minus sign before $\omega_{\rm R}$ in Eq.~\eqref{ABP_theta} was added so that the chiralities of this model is the same as in the chiral active Ornstein-Uhlenceck model below.

The stochastic process $\bm{n}_i(t)$ can be seen as a noise.
It satisfies $\langle\bm n_i(t)\rangle=0$ and
\begin{equation}
\langle\boldsymbol{n}_{i}(t)\boldsymbol{n}_{j}^{\mathsf{T}}(0)\rangle=\frac{\delta_{ij}}{2}\left(\begin{array}{cc}
\langle\cos(\Delta\theta_{i})\rangle & -\langle\sin(\Delta\theta_{i})\rangle\\
\langle\sin(\Delta\theta_{i})\rangle & \langle\cos(\Delta\theta_{i})\rangle
\end{array}\right)
\end{equation}
where $\Delta\theta_i \equiv \theta_i (t)-\theta_i (0)$ and we used the fact the system is rotationally invariant so $\langle\cos(\theta_i(t)+\theta_i(0))\rangle$ and $\langle\sin(\theta_i(t)+\theta_i(0))\rangle$ are zero.
It can be shown that $\langle e^{i\left(\theta_{i}\left(t\right)-\theta_{i}\left(0\right)\right)}\rangle=e^{-i\omega_{\rm R} t}e^{-|t|/\tau_{\rm R}}
$~\cite{avni2025dynamical}, and therefore
\begin{equation} \label{nn_corr}
\langle\boldsymbol{n}_{i}(t)\boldsymbol{n}^{\mathsf T}_{j}(0)\rangle=\frac{\delta_{ij}}{2}e^{-\left|t\right|/\tau_{\rm R}}
\begin{pmatrix}
\cos\left(\omega_{\rm R} t\right) & \sin\left(\omega_{\rm R} t\right)\\
-\sin\left(\omega_{\rm R} t\right) & \cos\left(\omega_{\rm R} t\right)
\end{pmatrix}
\end{equation}
Note that the process $\bm{n}_i(t)$
is not Gaussian, and our discussion only holds when it can be approximated as a Gaussian process. 
In general, one would have to take into account the non-Gaussian nature of the noise into account~\cite{vanKampen1961,Kanazawa2012,Kanazawa2015,Fodor2018,Lucente2023}.

\subsubsection{Chiral active Ornstein-Uhlenbeck particles}

Chiral active Ornstein-Uhlenbeck particles follow
\begin{equation}
\dot{\boldsymbol{n}}_{i}  =-\frac{\boldsymbol{n}_{i}}{\tau_{\rm R}}+\omega_{\rm R}\boldsymbol{n}_{i}\times\boldsymbol{z}+\sqrt{\frac{1}{\tau_{\rm R}}}\boldsymbol{\chi}^{\rm R}_{i}
\end{equation}
where ${\bm\chi}^{\rm R}_{i}$ is a vector Gaussian white noise with zero mean and $\langle \boldsymbol{\chi}^{\rm R}_{i}(t)(\boldsymbol{\chi}_{j}^{\rm R}(t'))^{\mathsf{T}}\rangle=\mathsf{I}\delta_{ij}\delta(t-t')$.

We can obtain an explicit expression for $\bm{n}_i(t)$, namely
\begin{subequations}
    \begin{align}
    \!\!\!
\boldsymbol{n}_{i}(t) & =\sqrt{\frac{1}{\tau_{{\rm R}}}}\int\limits _{-\infty}^{t}\exp\left[-\gamma^{-1}\mathsf{M}^{-1}_{{\rm R}}\frac{t-s}{\tau_{{\rm R}}}\right]\boldsymbol{\chi}^{\rm R}_{i}(s){\rm d}s
\end{align}
where initial conditions have been neglected, and in which
\begin{align}
\mathsf{M}_{{\rm R}} & \equiv\frac{1}{\gamma}
\begin{pmatrix}
1 & -\omega_{{\rm R}}\tau_{{\rm R}}\\
\omega_{{\rm R}}\tau_{{\rm R}} & 1
\end{pmatrix}^{-1}
\end{align}
\end{subequations}
from which it follows that $\bm n_i$ is a vector Gaussian noise with $\langle\bm n_i(t)\rangle=0$ and
\begin{equation}
\langle\boldsymbol{n}_{i}(t)\boldsymbol{n}_{j}^{\mathsf{T}}(0)\rangle=\frac{\delta_{ij}}{\tau_{{\rm R}}}\int\limits _{-\infty}^{{\rm min}(t,0)}e^{-\mathsf{M}_{{\rm R}}^{-1}\frac{t-s}{\gamma\tau_{{\rm R}}}}e^{(\mathsf{M}_{{\rm R}}^{\mathsf{T}})^{-1}\frac{s}{\gamma\tau_{{\rm R}}}}{\rm d}s.
 \end{equation}
Performing the same steps that were used to calculate $\langle\boldsymbol{\Xi}(\boldsymbol{x},t)\boldsymbol{\Xi}^{\mathsf{T}}(\boldsymbol{x}',t')\rangle$ in Eq.~\eqref{correlationfinitem}, we obtain exactly the same correlations as in Eq.~\eqref{nn_corr}, except that here the noise is Gaussian by construction.

\subsection{Odd SDFT for chiral active particles}

We now focus on the limit of small $\tau_{\rm R}$ of the above two models, in which $\tau_{\rm R}$ is much smaller than other relevant time scale of the dynamics, such as the one induced by the potential $U$. From literature on active models without chirality we know that $v_0^2 \tau_R$ is the coarse-grained diffusion coefficient~\cite{cates2013active,solon2015active}. Hence, we take the limit $\tau_R\to 0$ together with $v_0^2\tau_R = {\rm const}$, which amounts to very fast re-orientations and high speed.
We now distinguish two cases:
\begin{itemize}[nosep,left=0pt,label=--]
    \item the standard overdamped limit $\tau_{\rm R}\to 0$ at fixed $\omega_{\rm R}$, where chirality disappears in the limit.
    \item the chiral overdamped limit $\tau_{\rm R}\to0$ at fixed $\omega_{\rm R}\tau_R$, where chirality remains in the limit through a nonwhite noise.
\end{itemize}

Indeed, if the limit $\tau_{\rm R}\to 0$ is taken while $\omega_{\rm R}$ is held fixed,
 we obtain $\langle\boldsymbol{n}_{i}(t)\boldsymbol{n}_{j}^{\mathsf{T}}(t')\rangle\to\frac{\delta_{ij}\tau_{{\rm R}}}{2}\mathsf{I}\delta(t-t')$, in which case any trace of chirality vanishes. 
However, an interesting limit is $\tau_{\rm R}\to0$ while $\omega_{\rm R}\tau_R$ is held fixed. Then, Eq.~\eqref{nn_corr} reduces to
\begin{equation} \label{nn_limit}
\langle\boldsymbol{n}_{i}(t)\boldsymbol{n}_{j}^{\mathsf{T}}(t')\rangle\to\frac{\delta_{ij}\gamma\tau_{{\rm R}}}{2}\left[\mathsf{M}_{{\rm R}}\delta_{+}(t-t')+\mathsf{M}_{{\rm R}}^{\mathsf{T}}\delta_{-}(t-t')\right].
\end{equation}
Comparing Eqs.~\eqref{relation2} and~\eqref{nn_limit} with Eqs.~\eqref{Chun_result1}--\eqref{Chun_result3}, we find that the low-mass limit of particles moving under a Lorentz force, considered in the main text, can be mapped onto the chiral active matter models described above in the limit $\tau_{\rm R}\to 0$, with $v_0^2\tau_{\rm R}$ and $\omega_{\rm R}\tau_{\rm R}$ held constant, by making the substitutions
\begin{subequations}
    \begin{align}
        k_{B}T & \to\frac{\gamma v_{0}^{2}\tau_{\rm R}}{2}\\ \label{rotatedF}
\mathsf{M}\frac{\partial U}{\partial\boldsymbol{x}_{i}} & \to\frac{1}{\gamma}\frac{\partial U}{\partial\boldsymbol{x}_{i}}\\
\mathsf{M}&\to\mathsf{M}_{\rm R}.
    \end{align}
\end{subequations}
As a corollary, the odd Dean–Kawasaki equation \eqref{EOM} describes chiral active particles under these replacements (one can verify that no step in the derivation is sensitive to them), leading to
\begin{subequations}
\label{active_chiral_DK}
\begin{align} & \partial_{t}n(\boldsymbol{x},t)=-\boldsymbol{\nabla}\cdot\boldsymbol{J}\\
 & \boldsymbol{J}=-\frac{\gamma v_{0}^{2}\tau_{{\rm R}}}{2}\mathsf{M}_{{\rm R}}\boldsymbol{\nabla}n-\frac{1}{\gamma}n\nabla\frac{\delta U}{\delta n}+\boldsymbol{\chi}(\boldsymbol{x},t)\\
 & \langle\boldsymbol{\chi}(\boldsymbol{x},t)\boldsymbol{\chi}^{\mathsf{T}}(\boldsymbol{x}',t')\rangle=n(\boldsymbol{x},t)\delta(\boldsymbol{x}-\boldsymbol{x}')\mathsf{C}(t-t')\\
 & \mathsf{C}(t)=\frac{\gamma v_{0}^{2}\tau_{{\rm R}}}{2}\bigg[\mathsf{M}_{{\rm R}}\delta_{+}(t)+\mathsf{M}_{{\rm R}}^{\mathsf{T}}\delta_{-}(t)\bigg].
\end{align}
\end{subequations}

Equation~\eqref{active_chiral_DK} can be solved perturbatively and be analyzed similarly to the odd Dean-Kawasaki equation outlined in the main text. However, it does not satisfy the Green-Kubo relation in Eq.~\eqref{GK_M}. The reason is that, after rotating the force (Eq.~\eqref{rotatedF}), $\mathsf{M}_{\rm R}$ no longer plays the role of the mobility tensor, and the fluctuation-dissipation relation between the mobility and the noise correlations is lost. 

\section{Collective mobility} \label{CollectiveMobilitySec}
In this appendix, we calculate the response of a system described by the odd Dean-Kawasaki equation \eqref{EOM} to a spatiotemporal external force field, $\bm F(\bm x,t)=-\nabla V_{\rm ext}(\bm x,t)$. This extends the calculation of the collective mobility performed in Sec.~\ref{GK_nonint}, which is limited to noninteracting particles subject to a fixed uniform force.

For a spatiotemporal force, the definition of the collective mobility tensor in Eq.~\eqref{LF} generalizes to~\cite{Kubo1991}
\begin{equation} \label{M_general_def}
\langle\bm{J}(\boldsymbol{x},t)\rangle=\int{\rm d}t'\int{\rm d}\bm{x}'\mathsf{L}(\bm{x}-\bm{x}',t-t')\bm{F}(\bm{x}',t')
\end{equation}
and is more compactly written in Fourier space as,
\begin{equation} \label{def_Mkw}
\langle\hat{\bm{J}}_{\boldsymbol{k},\omega}\rangle=\hat{\mathsf{L}}_{\boldsymbol{k},\omega}\hat{\bm{F}}_{\boldsymbol{k},\omega}.
\end{equation}
Here, $\hat{f}_{\bm k,\omega}$ denotes the Fourier transform of $f(\bm x, t)$ in both space and time, $\hat{f}_{\boldsymbol{k},\omega}=\int{\rm d}t\int{\rm d}\boldsymbol{x}\,f(\boldsymbol{x},t)e^{-i(\boldsymbol{k}\cdot\boldsymbol{x}+\omega t)}
$. In the following, we derive $\hat{\mathsf{L}}$ for a system described by Eq.~\eqref{EOM} using direct linear response and Green-Kubo relation.

\subsection{Direct linear response} \label{DlRmobility}
In the direct response approach, we include the applied force and take the ensemble average of Eq.~\eqref{EOM2}, yielding
\begin{align}
\langle\boldsymbol{J}\rangle & =-k_{B}T\mathsf{M}\boldsymbol{\nabla}\langle n\rangle+\langle n\rangle\mathsf{M}\boldsymbol{F}(\boldsymbol{x},t)\\
 & -\mathsf{M}\boldsymbol{\nabla}\int{\rm d}\boldsymbol{x}'V(\boldsymbol{x}-\boldsymbol{x}')\langle n(\boldsymbol{x},t)n(\boldsymbol{x}',t)\rangle \nonumber.
\end{align}
While we cannot solve the odd Dean-Kawasaki equation (Eq.~\eqref{EOM}) exactly for $n(\bm x,t)$, we obtain an approximate solution by making two approximations. First, we apply the mean-field approximation, in which fluctuations about the mean are neglected, so that $n(\bm x,t)=\langle n(\bm x,t)\rangle$ and we can therefore replace $\langle n(\boldsymbol{x},t)n(\boldsymbol{x}',t)\rangle$ by $\langle n(\boldsymbol{x},t)\rangle\langle n(\boldsymbol{x}',t)\rangle
$ (this can be seen as the zeroth order in an expansion, see the main text for an example going beyond mean-field). Second, since we are are interested in the linear response, we expand $\langle n(\boldsymbol{x},t)\rangle$ and $\langle \bm J(\boldsymbol{x},t)\rangle$ in powers of $\bm F$, and truncate at linear order, yielding
\begin{subequations}
    \begin{align}
\langle n(\boldsymbol{x},t)\rangle & ={c}+n_{F}(\boldsymbol{x},t)+\mathcal{O}(F^{2})\\
\langle\bm{J}(\boldsymbol{x},t)\rangle & =\bm{J}_{F}(\boldsymbol{x},t)+\mathcal{O}(F^{2})
\end{align}
\end{subequations}
where $n_F,J^{\mu}_F\sim \mathcal{O}(F)$ with $F\equiv |\bm F|$, leading to
\begin{subequations}
\label{linearized_mf_eom}
\begin{align}\partial_{t}n_{F}(\boldsymbol{x},t) & =-\boldsymbol{\nabla}\cdot\boldsymbol{J}_{F}\\ \label{J1}
\boldsymbol{J}_{F}  =-k_{B}T&\mathsf{M}\boldsymbol{\nabla}n_{F}+{c}\mathsf{M}{\bm F}-{c}\mathsf{M}\boldsymbol{\nabla}V*n_{F}.
\end{align}
\end{subequations}
The solution of the linearized equation is given in Fourier space by
\begin{equation} \label{n1sol}
\hat{n}^{\boldsymbol{k},\omega}_{F}=-{c}\frac{ik_{\mu}M_{\mu\nu}\hat{F}_{\boldsymbol{k},\omega}^{\nu}}{M_{\rho\sigma}k_{\rho}k_{\sigma}(k_{B}T+{c}\tilde{V}_{\boldsymbol{k}})+i\omega}.
\end{equation}
Substituting Eq.~\eqref{n1sol} into the Fourier transform of Eq.~\eqref{J1}, and then inserting the resulting $\hat{\bm J}_F^{\bm k,\omega}$ into the definition of the collective mobility, Eq.~\eqref{def_Mkw}, yields
\begin{equation} \label{Mkw_result}
\hat{L}_{\boldsymbol{k},\omega}^{\mu\nu}={c}\left[M_{\mu\nu}-\frac{M_{\mu\rho}M_{\sigma\nu}k_{\rho}k_{\sigma}}{M_{\varrho\varsigma}k_{\varrho}k_{\varsigma}+\frac{i\omega}{k_{B}T+{c}\tilde{V}_{\boldsymbol{k}}}}\right].
\end{equation}
We note several things about Eq.~\eqref{Mkw_result}. First, it is evidently non-symmetric due to the non-symmetric structure of $\mathsf{M}$, but becomes symmetric in the absence of a magnetic field where $\mathsf{M}\to\gamma^{-1}\mathsf{I}$ leads to
\begin{equation}
\hat{L}_{\boldsymbol{k},\omega}^{\mu\nu}(B=0)=\frac{{c}}{\gamma}\left[\delta_{\mu\nu}-\frac{k_{\mu}k_{\nu}}{k^{2}+\frac{i\gamma\omega}{k_{B}T+{c}\tilde{V}_{\boldsymbol{k}}}}\right].
\end{equation}
Second, if the external force is spatially homogeneous, the response is described by the $\bm k\to0$ limit of $\mathsf{L}_{\bm k, \omega}$ and is simply
\begin{equation} \label{non_interacting_M}
\mathsf{L}_{\bm k \to 0, \omega}={c}\mathsf{M}=\frac{{c}}{\gamma\left(1+\frac{q^{2}B^{2}}{\gamma^{2}}\right)}\left(\begin{array}{ccc}
1 & \frac{qB}{\gamma} & 0\\
-\frac{qB}{\gamma} & 1 & 0\\
0 & 0 & 1+\frac{q^{2}B^{2}}{\gamma^{2}}
\end{array}\right).
\end{equation}
Notably, the collective mobility given by Eq.~\eqref{non_interacting_M} is independent of the inter-particle interactions. While we employed the mean-field and weak force approximations to get Eq.~\eqref{non_interacting_M}, this result is exact (i.e. valid beyond the mean-field approximation and for any force strength). To show that, we note that the average current due to a spatially homogeneous force is given by
\begin{equation} \label{delta_M}
\langle\bm{J}\rangle={c}\mathsf{M}\bm{F}-\mathsf{M}\int d\boldsymbol{x}'\boldsymbol{\nabla}V(\boldsymbol{x}-\boldsymbol{x}')\langle n(\boldsymbol{x},t)n(\boldsymbol{x}',t)\rangle.
\end{equation}
where we replaced $\langle n(\boldsymbol{x},t)\rangle= {c}$ due to the system's homogeneity. We notice that $\langle n(\boldsymbol{x},t)n(\boldsymbol{x}',t)\rangle$ is even in $\bm x-\bm x'$ (since it must equal $\langle n(\boldsymbol{x}',t)n(\boldsymbol{x},t)\rangle$) while $\boldsymbol{\nabla}V(\boldsymbol{x}-\boldsymbol{x}')$ is odd (as $V(\bm{x})=V(|\bm{x}|)$), hence the integral in Eq.~\eqref{delta_M} vanishes and we obtain Eq.~\eqref{non_interacting_M}. This result can be understood intuitively: since all particles experience the same force, they simply drift together, corresponding to a change of reference frame, while their relative separations remain unchanged. This is no longer the case when different forces act on different particle species, as for an electrolyte under electric field. Note that this result does not imply that the effective mobility or diffusion of a single particle is unaffected by interactions, since here we consider the response to a global force acting on all particles rather than to a force applied to a single particle. Interaction-induced renormalization of the single-particle mobility can instead be studied within a tracer-particle formalism~\cite{demery2014generalized}.

\subsection{Green-Kubo formula} \label{mobilityGK}
For a spatiotemporal response, the Green-Kubo relation in Eq.~\eqref{GK_M} generalizes to~\cite{Kubo1991}
\begin{equation} \label{GK_Mkw}
\hat{\mathsf{L}}_{\boldsymbol{k},\omega}  =\frac{1}{\mathcal{V}k_{B}T}\int_{0}^{\infty}{\rm d}t\,e^{-i\omega t}\langle\tilde{\boldsymbol{J}}_{\boldsymbol{k}}(t)\tilde{\boldsymbol{J}}_{-\boldsymbol{k}}(0)^{\mathsf{T}}\rangle_0.
\end{equation}
To find the current fluctuations in the unperturbed system we linearize the odd Dean-Kawasaki equation in small fluctuations about the mean field. To keep track of the orders in the expansion we multiply the noise by a constant $\bm \chi(\bm x,t)\to\epsilon \bm \chi(\bm x,t)$ (to be set to $1$ later on), and expand $n(\boldsymbol{x},t)\approx {c}+\epsilon\delta n(\boldsymbol{x},t)+\mathcal{O}(\epsilon^2)$ and $\epsilon \bm \chi(\bm x,t)\approx \epsilon\bm \chi_{0}(\bm x,t)+\mathcal{O}(\epsilon^2)$ where $\boldsymbol{\chi}_0(\boldsymbol{x},t)$ is an additive (rather than multiplicative) Gaussian noise with zero mean and the correlations
\begin{equation}
\langle\boldsymbol{\chi}_{0}(\boldsymbol{x},t)\boldsymbol{\chi}_{0}^{\mathsf{T}}(\boldsymbol{x}',t')\rangle={c}\delta(\boldsymbol{x}-\boldsymbol{x}')\mathsf{C}(t-t').
\end{equation}

Equating terms linear in $\epsilon$, we obtain
\begin{subequations}
\begin{align}
\partial_{t}\delta n(\boldsymbol{x},t)&=-\boldsymbol{\nabla}\cdot\boldsymbol{J}\\
\label{J_linearized}\boldsymbol{J}=-k_{B}T&\mathsf{M}\boldsymbol{\nabla}\delta n-{c}\mathsf{M}\boldsymbol{\nabla}V*\delta n+\boldsymbol{\chi}_{0}
\end{align}
\end{subequations}
which simplifies in Fourier space to
\begin{align}\nonumber
\partial_{t}\delta\tilde{n}_{\boldsymbol{k}}(t)=&-k_{\mu}k_{\nu}M_{\mu\nu}(k_{B}T+{c}\tilde{V}_{\boldsymbol{k}})\delta\tilde{n}_{\boldsymbol{k}}(t)\\ &-ik_{\mu}\tilde{\chi}_{0,\mu}^{\boldsymbol{k}}(t)
\end{align}
and has the steady state solution
\begin{equation}
\delta\tilde{n}_{\boldsymbol{k}}(t)=-ik_{\mu}\!\!\int\limits _{-\infty}^{t}\!{\rm d}s\,e^{-k_{\rho}k_{\sigma}M_{\rho\sigma}(k_{B}T+{c}\tilde{V}_{\boldsymbol{k}})(t-s)}\tilde{\chi}_{0,\mu}^{\boldsymbol{k}}(s).
\end{equation}
The steady-state density-density correlation function is easily calculated to be
\begin{equation} \label{nn}
\langle\delta\tilde{n}_{\boldsymbol{k}}(t)\,\delta\tilde{n}_{\boldsymbol{k}'}(t')\rangle_0  =(2\pi)^{3}\delta(\boldsymbol{k}+\boldsymbol{k}')A_{\boldsymbol{k}}^{nn}(t-t')
\end{equation}
with 
\begin{equation}
A_{\boldsymbol{k}}^{nn}(t)  \equiv\frac{{c}}{1+{c}\tilde{V}_{\boldsymbol{k}}/k_{B}T}e^{-k_{\mu}k_{\nu}M_{\mu\nu}(k_{B}T+{c}\tilde{V}_{\boldsymbol{k}})\left|t\right|}
\end{equation}
while the density-noise correlations are given by
\begin{equation}
 \label{nchi}
\langle\delta\tilde{n}_{\boldsymbol{k}}(t)\tilde{\chi}_{\boldsymbol{k}',\mu}^{0}(t')\rangle_0=(2\pi)^{3}\delta(\boldsymbol{k}+\boldsymbol{k}')A_{\boldsymbol{k},\mu}^{n\boldsymbol{\chi}}(t-t')
\end{equation}
with
\begin{align} \label{Anchi}
     & A_{\boldsymbol{k},\mu}^{n\boldsymbol{\chi}}(t)\equiv-ik_{\nu}k_{B}T{c}e^{-k_{\rho}k_{\sigma}M_{\rho\sigma}(k_{B}T+{c}\tilde{V}_{\bm{k}})t}\\
 & \quad\quad\quad\quad\quad\times(M_{\nu\mu}+M_{\mu\nu})\Theta(t)\nonumber.
\end{align}

The current-current correlations can be computed from Eq.~\eqref{J_linearized} in Fourier space as
\begin{align}\label{JJ_{c}orr}\nonumber\nonumber\langle\tilde{J}_{\boldsymbol{k}}^{\mu}(t)\tilde{J}_{-\boldsymbol{k}}^{\nu}(0)\rangle_{0} & =(k_{B}T+{c}\tilde{V}_{\bm{k}})^{2}\\
\nonumber & \times M_{\mu\rho}M_{\nu\sigma}k_{\rho}k_{\sigma}\langle\delta\tilde{n}_{\boldsymbol{k}}(t)\delta\tilde{n}_{-\boldsymbol{k}}(0)\rangle_{0}\\
\nonumber & -(k_{B}T+{c}\tilde{V}_{\bm{k}})\\
\nonumber & \times\bigg[iM_{\mu\rho}k_{\rho}\langle\delta\tilde{n}_{\boldsymbol{k}}(t)\tilde{\chi}_{-\boldsymbol{k},\nu}^{0}(0)\rangle_{0}\\
 & -iM_{\nu\sigma}k_{\sigma}\langle\delta\tilde{n}_{\boldsymbol{-k}}(0)\tilde{\chi}_{\boldsymbol{k},\mu}^{0}(t)\rangle_{0}\bigg]\\
\nonumber & +\langle\tilde{\chi}_{\boldsymbol{k},\mu}^{0}(t)\tilde{\chi}_{-\boldsymbol{k},\nu}^{0}(0)\rangle_{0}.
\end{align}
For $t>0$, substituting the correlations in Eqs.~\eqref{nn}-\eqref{Anchi}, we obtain
\begin{align}\label{JJMobility}\frac{\langle\tilde{J}_{\boldsymbol{k}}^{\mu}(t)\tilde{J}_{-\boldsymbol{k}}^{\nu}(0)\rangle_{0}}{{c}\mathcal{V}k_{B}T} & =\frac{C_{\mu\nu}(t)}{k_{B}T}-(k_{B}T+{c}\tilde{V}_{\boldsymbol{k}})\nonumber\\
 & \times M_{\mu\rho}M_{\sigma\nu}k_{\rho}k_{\sigma}e^{-k_{\varrho}k_{\varsigma}\mathsf{M}_{\varrho\varsigma}(k_{B}T+{c}\tilde{V}_{\bm{k}})t}.
\end{align}
Inserting Eq.~\eqref{JJMobility} into Eq.~\eqref{GK_Mkw} and performing the time integral over the second term we obtain 
\begin{align}\nonumber\hat{L}_{\boldsymbol{k},\omega}^{\mu\nu} & =\frac{{c}}{k_{B}T}\int_{0}^{\infty}{\rm d}t\,e^{-i\omega t}C_{\mu\nu}(t)\\
 & -{c}\frac{M_{\mu\rho}M_{\sigma\nu}k_{\rho}k_{\sigma}}{M_{\varrho\varsigma}k_{\varrho}k_{\varsigma}+\frac{i\omega}{k_{B}T+{c}\tilde{V}_{\bm{k}}}}.
\end{align}
Finally, using the relation $\int\limits _{0}^{\infty}{\rm d}t\,e^{-i\omega t}\,C_{\mu\nu}(t)=k_B T\, M_{\mu\nu}$ we recover the exact same $\hat{L}^{\mu\nu}_{\boldsymbol{k},\omega}$ as in Eq.~\eqref{Mkw_result}.

\section{Collective mobility of chiral active particles}
\label{mobility_active}

In this appendix, we calculate the response of a system described by the odd SDFT for chiral active particles, Eq.~\eqref{active_chiral_DK}, to a spatiotemporal external force field, $\bm F(\bm x,t)=-\nabla V_{\rm ext}(\bm x,t)$. The calculation is similar to the one carried out in Appendix~\ref{CollectiveMobilitySec} for the odd Dean-Kawasaki equation. However, since chiral active particles do not respect the Green-Kubo relation (see Appendix~\ref{active_chiral_DK} for discussion), we are restricted to the direct linear response route.

Repeating the same calculation as in Sec.~\ref{DlRmobility}, i.e., applying the mean-field approximation and truncating the equation of motion at linear order in the external force, we find the Fourier transformed number density to be
\begin{equation}
\hat{n}_{F}^{\boldsymbol{k},\omega}=-{c}\frac{ik_{\mu}\hat{F}_{\mu}^{\bm{k},\omega}}{\gamma^2 v_{0}^{2}\tau_{\rm R} k_{\rho}k_{\sigma}\mathsf{M}_{{\rm R}}^{\rho\sigma}/2+{c}k^{2}\tilde{V}_{\bm{k}}+i\gamma\omega}
\end{equation}
and the Fourier transformed density current to be
\begin{align}
\!\!
\hat{J}_{F,\mu}^{\bm{k},\omega} & =-\frac{\gamma v_{0}^{2}\tau_{\rm R}}{2}\mathsf{M}_{{\rm R}}^{\mu\nu}ik_{\nu}\hat{n}^{\boldsymbol{k},\omega}_{F}+\frac{{c}}{\gamma}\hat{F}_{\mu}^{\bm{k},\omega}
  -\frac{{c}}{\gamma}ik_{\mu}\tilde{V}_{\bm{k}}\hat{n}^{\boldsymbol{k},\omega}_{F}.
\end{align}
Inserting $\hat{J}_{F,\mu}^{\bm{k},\omega}$ into Eq.~\eqref{def_Mkw} yields the collective mobility of chiral active particles,
\begin{equation}
\hat{L}_{\boldsymbol{k},\omega}^{\mu\nu}=\frac{{c}}{\gamma}\delta_{\mu\nu}-\frac{{c}}{\gamma}\frac{\gamma^2 v_{0}^{2}\tau_{\rm R} k_{\rho}k_{\nu}\mathsf{M}_{{\rm R}}^{\mu\rho}/2+{c}k_{\mu}k_{\nu}\tilde{V}_{\bm{k}}}{\gamma^2 v_{0}^{2}\tau_{\rm R} k_{\varrho}k_{\varsigma}\mathsf{M}_{{\rm R}}^{\varrho\varsigma}/2+{c}k^{2}\tilde{V}_{\bm{k}}+i\gamma\omega}.
\end{equation}
Unlike the case of charged particles under Lorentz force (Eq.~\eqref{Mkw_result}), here the first term is symmetric, while the second term is not. Since the first term describes the response to a spatially uniform force (the $\bm k\to0$ limit), we conclude that transverse current is expected only in response to nonuniform perturbations.

\section{Derivation of the Green-Kubo relation} \label{GK_proof}
In this appendix we provide a derivation of the Green-Kubo relation for the mobility \eqref{GK_M} in the presence of a magnetic field. 
We refer to Refs.~\cite{Martin1967,Corkum1974,Pavliotis2010,Gaspard2013,Shalchi2011,Bonella2014,Bonella2017,faedi2026velocityforceautocorrelationsbrownian} for further discussions.
As we shall see, what preserves this relation is that the steady-state probability distribution is unaltered by the magnetic field (as in the Bohr-Van Leeuwen theorem).
A similar proof applies for the Green-Kubo relation for the conductivity~\eqref{GKconductivity}. 

We start from the Fokker-Planck equation of the underdamped dynamics in Eq.~\eqref{Langevin1}, perturbed by a uniform external force acting on each particle,
\begin{align}\label{FokkerPlanck}  \frac{\partial P\left({\bf X},{\bf V},t\right)}{\partial t}&=\sum_{i}\bigg[\frac{\gamma}{m}\nabla_{\boldsymbol{v}_{i}}\cdot\left(\boldsymbol{v}_{i}+\frac{k_{B}T}{m}\nabla_{\boldsymbol{v}_{i}}\right)-\boldsymbol{v}_{i}\cdot\nabla_{\boldsymbol{x}_{i}}\nonumber\\
  -\frac{\boldsymbol{F}_{i}\left({\bf X}\right)}{m}&\cdot\nabla_{\boldsymbol{v}_{i}}-\frac{q}{m}\nabla_{\boldsymbol{v}_{i}}\cdot\left(\boldsymbol{v}_{i}\times\boldsymbol{B}\right)\bigg]P\left({\bf X},{\bf V},t\right).
\end{align}
Here, $P$ is the system's probability density in phase space, with ${\bf X}\equiv\left\{ \boldsymbol{x}_{1},...,\boldsymbol{x}_{N}\right\} $ and
${\bf V}\equiv\left\{ \boldsymbol{v}_{1},...,\boldsymbol{v}_{N}\right\} $, and the force is decomposed as
\begin{equation}
\boldsymbol{F}_{i}\left({\bf X}\right)=\boldsymbol{F}_{i}^{\rm int}\left({\bf X}\right)+\boldsymbol{F}^{\rm ext}
\end{equation}
where ${\bm F}^{\rm int}_i=-\nabla_{\bm x_i}U_{\rm int}(\bm X)$ is the force acting on particle $i$ due to interactions with all the other particles, and  $\boldsymbol{F}^{\rm ext}=F_0\hat{n}$ is the external force.

Despite the presence of the Lorentz force, the steady-state solution of Eq.~\eqref{FokkerPlanck} with $F_0=0$ is independent of $\bm B$ and equals
\begin{equation} \label{Boltzmann}
    P_{\rm eq}\left({\bf X},{\bf V}\right)=Ce^{-\left(U_{{\rm int}}\left({\bf X}\right)+\sum_{i}\frac{1}{2}m\boldsymbol{v}_{i}^{2}\right)/k_{B}T}
\end{equation}
where $C$ is a normalization constant.
We can interpret the fact that the stationary distribution is independent of the magnetic field as a consequence of the fact that the Lorentz force does not perform work on the particles. 
Note, however, that the magnetic field gives rise to nonzero probability currents in phase space.

We can rewrite the Fokker-Planck equation as~\cite{risken1989fokker}
\begin{equation} \label{EqforP}
\partial_{t}P =\mathcal{L}_{{\rm eq}}^{\dagger}P+F_{0}\mathcal{L}_{{\rm pert}}^{\dagger}P
\end{equation}
where $\mathcal{L}_{{\rm eq}}^{\dagger}$ and $\mathcal{L}_{{\rm pert}}^{\dagger}$ are the operators
\begin{subequations}
    \begin{align}
\mathcal{L}_{{\rm eq}}^{\dagger} & \equiv\sum_{i}\bigg[\frac{\gamma}{m}\nabla_{\boldsymbol{v}_{i}}\cdot\left(\boldsymbol{v}_{i}+\frac{k_{B}T}{m}\nabla_{\boldsymbol{v}_{i}}\right)-\boldsymbol{v}_{i}\cdot\nabla_{\boldsymbol{x}_{i}}\\
 \nonumber& -\frac{\boldsymbol{F}_{i}^{{\rm int}}\left({\bf X}\right)}{m}\cdot\nabla_{\boldsymbol{v}_{i}}-\frac{q}{m}\nabla_{\boldsymbol{v}_{i}}\cdot\left(\boldsymbol{v}_{i}\times\boldsymbol{B}\right)\bigg]\\
\mathcal{L}_{{\rm pert}}^{\dagger} & \equiv-\hat{n}\cdot\sum_{i}\frac{1}{m}\nabla_{\boldsymbol{v}_{i}}. \label{LpertP}
\end{align}
\end{subequations}
We define the spatially-averaged current density as
\begin{equation}
    \label{J_def}
\bm{\overline{J}}\equiv\frac{1}{\mathcal{V}}\sum_{i}\boldsymbol{v}_{i}.
\end{equation}
We are interested in the steady-state expectation value
\begin{equation}
\langle\bm{\overline{J}}\rangle\equiv\int{\rm d}\Gamma\,\bm{\overline{J}}\,P_{\rm ss}({\bf X},{\bf V})
\end{equation}
where $P_{\rm ss}$ is the steady state probability distribution and we defined ${\rm d}\Gamma\equiv{\rm d}{\bf V}{\rm d}{\bf X}$.
Focusing on the linear response, we expand $P$ around $P_{\rm eq}$ to linear order in $F_{0}$ as
\begin{equation} \label{expansionP}
    P= P_{\rm eq}+F_{0}\delta P+\mathcal{O}(F_{0}^{2}).
\end{equation}
Substituting this expansion in Eq.~\eqref{EqforP} and retaining terms to linear order in $F_0$~\cite{risken1989fokker}, we obtain $\partial_{t}\delta P=\mathcal{L}_{{\rm eq}}^{\dagger}\delta P+\mathcal{L}_{{\rm pert}}^{\dagger}P_{{\rm eq}}
$
whose stationary solution can be written as
\begin{align} \nonumber
\delta P_{{\rm ss}} & =\int\limits _{-\infty}^{t}e^{\mathcal{L}_{{\rm eq}}^{\dagger}\left(t-s\right)}\mathcal{L}_{{\rm pert}}^{\dagger}P_{{\rm eq}}{\rm d}s\\
 & =-\frac{1}{m}\hat{n}\cdot\int\limits _{0}^{\infty}e^{\mathcal{L}_{{\rm eq}}^{\dagger}u}\sum_{i}\nabla_{\bm{v}_{i}}P_{{\rm eq}}{\rm d}u
\end{align}
where in the second line we used Eq.~\eqref{LpertP} and the change of variables $t-s=u$.
Using Eqs.~\eqref{Boltzmann} and~\eqref{J_def}, we find $\sum_{i}\nabla_{\bm{v}_{i}}P_{{\rm eq}}=-\frac{m\mathcal{V}}{k_{B}T}\overline{\bm{J}}P_{{\rm eq}}
$, so that
\begin{equation}
\delta P_{{\rm ss}}=\frac{\mathcal{V}}{k_{B}T}\int\limits _{0}^{\infty}e^{\mathcal{L}_{{\rm eq}}^{\dagger}t}[(\overline{\bm{J}}\cdot\hat{n})P_{{\rm eq}}]{\rm d}t.
\end{equation}

From symmetry, $\int{\rm d}\Gamma\,\bm{\overline{J}}\,P_{\rm eq}({\bf X},{\bf V})=0$, and therefore, to linear order in $F_0$
\begin{align} \label{JPss}
\langle\boldsymbol{\overline{J}}\rangle&= F_{0}\int{\rm d}\Gamma\,\bm{\overline{J}}\,\delta P_{\rm ss}({\bf X},{\bf V}) \nonumber\\
&=\frac{F_{0}\mathcal{V}}{k_{B}T}\int\limits _{0}^{\infty}{\rm d}t\int{\rm d}\Gamma\,\bm{\overline{J}}e^{\mathcal{L}_{{\rm eq}}^{\dagger}t}[(\bm{\overline{J}}\cdot\hat{n})P_{{\rm eq}}].
\end{align}
Using the adjoint relation
$\int {\rm d}\Gamma\,
f\,(e^{\mathcal{L}_{\rm eq}^{\dagger}t}h)
=
\int {\rm d}\Gamma\,
(e^{\mathcal{L}_{\rm eq}t}f)h$
where $\mathcal{L}_{\rm eq}$ is the backward generator acting on observables, we obtain
\begin{equation}
\langle\bm{\overline{J}}\rangle  =\frac{F_{0}\mathcal{V}}{k_{B}T}\int\limits _{0}^{\infty}{\rm d}t\int{\rm d}\Gamma\,(e^{\mathcal{L}_{{\rm eq}}t}\bm{\overline{J}})[(\bm{\overline{J}}\cdot\hat{n})P_{{\rm eq}}].
\end{equation}
Since $e^{\mathcal{L}_{{\rm eq}}t}\bm{\overline{J}}$ gives the equilibrium time evolution of the current observable, the phase-space integral can be identified with the corresponding two-time correlation function,
\begin{equation}
    \int{\rm d}\Gamma\,(e^{\mathcal{L}_{{\rm eq}}t}\bm{\overline{J}})[(\bm{\overline{J}}\cdot\hat{n})P_{{\rm eq}}]=\langle\bm{\overline{J}}(t)(\bm{\overline{J}}(0)\cdot\hat{n})\rangle_{0},
\end{equation}
yielding
\begin{equation}  \langle\bm{\overline{J}}\rangle=\frac{F_{0}\mathcal{V}}{k_{B}T}\int\limits _{0}^{\infty}{\rm d}t\,\langle\bm{\overline{J}}(t)(\bm{\overline{J}}(0)\cdot\hat{n})\rangle_{0}.
\end{equation}
Since $\bm{F}=F_{0}\hat{\bm{n}}$ and
$\langle\overline{\bm{J}}\rangle=\mathsf{L}\bm{F}$, this leads to
\begin{equation}
    \mathsf{L}=\frac{\mathcal{V}}{k_{B}T}\int_{0}^{\infty}\langle\bm{\overline{J}}(t)\bm{\overline{J}}(0)^{\mathsf T}\rangle_0{\rm d}t.
\end{equation}
Finally, expressing the spatially averaged current in terms of the local density current
\begin{equation}
    \overline{\bm J}(t)=\frac{1}{\mathcal{V}}\int{\rm d}\bm x\,\boldsymbol{J}(\bm x,t)
\end{equation}
 yields
\begin{equation}
    \mathsf{L}=\frac{1}{\mathcal{V}k_{B}T}
    \int_{0}^{\infty}\!\!{\rm d}t
    \int\!{\rm d}\bm x\!\!
    \int\!{\rm d}\bm x'
    \langle\boldsymbol{J}(\bm x,t)\boldsymbol{J}(\bm x',0)^{\mathsf{T}}\rangle_0.
\end{equation}

\section{Derivation of the hybrid fluctuation-dissipation relation} \label{hybrid_proof}
In this appendix, we derive the hybrid fluctuation-dissipation relation for the conductivity in Eq.~\eqref{hybrid}, which contains both a direct response term and a modified Green-Kubo-like term. The idea is to perform a derivation similar to that in Appendix~\ref{GK_proof}, but for the low-mass limit dynamics~\cite{Felderhof1983,hoang2023frequency}. We focus on the conductivity of a symmetric $q:q$ electrolyte as studied in Sec.~\ref{SecHall}, although an analogous formula for the mobility or an asymmetric electrolyte can be obtained similarly.

Chun et al.~\cite{chun2018emergence} showed that the low-mass Langevin dynamics of a single charged particle under magnetic field, $\dot{\boldsymbol{x}}=\mathsf{M}\bm{F}(\bm{x})+\boldsymbol{\eta}(t)
$, with $\bm \eta (t)$ being the Gaussian zero-mean noise with $\langle\boldsymbol{\eta}(t)\boldsymbol{\eta}^{\mathsf{T}}(t')\rangle=\mathsf{C}(t-t')
$, corresponds to the Fokker-Planck equation 
\begin{equation}
    \frac{\partial Q(\bm{x},t)}{\partial t}=-\nabla_{\bm{x}}\cdot\bigg[\mathsf{M}\boldsymbol{F}(\bm{x})-k_{B}T\mathsf{M}\nabla_{\bm{x}}\bigg]Q(\bm{x},t)
\end{equation}
where $Q({\bm x},t)$ is the marginal probability distribution
\begin{equation}
Q(\bm{x},t)=\int{\rm d}\bm{v}\,P(\bm{x},\bm{v},t).
\end{equation}
It can be straightforwardly generalized to multiple particles and two ionic species as,
\begin{align} \label{FPQ}
\frac{\partial Q({\bf X},t)}{\partial t} & =-\sum_{\alpha,i}\nabla_{\boldsymbol{x}_{\alpha,i}}\cdot\bigg[\mathsf{M}_{\alpha}\boldsymbol{F}_{\alpha,i}({\bf X})-k_{B}T\mathsf{M}_{\alpha}\nabla_{\boldsymbol{x}_{\alpha,i}}\bigg]\nonumber\\
 & \quad\times Q({\bf X},t),
\end{align}
which corresponds to the Langevin dynamics
\begin{equation} \label{chunmanybody}
\dot{\boldsymbol{x}}_{\alpha,i}=\mathsf{M}_\alpha \boldsymbol{F}_{\alpha,i}+\boldsymbol{\eta}_{\alpha,i}(t).
\end{equation}
We decompose $\boldsymbol{F}_{\alpha,i}({\bf X})=\boldsymbol{F}_{\alpha,i}^{\rm int}({\bf X})+\boldsymbol{F}_{\alpha}^{\rm ext}
$ where ${\bm F}^{\rm int}_{\alpha,i}=-\nabla_{\bm x_{\alpha,i}}U_{\rm int}(\bm X)$ and $\boldsymbol{F}^{\rm ext}_{\alpha}=q_{\alpha}E_0\hat{n}$ is the perturbation.
Without $E_0$, the steady-state solution of Eq.~\eqref{FPQ} is
\begin{equation} \label{QEq}
Q_{\rm eq}({\bf X})=Ce^{-U_{\rm int}({\bf X})/k_BT}    
\end{equation}
where $C$ is a constant.

Similarly to Appendix~\ref{GK_proof}, the Fokker-Planck equation can be written as
\begin{equation} \label{Qdynamic}
\partial_{t}Q=\mathcal{L}_{{\rm eq}}^{\dagger}Q+E_{0}\mathcal{L}_{{\rm pert}}^{\dagger}Q
\end{equation}
but here the operators are
\begin{subequations}
    \begin{align}
\mathcal{L}_{{\rm eq}}^{\dagger} & =-\sum_{\alpha,i}\nabla_{\boldsymbol{x}_{\alpha,i}}\cdot\bigg[\mathsf{M}_{\alpha}\boldsymbol{F}_{\alpha,i}^{{\rm int}}({\bf X})-k_{B}T\mathsf{M}_{\alpha}\nabla_{\boldsymbol{x}_{\alpha,i}}\bigg]\\
\mathcal{L}_{{\rm pert}}^{\dagger} & =-\sum_{\alpha,i}q_{\alpha}\mathsf{M}_{\alpha}^{\mu\nu}\hat{n}_{\nu}\nabla_{x_{\alpha,i}^{\mu}}. \label{LpertQ}
\end{align}
\end{subequations}
We define $\overline{\bm J}_c$ analogously to Eq.~\eqref{J_def} as 
\begin{equation} \label{JDef2}
\overline{\bm{J}}_{c}=\frac{1}{\mathcal{V}}\sum_{i,\alpha}q_{\alpha}\bm{v}_{\alpha,i}.    
\end{equation}
Note that $\bm v_i$ is no longer a phase space coordinate in Eq.~\eqref{FPQ}. However, we can evaluate it from the Langevin dynamics, Eq.~\eqref{chunmanybody} (${\bm v}_{\alpha,i}=\dot{\bm x}_{\alpha,i}$), which leads to
\begin{equation}
\overline{\bm{J}}_{c} =cq^{2}\sum_{\alpha}\mathsf{M}_{\alpha}\bm{E}+\frac{1}{\mathcal{V}}\sum_{\alpha,i}q_{\alpha}\bm{\eta}_{\alpha,i}(t) +\overline{\bm{J}}_{{\rm i}}\nonumber
\end{equation}
where
\begin{equation}\label{Jibar}
\overline{\boldsymbol{J}}_{\rm i}=\frac{1}{\mathcal{V}}\sum_{\alpha,i}q_{\alpha}\mathsf{M}_{\alpha}\boldsymbol{F}_{\alpha,i}^{\rm int}
\end{equation}
which is the spatial average of the interaction current defined in Eq.~\eqref{Ji} (written with the Coulomb potential $V_{\alpha\beta}$), i.e., $\overline{\boldsymbol{J}}_{\rm i}=\frac{1}{\mathcal{V}}\int{\rm d}\bm{x}\bm{J}_{{\rm i}}(\bm{x},t)$.

Taking the average of the current and using the zero mean of the noise, we obtain
\begin{equation}
\langle\bm{\overline{J}}_{c}\rangle={c}q^{2}\sum_{\alpha}\mathsf{M}_{\alpha}\boldsymbol{E}+\int{\rm d}{\bf X}\boldsymbol{\overline J}_{{\rm i}}Q_{{\rm ss}}({\bf X})
\end{equation}
where $Q_{\rm ss}$ is the steady state probability distribution.

We now expand $Q$ around $Q_{\rm eq}$ in powers of $E_0$,
\begin{equation} \label{Qexpand}
    Q\equiv Q_{\rm eq}+ E_{0}\delta Q+\mathcal{O}(E_{0}^{2}).
\end{equation}
Inserting Eq.~\eqref{Qexpand} into Eq.~\eqref{Qdynamic} and retaining terms at linear order in $E_0$~\cite{risken1989fokker}, we obtain $\partial_{t}\delta Q=\mathcal{L}_{{\rm eq}}^{\dagger}\delta Q+\mathcal{L}_{{\rm pert}}^{\dagger}Q_{{\rm eq}}
$ with the stationary solution
\begin{align}\nonumber\delta Q_{{\rm ss}} & =\int\limits _{-\infty}^{t}e^{\mathcal{L}_{{\rm eq}}^{\dagger}\left(t-s\right)}\mathcal{L}_{{\rm pert}}^{\dagger}Q_{{\rm eq}}{\rm d}s\\
 & =-\int\limits _{0}^{\infty}e^{\mathcal{L}_{{\rm eq}}^{\dagger}u}\sum_{\alpha,i}q_{\alpha}\mathsf{M}_{\alpha}^{\mu\nu}\hat{n}_{\nu}\nabla_{x_{\alpha,i}^{\mu}}Q_{{\rm eq}}{\rm d}u
\end{align}
where in the second line we used Eq.~\eqref{LpertQ} and the change of variables $t-s=u$.
Substituting Eq.~\eqref{QEq} into the integral, we obtain
\begin{equation} \label{deltaQ_solution}
 \delta Q_{{\rm ss}}=-\frac{\mathcal{V}}{k_{B}T}\int\limits _{0}^{\infty}e^{\mathcal{L}_{{\rm eq}}^{\dagger}t}[(\overline{\overleftarrow{\boldsymbol{J}}}_{{\rm i}}\cdot\hat{n})Q_{{\rm eq}}]{\rm d}t
\end{equation}
where we defined
\begin{equation}
\ensuremath{\overline{\overleftarrow{\boldsymbol{J}}}_{{\rm i}}\equiv\frac{1}{\mathcal{V}}\sum_{\alpha,i}q_{\alpha}\mathsf{M}_{\alpha}^{\mathsf{T}}\boldsymbol{F}_{\alpha,i}^{{\rm int}}},
\end{equation}
to be distinguished from $\overline{\boldsymbol{J}}_{\rm i}$ (Eq.~\eqref{Jibar}), since $\mathsf{M}_{\alpha}$ is transposed.

Since $\langle\boldsymbol{\overline{J}}_{{\rm i}}\rangle_0=0
$, we obtain to linear order in $E_0$
\begin{align}
\langle&\bm{\overline{J}}_{c}\rangle  ={c}q^{2}\sum_{\alpha}\mathsf{M}_{\alpha}\boldsymbol{E}+E_{0}\int{\rm d}{\bf X}\boldsymbol{\overline{J}}_{{\rm i}}\delta Q_{{\rm ss}}({\bf X})\\
 & ={c}q^{2}\sum_{\alpha}\mathsf{M}_{\alpha}\boldsymbol{E}-\frac{\mathcal{V}E_{0}}{k_{B}T}\int\limits _{0}^{\infty}{\rm d}t\int{\rm d}{\bf X}\boldsymbol{\overline{J}}_{{\rm i}}e^{\mathcal{L}_{{\rm eq}}^{\dagger}t}[(\overline{\overleftarrow{\boldsymbol{J}}}_{{\rm i}}\cdot\hat{n})Q_{{\rm eq}}]. \nonumber
\end{align}
Using the adjoint relation as in Appendix~\ref{GK_proof} we obtain
\begin{equation}
\langle\overline{\bm{J}}_{c}\rangle={c}q^{2}\sum_{\alpha}\mathsf{M}_{\alpha}\bm{E}-\frac{\mathcal{V}E_{0}}{k_{B}T}\int\limits _{0}^{\infty}{\rm d}t\langle\overline{\bm{J}}_{i}(t)(\overline{\overleftarrow{\bm{J}}}_{{\rm i}}(0)\cdot\hat{n})\rangle_0.
\end{equation}
Since $\bm{E}=E_{0}\hat{\bm{n}}$ and
$\langle\overline{\bm{J}}_{c}\rangle=\mathsf{K}\bm{E}$, this can equivalently
be written as
\begin{equation}
\mathsf{K}={c}q^{2}\sum_{\alpha}\mathsf{M}_{\alpha}-\frac{\mathcal{V}}{k_{B}T}\int\limits _{0}^{\infty}{\rm d}t\langle\boldsymbol{\overline{J}}_{{\rm i}}(t)\overline{\overleftarrow{\bm{J}}}_{{\rm i}}^{\mathsf{T}}(0)\rangle_0.
\end{equation}
Finally, substituting $\overline{\boldsymbol{J}}_{\rm i}=\frac{1}{\mathcal{V}}\int{\rm d}\bm{x}\bm{J}_{{\rm i}}(\bm{x},t)$ and $\ensuremath{\overline{\overleftarrow{\boldsymbol{J}}}_{{\rm i}}=\frac{1}{\mathcal{V}}\int{\rm d}\bm{x}\overleftarrow{\bm{J}}_{{\rm i}}(\bm{x},t)}$  (defined in Eq.~\eqref{Jtilde}) we obtain Eq.~\eqref{hybrid}.

\section{Comparison with the cluster expansion results of \texorpdfstring{Ref.~\cite{friedman1965calculation}}{Ref.~[\ref{bib-friedman1965calculation}]}}
\label{FridmanComparison2}

\let\modifhat\hat
\let\hat\orighat

\def\eel{e_{\text{el}}}
In this Appendix, we summarize results from Ref.~\cite{friedman1965calculation}, where the Hall conductance of electrolytes is computed using a cluster expansion. We then compare these to our results.

\subsection{Analytical comparison}

In Ref.~\cite{friedman1965calculation}, the Hall conductance $\sigma'$ (called $\kappa_{\text{H}}$ in our main text) is expressed as (Eq.~(1.2) of Ref.~\cite{friedman1965calculation})
\begin{equation}
    \sigma' = A' c + S' c^{3/2} + \cdots 
\end{equation}
in which $c$ is the salt concentration, and in which
\begin{equation}
    \label{AprimeoverB}
    \frac{A'}{B} = \eel^3 \sum_{s} \frac{x_s z_s^3}{\zeta_s^2}
\end{equation}
while
\begin{equation}
    \label{SprimeoverB}
    \frac{S'}{B} = \frac{2 \sqrt{\pi}}{3} \ell_{\text{B}}^{3/2} \eel^3
    \bigg[
    \sum_s x_s z_s^2
    \bigg]^{1/2}
    \bigg[\sum_{s} \frac{x_s z_s^3}{\zeta_s^2} S_s'\bigg].
\end{equation}
In these expressions, $s$ labels the ionic species, $x_s \equiv c_s/c$ and $z_s \equiv e_s/q$, where $c_s$ is the concentration of ions $s$, $e_s$ is the charge of ions $s$, $\zeta_s$ is the friction coefficient of ions $s$ (called $\gamma_s$ in the main text) $\eel$ is the elementary charge (called $e$ in the main text), $B$ is the applied magnetic field, and $\ell_{\text{B}} \equiv \eel^{2}/(4\pi\varepsilon_{0} \varepsilon_r k_{\text{B}} T)$ is the Bjerrum length, where $\varepsilon_{0}$ is the vacuum permittivity, $\varepsilon_r$ is the relative permittivity (dielectric constant) of the medium, $k_{\text{B}}$ is the Boltzmann constant, $T$ is the temperature.
Please note that unless otherwise specified, we use SI units; in contrast, Ref.~\cite{friedman1965calculation} uses CGS units (in which, in particular, $\ell_{\text{B}} = \eel^{2}/(\varepsilon_r k_{B}T)$ [CGS]). 
The magnetic field $B$ used here corresponds to the quantity $H/\hat{c}$ in Ref.~\cite{friedman1965calculation}.
The quantity $S_s'$ is expressed as the sum
\begin{equation}
    S_s' \equiv \sum_{M=1}^{13} S_s'(M)
\end{equation}
in which the $M$ labels various cluster diagrams given in Fig.~4 of Ref.~\cite{friedman1965calculation}.
The $S'(M)$ are given in Ref.~\cite{friedman1965calculation} by
\begin{subequations} \label{crazycrazy}
\allowdisplaybreaks
\begin{align}
S'_a(1) &= -\sum _{b=1}^N \frac{z_a \mu _b \left(u_a z_a-u_b z_b\right)}{\left(\hat{q}_a+1\right) \left(u_a+u_b\right)}\\ 
S'_a(2) &= \sum _{b=1}^N \sum _{c=1}^N z_a \mu _b z_b \mu _c I_{a b c}\\ 
S'_a(3) &= -\frac{2 u_{\text{el}}}{u_a}\\ 
S'_a(4) &= {\color{tab_red}+} \sum _{b=1}^N \frac{\mu _b u_{\text{el}}}{\hat{q}_a \left(\hat{q}_a+1\right) \left(u_a+u_b\right)}\\ 
S'_a(5) &= -\sum _{b=1}^N \frac{u_a z_a^2 \mu _b}{\left(\hat{q}_a+1\right) \left(u_a+u_b\right)}\\ 
S'_a(6) &= -\frac{\hat{q}_a^2 u_{\text{el}}}{u_a}\\ 
S'_a(7) &= \sum _{b=1}^N \frac{\mu _b u_b^2 z_b^2}{\left(\hat{q}_a+1\right) u_a \left(u_a+u_b\right)}\\ 
S'_a(8) &= -\sum _{b=1}^N \frac{2 \mu _b u_b z_b u_{\text{el}}}{u_a^2 z_a}\\ 
S'_a(9) &= -\sum _{b=1}^N \frac{\mu _b u_b^2 z_b u_{\text{el}}}{u_a^2 z_a \left(\hat{q}_b+1\right) \left(u_a+u_b\right)}\\ 
S'_a(10) &= \sum _{b=1}^N \sum _{c=1}^N \frac{\mu _b u_b z_b^2 \mu _c I_{a b c}}{u_a}\\ 
S'_a(11) &= \sum _{b=1}^N \sum _{c=1}^N \frac{\mu _b u_b z_b \mu _c u_c u_{\text{el}}}{u_a z_a \hat{q}_c \left(\hat{q}_c+1\right) \left(u_a+u_c\right) \left(u_b+u_c\right)}\\ 
S'_a(12) &= \sum _{b=1}^N \sum _{c=1}^N \frac{\mu _b u_b \mu _c u_{\text{el}} \left(\frac{u_b}{u_a+u_b}+1\right) \left(u_c z_c-u_a z_a\right)}{\hat{q}_a \left(\hat{q}_a+1\right) u_a^2 z_a \left(u_a+u_c\right)}\\ 
S'_a(13) &= \sum _{b,c,d=1}^N  
\frac{\mu _b u_b \mu _c \mu _d u_{\text{el}} \left(u_d-u_c\right) \left(u_c z_c-u_d z_d\right) J_{a b c d}}{2 u_a z_a}
\end{align}
\end{subequations}
where we have added the prime symbols missing in the unnumbered list on page 2624 of Ref.~\cite{friedman1965calculation}, and where the sign of $S_a'(4)$ has been adjusted with respect to Ref.~\cite{friedman1965calculation} (our change is in red) in order to match the numerical results reported in Table I of Ref.~\cite{friedman1965calculation} for system I (\ce{LiCl}).

In these expressions, 
\begin{equation}
\!\!\!\!
    \mu_s \equiv \frac{c_s e_s^2}{\sum_a c_a e_a^2}
    \qquad
    u_s \equiv 1/\zeta_s
    \qquad
    \hat{q}_a^2 \equiv \sum_s \frac{\mu_s u_s}{u_a + u_s}
\end{equation}
while $u_{\text{el}}$ is a characteristic mobility that we set to zero to remove the electrophoretic term (not computed in our work), and is defined in CGS units by Ref.~\cite{friedman1965calculation} to be $u_{\text{el}} \equiv \epsilon/[4\pi \eta \beta \eel^2]$ [CGS] in which $\eta$ is the viscosity of the solvent and $\beta = 1/k_{\text{B}} T$.
In addition, the reciprocal $\kappa$ of the Debye screening length is defined by
\begin{equation}
    \kappa^2 \equiv \frac{4\pi\beta}{\epsilon} \sum_a c_s e_s^2.
\end{equation}
The quantity $J_{abcd}$ is an integral that is irrelevant to our current concerns, because it only enters in an electrophoretic contribution, see Ref.~\cite{friedman1965calculation} for its expression. 

Finally, the quantity $I_{abc}$ is given by
\begin{equation}
    \label{app_friedman_Iabc}
    I_{abc} \equiv \frac{1}{\pi^2}
    \int_{0}^{\infty}
    \frac{q^2}{1+q^2}
    [ \tilde{\imath}_{abc}(q) - \tilde{\imath}_{acb}(q) ] \dd q
\end{equation}
where
\begin{equation}
\begin{split}
    \label{app_friedman_imathabc}
    \tilde{\imath}_{abc}(q)
    \equiv
    \ii
    \int_{-\infty}^{\infty}
    \frac{1}{
    [z + \ii q^2 u_a]
    [z + \ii q^2 u_b]
    [z -\ii q^2 u_c]
    }
    \\
    \times
    \frac{1}{\varphi(z, q)
    \varphi(-z, q)}
    \dd z
\end{split}
\end{equation}
and 
\begin{equation}
    \label{app_friedman_varphi}
    \varphi(z, q) \equiv 1 + \sum_{s=1}^{N} 
    \frac{\mu_s u_s}{q^2 u_s + \ii z}.
\end{equation}
This expression was obtained as follows: $I_{abc}$ is defined in the Appendix \enquote{Cluster Integrals Involving More Than Two Ions, aside from Chain Sums} of Ref.~\cite{friedman1965calculation} as
\begin{equation}
    \label{Iabc_friedman}
    I_{abc} \equiv 
    \frac{\color{tab_red}\kappa^3}{{\color{tab_red} \zeta_a \zeta_b}\beta^2 \pi^2}
    \int_{0}^{\infty}
    \frac{k^2}{k^2+\kappa^2} [i_{abc}(k) - i_{acb}(k)] \dd k
\end{equation}
with a $\kappa^{5}$ instead of the red $\kappa^{3}$ and without the red $\zeta_a$ and $\zeta_b$.
The changes we made are required so that (i) the expression is dimensionally correct, in particular to get a $c^{3/2}$ behavior, (ii) the comment at the end of Sec.~6 of Ref.~\cite{friedman1965calculation} comparing the $S'(M)$ with quantities $S(1)$, $S(I1)$, $S(I3)$ defined in Ref.~\cite{Friedman1965a} (Ref.~16 in Ref.~\cite{friedman1965calculation}) holds, and (iii) we could reproduce the numerical results reported in Table I of Ref.~\cite{friedman1965calculation}.
The quantity $i_{abc}$ in Eq.~\eqref{Iabc_friedman} is defined in Ref.~\cite{friedman1965calculation} as 
\begin{equation}
\begin{split}
    i_{abc}(k)
    \equiv
    \int_{-\infty}^{\infty}
    \frac{1}{
    [H_c + \ii z]
    [H_a - \ii z]
    [H_b - \ii z]
    }
    \\
    \times \frac{1}{\phi(z, k)
    \phi(-z, k)} \dd z
\end{split}
\end{equation}
with $H_s \equiv k^2 /\beta\zeta_s$ and
\begin{equation}
    \phi(z, k) \equiv 1 + \sum_{s=1}^{N} 
    \frac{4\pi c_s e_s^2/\epsilon\zeta_s}{k^2/\beta\zeta_s + \ii z}
\end{equation}
Applying the changes of variable $k \to k/\kappa=q$ and $z \to z/[\kappa^2/\beta]$ along with some reorganization of the expressions then yields Eqs.~(\ref{app_friedman_Iabc}--\ref{app_friedman_imathabc}).

We were not able to reliably extract closed-form expressions of $I_{abc}$ from these references, and so we now compute these starting from Eqs.~(\ref{app_friedman_Iabc}--\ref{app_friedman_imathabc}) in the case of $N=2$ species (intermediate calculations can be found in a Mathematica notebook in the Supplemental Material).
In this situation, we only need to compute $I_{aab} = - I_{aba}$ with $a \neq b$, in which case
\begin{equation}
    \varphi(z, q) \equiv 1 +
    \frac{\mu_a u_a}{q^2 u_a + \ii z}
    +
    \frac{\mu_b u_b}{q^2 u_b + \ii z}.
\end{equation}
To compute the $I_{aab}$, we put together the integrals in $\tilde{\imath}_{aab}(q) - \tilde{\imath}_{aba}(q)$ to obtain a single $z$ integral that we evaluate using the residue theorem (closing e.g. the real line with an upper semi-circle; there are no poles on the real axis). 
Expanding the result in partial fractions and performing the remaining integral yields three identical integrals of the form $\int_0^{\infty} \dd q/[q^2+\lambda]=\pi/[2\sqrt{\lambda}]$. Tedious algebraic manipulations then yield the result
\begin{equation}
    I_{aab} = \frac{u_a^2 \left(u_a-u_b\right)}{(A+B) (C+D) (A C+B D)}
\end{equation}
where (here $B$ is not the magnetic field)
\begin{equation*}
    A =\sqrt{2 u_a (u_a+u_b)}
    \qquad
    B =\sqrt{\mu _a u_a \left(u_a+u_b\right)+2 u_a \mu _b u_b}
\end{equation*}
\begin{equation*}
    C =\sqrt{\mu_a u_a+\mu_b u_b}
    \qquad
    D =\sqrt{u_a+u_b}
\end{equation*}

Lastly, we evaluate Eqs.~\eqref{AprimeoverB} and \eqref{SprimeoverB} for symmetric electrolytes with $x_{+}=x_{-}=1$ and $z_{\pm}=\pm Z$ (labeling with $s=\pm$ the ions with positive (negative) charge), and with $u_{\text{el}} \equiv 0$ to remove the electrophoretic contribution, which amounts to summing over $M\in(1, 2, 5, 7, 10)$.
We first get
\begin{equation}
    \frac{A'}{B} = Z^3 \eel^3 \left[ \frac{1}{\zeta_+^2} - \frac{1}{\zeta_-^2}
    \right]
\end{equation}
which matches $\mathcal{A}_\text{H}$ in Eq.~\eqref{kappa_H_discussion} (recall that $\zeta_s \equiv \gamma_s$ and $Z \eel \equiv q$).
The first square bracket Eq.~\eqref{SprimeoverB} yields
\begin{equation}
    \bigg[
    \sum_s x_s z_s^2
    \bigg]^{1/2} = Z \sqrt{2}
\end{equation}
and for the second square bracket, a tedious but straightforward resummation of all terms (see Supplemental Material) yields
\begin{equation}
    \sum_{s} \frac{x_s \hat{z}_s^3}{\zeta_s^2} S_s'
    =
    \frac{S_{+}}{\zeta_{+}^2}-\frac{S_{-}}{\zeta_{-}^2}
    =
    -(2-\sqrt{2}) 
    \left(\frac{1}{\zeta_{+}^2}-\frac{1}{\zeta_{-}^2}\right)
\end{equation}
where $\hat{z}_s \equiv z_s/Z$ (so $\hat{z}_\pm = \pm 1$), eventually giving
\begin{equation}
    \label{SprimeoverB_final}
    \frac{S'}{B} = - \frac{2 \sqrt{2\pi}}{3} \ell_{\text{B}}^{3/2} Z^6 \eel^3
    (2-\sqrt{2}) 
    \left(\frac{1}{\zeta_{+}^2}-\frac{1}{\zeta_{-}^2}\right)
\end{equation}
which matches $\mathcal{B}_{\text{H}}$ in Eq.~\eqref{kappa_H_discussion} (recall again that $\zeta_s \equiv \gamma_s$ and $Z \eel \equiv q$).

\subsection{Numerical comparison}

We also provide an independent numerical check that the results by Friedman match ours, using Table I in Ref.~\cite{friedman1965calculation}, which details the numerical values of the 13 $S'_{\pm}(M)$ terms for LiCl in water at $25\,^\circ\mathrm{C}$ (i.e., here we do not rely on the analytical expressions given by Eq.~\eqref{crazycrazy}).

We work in SI units in which the conductance $\mathsf{K}$ has units of
${\rm S}/{\rm m}$. For monovalent ions ($q=e$) in water at $25\,^\circ\mathrm{C}$, the Bjerrum length is $\ell_{B}=7.16\times10^{-10}\,\mathrm{m}$.
Moreover, the friction coefficients of $\rm Li^+$ and $\rm Cl^-$ in water at $25\,^\circ\mathrm{C}$ are~\cite{haynes2016crc} $\gamma_{+}=4.000\times10^{-12}{\rm kg}/{\rm s}$ and
$\gamma_{-}=2.026\times10^{-12}{\rm kg}/{\rm s}$, respectively.
Substituting these values to Eqs.~\eqref{K0_approx} and~\eqref{deltaK_linres}, restricted to linear order in $B$, we obtain $\kappa_0^{\rm H}[{\rm S}/{\rm m}]=-4.49\times10^{-10}{c}[{\rm M}]B[{\rm T}]
$ and
\begin{equation} \label{ourdeltaKnumeric}
\Delta\kappa^{\rm H}[{\rm S}/{\rm m}]=6.52\times10^{-12}{c}[{\rm M}]^{3/2}B[{\rm T}]
\end{equation}
where ${c}[\rm M]$ is concentration in molars and $B[\rm T]$ is magnetic field in Tesla.
On the other hand, the result of Friedman, given by Eqs.~6.1 and 6.4 of Ref.~\cite{friedman1965calculation}, with the notation adapted to ours and specialized to two species of monovalent ions, is
\begin{equation}\label{K_Fried}
\Delta \kappa^{\rm H,\,Fried}=\frac{2\sqrt{2\pi}}{3}\ell_{B}^{3/2}{c}^{3/2}e^{3}B\left(\frac{S'_{+}}{\gamma_{+}^{2}}-\frac{S'_{-}}{\gamma_{-}^{2}}\right)
\end{equation}
with $S'_{\pm}=\sum_{M=1}^{13} S_{\pm}(M)$.
Substituting the values from Table I of Ref.~\cite{friedman1965calculation} that contribute only to the relaxation effect ($M=1,2,5,7,10$) and neglecting the electrophoretic terms, we obtain
\begin{align}
10^{3}S'_{+,{\rm rel}} & =-284-9-237+513+9=-8\\
10^{3}S'_{-,{\rm rel}} & =-304+11-353+204+5=-437,
\end{align}
which when inserted into Eq.~\eqref{K_Fried} leads to
\begin{equation}
\Delta \kappa^{\rm H,Fried}=6.51\times10^{-12}{c}[{\rm M}]^{3/2}B[{\rm T}],
\end{equation}
in very good agreement with Eq.~\eqref{ourdeltaKnumeric}.

\bibliographystyle{apsrev4-2}
\bibliography{bibliography}

@article{dean1996langevin,
  title={Langevin equation for the density of a system of interacting Langevin processes},
  author={Dean, David S},
  journal={Journal of Physics A: Mathematical and General},
  volume={29},
  number={24},
  pages={L613},
  year={1996},
  publisher={IOP Publishing},
  doi={10.1088/0305-4470/29/24/001}
}

@article{bernard1992conductance,
  doi = {10.1021/j100188a049},
  title={Conductance in electrolyte solutions using the mean spherical approximation},
  author={Bernard, Olivier and Kunz, Werner and Turq, Pierre and Blum, Lesser},
  journal={The Journal of Physical Chemistry},
  volume={96},
  number={9},
  pages={3833--3840},
  year={1992},
  publisher={ACS Publications}
}

@article{demery2016conductivity,
  doi = {10.1088/1742-5468/2016/02/023106},
  title={The conductivity of strong electrolytes from stochastic density functional theory},
  author={D{\'e}mery, Vincent and Dean, David S},
  journal={Journal of Statistical Mechanics: Theory and Experiment},
  volume={2016},
  number={2},
  pages={023106},
  year={2016},
  publisher={IOP Publishing}
}

@article{kawasaki1994stochastic,
  title={Stochastic model of slow dynamics in supercooled liquids and dense colloidal suspensions},
  author={Kawasaki, Kyozi},
  journal={Physica A: Statistical Mechanics and its Applications},
  volume={208},
  number={1},
  pages={35--64},
  year={1994},
  publisher={Elsevier}
}

@article{peraud2017fluctuation,
doi = {10.1073/pnas.1714464114},
 title = {Fluctuation-enhanced electric conductivity in electrolyte solutions},
 author = {Péraud, Jean-Philippe and Nonaka, Andrew J. and Bell, John B. and Donev, Aleksandar and Garcia, Alejandro L.},
 journal = {Proceedings of the National Academy of Sciences},
 volume = {114},
 number = {41},
 pages = {10829–10833},
 year = {2017},
 month = {sept},
 issn = {1091-6490},
 publisher = {National Academy of Sciences}
}

@article{avni2022conductivity,
  doi = {10.1103/physrevlett.128.098002},
  title={Conductivity of concentrated electrolytes},
  author={Avni, Yael and Adar, Ram M and Andelman, David and Orland, Henri},
  journal={Physical Review Letters},
  volume={128},
  number={9},
  pages={098002},
  year={2022},
  publisher={APS}
}

@article{altenberger1983theory,
  doi = {10.1063/1.445093},
  title={Theory of conductance and related isothermal transport coefficients in electrolytes},
  author={Altenberger, AR and Friedman, Harold L},
  journal={Journal of Chemical Physics},
  volume={78},
  number={6},
  pages={4162--4173},
  year={1983}
}

@article{chandra1999ion,
  doi = {10.1063/1.478876},
  title={Ion conductance in electrolyte solutions},
  author={Chandra, Amalendu and Bagchi, Biman},
  journal={The Journal of chemical physics},
  volume={110},
  number={20},
  pages={10024--10034},
  year={1999},
  publisher={American Institute of Physics}
}

@article{han2021fluctuating,
  doi = {10.1038/s41567-021-01360-7},
  title={Fluctuating hydrodynamics of chiral active fluids},
  author={Han, Ming and Fruchart, Michel and Scheibner, Colin and Vaikuntanathan, Suriyanarayanan and De Pablo, Juan J and Vitelli, Vincenzo},
  journal={Nature Physics},
  volume={17},
  number={11},
  pages={1260--1269},
  year={2021},
  publisher={Nature Publishing Group UK London}
}

@article{robin2024correlation,
  doi = {10.1063/5.0188215},
  title={Correlation-induced viscous dissipation in concentrated electrolytes},
  author={Robin, Paul},
  journal={The Journal of Chemical Physics},
  volume={160},
  number={6},
  year={2024},
  pages={064503},
  publisher={AIP Publishing}
}

@article{benois2023enhanced,
  doi = {10.1103/physreve.108.054606},
  title={Enhanced diffusion of tracer particles in nonreciprocal mixtures},
  author={Benois, Anthony and Jardat, Marie and Dahirel, Vincent and D{\'e}mery, Vincent and Agudo-Canalejo, Jaime and Golestanian, Ramin and Illien, Pierre},
  journal={Physical Review E},
  volume={108},
  number={5},
  pages={054606},
  year={2023},
  publisher={APS}
}

@article{chun2018emergence,
  doi = {10.1103/physreve.97.032117},
  title={Emergence of nonwhite noise in Langevin dynamics with magnetic Lorentz force},
  author={Chun, Hyun-Myung and Durang, Xavier and Noh, Jae Dong},
  journal={Physical Review E},
  volume={97},
  number={3},
  pages={032117},
  year={2018},
  publisher={APS}
}

@article{chun2021nonequilibrium,
  doi = {10.1103/physrevresearch.3.043172},
  title={Nonequilibrium Green-Kubo relations for hydrodynamic transport from an equilibrium-like fluctuation-response equality},
  author={Chun, Hyun-Myung and Gao, Qi and Horowitz, Jordan M},
  journal={Physical Review Research},
  volume={3},
  number={4},
  pages={043172},
  year={2021},
  publisher={APS}
}

@article{nakamura2009derivation,
  doi = {10.1088/1751-8113/42/6/065001},
  title={Derivation of the nonlinear fluctuating hydrodynamic equation from the underdamped Langevin equation},
  author={Nakamura, Takenobu and Yoshimori, Akira},
  journal={Journal of Physics A: Mathematical and Theoretical},
  volume={42},
  number={6},
  pages={065001},
  year={2009},
  publisher={IOP Publishing}
}

@article{kuroiwa2013brownian,
  doi = {10.1088/1751-8113/47/1/012001},
  title={Brownian motion with multiplicative noises revisited},
  author={Kuroiwa, Takeshi and Miyazaki, Kunimasa},
  journal={Journal of Physics A: Mathematical and Theoretical},
  volume={47},
  number={1},
  pages={012001},
  year={2013},
  publisher={IOP Publishing}
}

@article{hoang2023frequency,
 doi = {10.1063/5.0139258},
 title = {Frequency and field-dependent response of confined electrolytes from Brownian dynamics simulations},
 author = {Hoang Ngoc Minh, Thê and Stoltz, Gabriel and Rotenberg, Benjamin},
 journal = {The Journal of Chemical Physics},
 volume = {158},
 number = {10},
 year = {2023},
 month = {mar},
 pages = {104103},
 issn = {1089-7690},
 publisher = {AIP Publishing}
}

@incollection{risken1989fokker,
  title={Fokker-planck equation},
  author={Risken, Hannes},
  booktitle={The Fokker-Planck equation: methods of solution and applications},
  pages={63--95},
  year={1989},
  publisher={Springer}
}

@misc{dean2026dielectric,
 eprint = {2604.00262},
 title = {Dielectric response as a source of viscosity in polar liquids},
 author = {David S. Dean and Haim Diamant},
 year = {2026},
 eprinttype = {arXiv},
 archivePrefix = {arXiv}
}

@article{kruger2018stresses,
  doi = {10.1063/1.5019424},
  title={Stresses in non-equilibrium fluids: Exact formulation and coarse-grained theory},
  author={Kr{\"u}ger, Matthias and Solon, Alexandre and D{\'e}mery, Vincent and Rohwer, Christian M and Dean, David S},
  journal={The Journal of chemical physics},
  volume={148},
  number={8},
  year={2018},
  pages={084503},
  publisher={AIP Publishing}
}

@article{avni2022conductance,
  doi = {10.1063/5.0111645},
  title={Conductance of concentrated electrolytes: Multivalency and the Wien effect},
  author={Avni, Yael and Andelman, David and Orland, Henri},
  journal={The Journal of Chemical Physics},
  volume={157},
  number={15},
  year={2022},
  pages={154502},
  publisher={AIP Publishing}
}

@article{abdoli2026dynamical,
 doi = {10.1021/acs.jpcb.6c00652},
 title = {Dynamical Density Functional Theory for Dense Odd-Diffusive Fluids},
 author = {Abdoli, Iman and Wittmann, René and Löwen, Hartmut},
 journal = {The Journal of Physical Chemistry B},
 volume = {130},
 number = {17},
 pages = {4672–4682},
 year = {2026},
 month = {apr},
 issn = {1520-5207},
 publisher = {American Chemical Society (ACS)}
}

@article{bazant2004diffuse,
  title={Diffuse-charge dynamics in electrochemical systems},
  author={Bazant, Martin Z and Thornton, Katsuyo and Ajdari, Armand},
  journal={Physical Review E—Statistical, Nonlinear, and Soft Matter Physics},
  volume={70},
  number={2},
  pages={021506},
  year={2004},
  publisher={APS},
   doi = {10.1103/physreve.70.021506},
}

@article{novikov1965functionals,
  title={Functionals and the random-force method in turbulence theory},
  author={Novikov, Evgenii A},
  journal={Sov. Phys. JETP},
  volume={20},
  number={5},
  pages={1290--1294},
  year={1965},
  url={https://www.jetp.ras.ru/cgi-bin/e/index/e/20/5/p1290?a=list}
}

@article{onsager1927report,
  doi = {10.1039/tf9272300341},
  title={Report on a revision of the conductivity theory},
  author={Onsager, Lars},
  journal={Transactions of the Faraday Society},
  volume={23},
  pages={341--349},
  year={1927},
  publisher={Royal Society of Chemistry}
}

@article{onsager2002irreversible,
  doi={10.1021/j150341a001},
  title={Irreversible processes in electrolytes. Diffusion, conductance and viscous flow in arbitrary mixtures of strong electrolytes},
  author={Onsager, Lars and Fuoss, Raymond Matthew},
  journal={The Journal of Physical Chemistry},
  volume={36},
  number={11},
  pages={2689--2778},
  year={2002},
  publisher={ACS Publications}
}

@article{caprini2023chiral,
  title={Chiral active matter in external potentials},
  author={Caprini, Lorenzo and L{\"o}wen, Hartmut and Marconi, Umberto Marini Bettolo},
  journal={Soft Matter},
  volume={19},
  number={33},
  pages={6234--6246},
  year={2023},
  publisher={Royal Society of Chemistry},
  doi={10.1039/d3sm00793f}
}

@article{demery2014generalized,
  doi = {10.1088/1367-2630/16/5/053032},
  title={Generalized Langevin equations for a driven tracer in dense soft colloids: construction and applications},
  author={D{\'e}mery, Vincent and B{\'e}nichou, Olivier and Jacquin, Hugo},
  journal={New Journal of Physics},
  volume={16},
  number={5},
  pages={053032},
  year={2014},
  publisher={IOP Publishing}
}

@misc{goerlich2026particleresolvedrheologicalstudychirality,
      title={A particle-resolved rheological study of chirality transfer and odd transport}, 
      author={Rémi Goerlich and Alexander P. Antonov and Kristian Stølevik Olsen and Lorenzo Caprini and Christian Scholz and Hartmut Löwen and Yael Roichman},
      year={2026},
      eprint={2605.25136},
      archivePrefix={arXiv},
}

@article{friedman1965calculation,
 doi = {10.1021/j100892a023},
 title = {Calculation of the Hall Effect in Ionic Solutions},
 author = {Friedman, Harold L.},
 journal = {The Journal of Physical Chemistry},
 volume = {69},
 number = {8},
 pages = {2617–2628},
 year = {1965},
 month = {aug},
 issn = {1541-5740},
 publisher = {American Chemical Society (ACS)}
}

@article{meton1976hall,
 doi = {10.1016/0009-2614(76)80733-6},
 title = {Hall effect in dilute electrolytes},
 author = {Meton, Maurice and Gerard, Paul},
 journal = {Chemical Physics Letters},
 volume = {44},
 number = {3},
 pages = {582–585},
 year = {1976},
 month = {dec},
 issn = {0009-2614},
 publisher = {Elsevier}
}

@article{avni2025dynamical,
  title={Dynamical phase transitions in the nonreciprocal Ising model},
  author={Avni, Yael and Fruchart, Michel and Martin, David and Seara, Daniel and Vitelli, Vincenzo},
  journal={Physical Review E},
  volume={111},
  number={3},
  pages={034124},
  year={2025},
  publisher={APS},
  doi = {10.1103/physreve.111.034124}
}

@book{haynes2016crc,
  title={CRC handbook of chemistry and physics},
  author={Haynes, William M},
  year={2016},
  publisher={CRC press}
}

@article{cates2013active,
  doi={10.1209/0295-5075/101/20010},
  title={When are active Brownian particles and run-and-tumble particles equivalent? Consequences for motility-induced phase separation},
  author={Cates, Michael E and Tailleur, Julien},
  journal={Europhysics Letters},
  volume={101},
  number={2},
  pages={20010},
  year={2013},
  publisher={EDP Sciences, IOP Publishing and Societ{\`a} Italiana di Fisica}
}

@article{solon2015active,
  doi = {10.1140/epjst/e2015-02457-0},
  title={Active brownian particles and run-and-tumble particles: A comparative study},
  author={Solon, Alexandre P and Cates, Michael E and Tailleur, Julien},
  journal={The European Physical Journal Special Topics},
  volume={224},
  number={7},
  pages={1231--1262},
  year={2015},
  publisher={Springer}
}

@misc{faedi2026velocityforceautocorrelationsbrownian,
      title={Velocity and force autocorrelations in Brownian dynamics with a Lorentz force}, 
      author={Filippo Faedi and Abhinav Sharma},
      year={2026},
      eprint={2607.02282},
      archivePrefix={arXiv},
      primaryClass={cond-mat.stat-mech},
      url={https://arxiv.org/abs/2607.02282}, 
}

@article{sung1987microscopic,
 doi = {10.1063/1.453559},
 title = {The microscopic theory of the Hall effect in ionic solutions. I. General formulation and the Brownian ion-continuum solvent limit},
 author = {Sung, Wokyung and Friedman, Harold L.},
 journal = {The Journal of Chemical Physics},
 volume = {87},
 number = {1},
 pages = {643–653},
 year = {1987},
 month = {july},
 issn = {1089-7690},
 publisher = {AIP Publishing}
}

@article{hubbard1981electrohydrodynamic,
 doi = {10.1063/1.442400},
 title = {An electrohydrodynamic contribution to the Hall effect in electrolyte solutions},
 author = {Hubbard, J. B. and Wolynes, P. G.},
 journal = {The Journal of Chemical Physics},
 volume = {75},
 number = {6},
 pages = {3051–3054},
 year = {1981},
 month = {sept},
 issn = {1089-7690},
 publisher = {AIP Publishing}
}

@article{kroh1990hall,
 doi = {10.1080/00268979000100881},
 title = {The Hall effect in dilute ionic solutions},
 author = {Kroh, H.J. and Felderhof, B.U.},
 journal = {Molecular Physics},
 volume = {70},
 number = {1},
 pages = {119–128},
 year = {1990},
 month = {may},
 issn = {1362-3028},
 publisher = {Informa UK Limited}
}

@article{gerard1990hall,
  title={Hall effect in aqueous acid solutions at different concentrations},
  author={Gerard, P and Gerard, R and Meton, M and Picard, EJ},
  journal={Journal of the Electrochemical Society},
  volume={137},
  number={12},
  pages={3873--3875},
  year={1990},
  publisher={The Electrochemical Society, Inc.},
  doi = {10.1149/1.2086318},
}

@article{wolynes1980dynamics,
 doi = {10.1146/annurev.pc.31.100180.002021},
 title = {Dynamics of Electrolyte Solutions},
 author = {Wolynes, P G},
 journal = {Annual Review of Physical Chemistry},
 volume = {31},
 number = {1},
 pages = {345–376},
 year = {1980},
 month = {oct},
 issn = {1545-1593},
 publisher = {Annual Reviews}
}

@book{robinson2002electrolyte,
  title={Electrolyte solutions},
  author={Robinson, Robert Anthony and Stokes, Robert Harold},
  year={2002},
  publisher={Courier Corporation}
}

@book{bockris1998modern,
  author    = {Bockris, John O'M. and Reddy, Amulya K. N.},
  title     = {Modern Electrochemistry 1: Ionics},
  edition   = {2},
  publisher = {Kluwer Academic/Plenum Publishers},
  address   = {New York},
  year      = {1998},
  isbn      = {0-306-45554-4}
}

@misc{ashcroft1976solid,
  title={Solid state physics},
  author={Ashcroft, Neil W and Mermin, N David and others},
  year={1976},
  publisher={holt, rinehart and winston, new york London}
}

@article{Mergault1964,
title={Mesure d'effets magnétoélectriques dans des solutions aqueuses d'électrolytes},
author={Pierre Mergault and Josette Pagès-Nelson},
journal={Comptes Rendus de l'Académie des Sciences},
volume={259},
pages={4656--4659},
year=1964,
url={https://gallica.bnf.fr/ark:/12148/bpt6k4015m/f1786.item}
}

@article{Gerard1990,
 doi = {10.1149/1.2086318},
 title = {Hall Effect in Aqueous Acid Solutions at Different Concentrations},
 author = {Gérard, P. and Gérard, R. and Meton, M. and Picard, E. J.},
 journal = {Journal of The Electrochemical Society},
 volume = {137},
 number = {12},
 pages = {3873–3875},
 year = {1990},
 month = {dec},
 issn = {1945-7111},
 publisher = {The Electrochemical Society}
}

@article{Frank1972,
 doi = {10.1080/00319107208084099},
 title = {Hall voltage in electrolyte solutions},
 author = {Frank, Ronald L. and Hoffman, Joseph G.},
 journal = {Physics and Chemistry of Liquids},
 volume = {3},
 number = {4},
 pages = {191–204},
 year = {1972},
 month = {jan},
 issn = {1029-0451},
 publisher = {Informa UK Limited}
}

@article{Abbes1982,
 doi = {10.1051/jphyslet:01982004305012700},
 title = {Effet Hall de chlorures et d’iodures en solutions diluées dans le méthanol et dans l’eau},
 author = {Abbes, M. and Gérard, R. and Gérard, P. and Meton, M. and Picard, E.J.},
 journal = {Journal de Physique Lettres},
 volume = {43},
 number = {5},
 pages = {127–132},
 year = {1982},
 issn = {0302-072X},
 publisher = {EDP Sciences}
}

@article{Bellissent1971,
 doi = {10.1149/1.2407872},
 title = {Eperimental Measurements of the Hall Effect in Very Dilute Solutions},
 author = {Bellissent, Marie-Claire and Gerard, Paul and Longevialle, Christian and Meton, Maurice and Pich, Michel and Morand, Geneviève},
 journal = {Journal of The Electrochemical Society},
 volume = {118},
 number = {12},
 pages = {1944},
 year = {1971},
 issn = {0013-4651},
 publisher = {The Electrochemical Society}
}

@article{Meton1976b,
 doi = {10.1051/jphyslet:019760037010024700},
 title = {Effet hall dans les solutions électrolytiques diluées, solvatation et dynamique des ions},
 author = {Meton, M. and Gerard, P. and Picard, E.J.},
 journal = {Journal de Physique Lettres},
 volume = {37},
 number = {10},
 pages = {247–250},
 year = {1976},
 issn = {0302-072X},
 publisher = {EDP Sciences}
}

@article{kuroda2023microscopic,
  doi = {10.1088/1742-5468/ad0639},
  title={Microscopic theory for hyperuniformity in two-dimensional chiral active fluid},
  author={Kuroda, Yuta and Miyazaki, Kunimasa},
  journal={Journal of Statistical Mechanics: Theory and Experiment},
  volume={2023},
  number={10},
  pages={103203},
  year={2023},
  publisher={IOP Publishing}
}

@article{Abbes1980,
 doi = {10.1051/jphyslet:019800041023057500},
 title = {Effet Hall dans des solutions électrolytiques de concentration variable},
 author = {Abbes, M. and Gérard, R. and Gérard, P. and Meton, M. and Picard, E.J.},
 journal = {Journal de Physique Lettres},
 volume = {41},
 number = {23},
 pages = {575–580},
 year = {1980},
 issn = {0302-072X},
 publisher = {EDP Sciences}
}

@article{LaforgueKantzer1965,
 doi = {10.1016/0013-4686(65)87038-4},
 title = {Effet magnetoelectrique des solutions d’acides mineraux},
 author = {Laforgue-Kantzer, D.},
 journal = {Electrochimica Acta},
 volume = {10},
 number = {6},
 pages = {585–603},
 year = {1965},
 month = {june},
 issn = {0013-4686},
 publisher = {Elsevier BV}
}

@article{kalz2022collisions,
 doi = {10.1103/physrevlett.129.090601},
  title={Collisions enhance self-diffusion in odd-diffusive systems},
  author={Kalz, Erik and Vuijk, Hidde Derk and Abdoli, Iman and Sommer, Jens-Uwe and L{\"o}wen, Hartmut and Sharma, Abhinav},
  journal={Physical Review Letters},
  volume={129},
  number={9},
  pages={090601},
  year={2022},
  publisher={APS}
}

@article{kalz2026reversal,
  doi = {10.1103/k8vq-dw2w},
  title={Reversal of tracer advection and Hall drift in an interacting chiral fluid},
  author={Kalz, Erik and Ravichandir, Shashank and Birkenmeier, Johannes and Metzler, Ralf and Sharma, Abhinav},
  journal={Physical Review E},
  volume={113},
  number={4},
  pages={L042104},
  year={2026},
  publisher={APS}
}

@article{brinkman1956brownian,
  title={Brownian motion in a field of force and the diffusion theory of chemical reactions},
  author={Brinkman, HC},
  journal={Physica},
  volume={22},
  number={1-5},
  pages={29--34},
  year={1956},
  publisher={Elsevier},
   doi = {10.1016/s0031-8914(56)80006-2},
}

@article{hargus2025odd,
 doi = {10.1103/pdpf-sd9q},
  title={Odd dynamics of passive objects in a chiral active bath},
  author={Hargus, Cory and Ghimenti, Federico and Tailleur, Julien and van Wijland, Fr{\'e}d{\'e}ric},
  journal={Physical Review Letters},
  volume={135},
  number={16},
  pages={167102},
  year={2025},
  publisher={APS}
}

@article{Hanebeck1974,
 doi = {10.1002/bbpc.19740780405},
 title = {Über den Hall‐Effekt an einigen Elektrolytlösungen und Membranen},
 author = {Hanebeck, N. and Schmid, G.},
 journal = {Berichte der Bunsengesellschaft für physikalische Chemie},
 volume = {78},
 number = {4},
 pages = {325–331},
 year = {1974},
 month = {apr},
 issn = {0005-9021},
 publisher = {Wiley}
}

@article{Wendhausen1968,
 doi = {10.1524/zpch.1968.58.5_6.325},
 title = {Messungen des Hall-Effektes in wäßrigen Elektrolytlösungen},
 author = {Wendhausen, Henning},
 journal = {Zeitschrift für Physikalische Chemie},
 volume = {58},
 number = {5_6},
 pages = {325–326},
 year = {1968},
 month = {apr},
 issn = {0942-9352},
 publisher = {Walter de Gruyter GmbH}
}

@article{Fahidy1983,
 doi = {10.1007/bf00617811},
 title = {Magnetoelectrolysis},
 author = {Fahidy, T. Z.},
 journal = {Journal of Applied Electrochemistry},
 volume = {13},
 number = {5},
 pages = {553–563},
 year = {1983},
 month = {sept},
 issn = {1572-8838},
 publisher = {Springer Science and Business Media LLC}
}

@article{Plummer1972,
 doi = {10.1080/00319107208084093},
 title = {The hall effect in molten salts},
 author = {Plummer, Mitty C. and Ham, Joe S.},
 journal = {Physics and Chemistry of Liquids},
 volume = {3},
 number = {3},
 pages = {127–135},
 year = {1972},
 month = {jan},
 issn = {1029-0451},
 publisher = {Informa UK Limited}
}

@article{GuerinOuler1982,
 doi = {10.1016/0013-4686(82)80215-6},
 title = {The magneto-electric effect in molten salts—1. Visualization and measurement of the magneto-electric voltage in 1-ethyl-pyridinium bromide; comparison of results in the molten salt at 130 °C and in solutions in water and ethanol},
 author = {Guerin—Ouler, Daniele and Nicollin, Camille},
 journal = {Electrochimica Acta},
 volume = {27},
 number = {7},
 pages = {909–917},
 year = {1982},
 month = {july},
 issn = {0013-4686},
 publisher = {Elsevier BV}
}

@article{GuerinOuler1983,
 doi = {10.1016/0013-4686(83)85069-5},
 title = {The magneto-electric effect in molten salts—2. Measurement of the magneto-electric voltage in some molten I-I inorganic salts: AgCl, KNO3 and NaNO3},
 author = {Guerin-Ouler, Danièle and Nicollin, Camille},
 journal = {Electrochimica Acta},
 volume = {28},
 number = {5},
 pages = {709–714},
 year = {1983},
 month = {may},
 issn = {0013-4686},
 publisher = {Elsevier BV}
}

@article{Blossey2022,
 doi = {10.1103/physrevresearch.4.023033},
 title = {Field theory of structured liquid dielectrics},
 author = {Blossey, Ralf and Podgornik, Rudolf},
 journal = {Physical Review Research},
 volume = {4},
 number = {2},
 pages = {023033},
 year = {2022},
 month = {apr},
 issn = {2643-1564},
 publisher = {American Physical Society (APS)}
}

@article{Berthoumieux2021,
 doi = {10.1063/5.0056430},
 title = {Dipolar Poisson models in a dual view},
 author = {Berthoumieux, Hélène and Monet, Geoffrey and Blossey, Ralf},
 journal = {The Journal of Chemical Physics},
 volume = {155},
 number = {2},
 year = {2021},
 month = {july},
 issn = {1089-7690},
 pages={024112},
 publisher = {AIP Publishing}
}

@article{Berthoumieux2019,
 doi = {10.1063/1.5080183},
 title = {Dielectric response in the vicinity of an ion: A nonlocal and nonlinear model of the dielectric properties of water},
 author = {Berthoumieux, H. and Paillusson, F.},
 journal = {The Journal of Chemical Physics},
 volume = {150},
 number = {9},
 year = {2019},
 month = {mar},
 pages = {094507},
 issn = {1089-7690},
 publisher = {AIP Publishing}
}

@article{Levy2012,
 doi = {10.1103/physrevlett.108.227801},
 title = {Dielectric Constant of Ionic Solutions: A Field-Theory Approach},
 author = {Levy, Amir and Andelman, David and Orland, Henri},
 journal = {Physical Review Letters},
 volume = {108},
 number = {22},
 pages = {227801},
 year = {2012},
 month = {may},
 issn = {1079-7114},
 publisher = {American Physical Society (APS)}
}

@article{Kornyshev1986,
 doi = {10.1016/0022-0728(86)80509-5},
 title = {On the non-local electrostatic theory of hydration force},
 author = {Kornyshev, A.A.},
 journal = {Journal of Electroanalytical Chemistry and Interfacial Electrochemistry},
 volume = {204},
 number = {1-2},
 pages = {79–84},
 year = {1986},
 month = {june},
 issn = {0022-0728},
 publisher = {Elsevier BV}
}

@article{Belaya1986,
 doi = {10.1016/s0009-2614(86)80099-9},
 title = {Hydration forces as a result of non-local water polarizability},
 author = {Belaya, M.L. and Feigel’man, M.V. and Levadny, V.G.},
 journal = {Chemical Physics Letters},
 volume = {126},
 number = {3-4},
 pages = {361–364},
 year = {1986},
 month = {may},
 issn = {0009-2614},
 publisher = {Elsevier BV}
}

@article{Illien2024,
 doi = {10.1103/physrevlett.133.268002},
 title = {Stochastic Density Functional Theory for Ions in a Polar Solvent},
 author = {Illien, Pierre and Carof, Antoine and Rotenberg, Benjamin},
 journal = {Physical Review Letters},
 volume = {133},
 number = {26},
 pages = {268002},
 year = {2024},
 month = {dec},
 issn = {1079-7114},
 publisher = {American Physical Society (APS)}
}

@article{Blossey2022b,
 doi = {10.1209/0295-5075/ac7d0a},
 title = {Continuum theories of structured dielectrics},
 author = {Blossey, Ralf and Podgornik, Rudolf},
 journal = {Europhysics Letters},
 volume = {139},
 number = {2},
 pages = {27002},
 year = {2022},
 month = {july},
 issn = {1286-4854},
 publisher = {IOP Publishing}
}

@article{Berthoumieux2024,
 doi = {10.1063/5.0226773},
 title = {Nonlinear conductivity of aqueous electrolytes: Beyond the first Wien effect},
 author = {Berthoumieux, Hélène and Démery, Vincent and Maggs, Anthony C.},
 journal = {The Journal of Chemical Physics},
 volume = {161},
 number = {18},
 year = {2024},
 month = {nov},
 pages={184504},
 issn = {1089-7690},
 publisher = {AIP Publishing}
}

@article{Harris1970,
 doi = {10.1007/bf01020444},
 title = {Hall coefficient in solution for a simple dynamical model},
 author = {Harris, S.},
 journal = {Journal of Statistical Physics},
 volume = {2},
 number = {4},
 pages = {379–385},
 year = {1970},
 issn = {1572-9613},
 publisher = {Springer Science and Business Media LLC}
}

@article{Friedman1968,
 doi = {10.1063/1.1664456},
 title = {Calculation of the Effect of Non-Brownian Motion on Some dc Transport Coefficients in Solution},
 author = {Friedman, Harold L. and Ben-Naim, Arieh},
 journal = {The Journal of Chemical Physics},
 volume = {48},
 number = {1},
 pages = {120–127},
 year = {1968},
 month = {jan},
 issn = {1089-7690},
 publisher = {AIP Publishing}
}

@article{Harris1972,
 doi = {10.1016/0009-2614(72)90015-2},
 title = {Hall coefficient in brownian-like solutions},
 author = {Harris, S.},
 journal = {Chemical Physics Letters},
 volume = {12},
 number = {3},
 pages = {493–494},
 year = {1972},
 month = {jan},
 issn = {0009-2614},
 publisher = {Elsevier BV}
}

@article{LaforgueKantzer1964,
 doi = {10.1051/jphys:01964002508-9084000},
 title = {Mise en évidence, sur les conducteurs ioniques, d’un effet magnétoélectrique semblable à l’effet Hall},
 author = {Laforgue-Kantzer, Denise},
 journal = {Journal de Physique},
 volume = {25},
 number = {8-9},
 pages = {840–842},
 year = {1964},
 issn = {0302-0738},
 publisher = {EDP Sciences}
}

@article{Chechetkin1989,
 doi = {10.1080/00268978900102681},
 title = {On the Hall number in dilute electrolyte solutions},
 author = {Chechetkin, V.R. and Lutovinov, V.S.},
 journal = {Molecular Physics},
 volume = {68},
 number = {4},
 pages = {979–981},
 year = {1989},
 month = {nov},
 issn = {1362-3028},
 publisher = {Informa UK Limited}
}

@article{Kroh1988,
 doi = {10.1016/0378-4371(88)90103-3},
 title = {On the theory of the hall effect in ionic solutions},
 author = {Kroh, H.J. and Felderhof, B.U.},
 journal = {Physica A: Statistical Mechanics and its Applications},
 volume = {153},
 number = {1},
 pages = {73–83},
 year = {1988},
 month = {nov},
 issn = {0378-4371},
 publisher = {Elsevier BV}
}

@article{Khanh1972,
 doi = {10.1016/0013-4686(72)85015-1},
 title = {Application de la statistique classique à l’étude de la conductibilité électrique et électromagnétique des solutions ioniques diluées},
 author = {Khanh, Tran Cong and Laforgue, A. and Laforgue-Kantzer, D.},
 journal = {Electrochimica Acta},
 volume = {17},
 number = {1},
 pages = {143–149},
 year = {1972},
 month = {jan},
 issn = {0013-4686},
 publisher = {Elsevier BV}
}

@book{gardiner2009stochastic,
  title={Stochastic methods},
  author={Gardiner, Crispin},
  volume={4},
  year={2009},
  publisher={Springer Berlin Heidelberg}
}

@article{adar2018dielectric,
  doi = {10.1063/1.5042235},
  title={Dielectric constant of ionic solutions: Combined effects of correlations and excluded volume},
  author={Adar, Ram M and Markovich, Tomer and Levy, Amir and Orland, Henri and Andelman, David},
  journal={The Journal of chemical physics},
  volume={149},
  number={5},
  pages={054504},
  year={2018},
  publisher={AIP Publishing}
}

@article{Kummel2013,
 doi = {10.1103/physrevlett.110.198302},
 title = {Circular Motion of Asymmetric Self-Propelling Particles},
 author = {Kümmel, Felix and ten Hagen, Borge and Wittkowski, Raphael and Buttinoni, Ivo and Eichhorn, Ralf and Volpe, Giovanni and Löwen, Hartmut and Bechinger, Clemens},
 journal = {Physical Review Letters},
 volume = {110},
 number = {19},
 pages = {198302},
 year = {2013},
 month = {may},
 issn = {1079-7114},
 publisher = {American Physical Society (APS)}
}

@article{Sevilla2016,
 doi = {10.1103/physreve.94.062120},
 title = {Diffusion of active chiral particles},
 author = {Sevilla, Francisco J.},
 journal = {Physical Review E},
 volume = {94},
 number = {6},
 pages = {062120},
 year = {2016},
 month = {dec},
 issn = {2470-0053},
 publisher = {American Physical Society (APS)}
}

@article{Kuroda2025,
 doi = {10.1103/25s2-6y3m},
 title = {Singular density correlations in chiral active fluids in three dimensions},
 author = {Kuroda, Yuta and Kawasaki, Takeshi and Miyazaki, Kunimasa},
 journal = {Physical Review E},
 volume = {112},
 number = {4},
 pages = {1103/25s2-6y3m},
 year = {2025},
 month = {oct},
 issn = {2470-0053},
 publisher = {American Physical Society (APS)}
}

@article{Jardat1999,
 doi = {10.1063/1.478703},
 title = {Transport coefficients of electrolyte solutions from Smart Brownian dynamics simulations},
 author = {Jardat, M. and Bernard, O. and Turq, P. and Kneller, G. R.},
 journal = {The Journal of Chemical Physics},
 volume = {110},
 number = {16},
 pages = {7993–7999},
 year = {1999},
 month = {apr},
 issn = {1089-7690},
 publisher = {AIP Publishing}
}

@article{Felderhof1983,
 doi = {10.1016/0378-4371(83)90111-5},
 title = {Linear response theory of sedimentation and diffusion in a suspension of spherical particles},
 author = {Felderhof, B.U. and Jones, R.B.},
 journal = {Physica A: Statistical Mechanics and its Applications},
 volume = {119},
 number = {3},
 pages = {591–608},
 year = {1983},
 month = {may},
 issn = {0378-4371},
 publisher = {Elsevier BV}
}

@article{Felderhof1987,
 doi = {10.1016/0378-4371(87)90278-0},
 title = {Linear response theory of the viscosity of suspensions of spherical brownian particles},
 author = {Felderhof, B.U. and Jones, R.B.},
 journal = {Physica A: Statistical Mechanics and its Applications},
 volume = {146},
 number = {3},
 pages = {417–432},
 year = {1987},
 month = {dec},
 issn = {0378-4371},
 publisher = {Elsevier BV}
}

@article{Friedman1965a,
 doi = {10.1063/1.1695956},
 title = {Relaxation Term of the Limiting Law of the Conductance of Electrolyte Mixtures},
 author = {Friedman, Harold L.},
 journal = {The Journal of Chemical Physics},
 volume = {42},
 number = {2},
 pages = {462–469},
 year = {1965},
 month = {jan},
 issn = {1089-7690},
 publisher = {AIP Publishing}
}

@article{Budkov2026,
 doi = {10.1063/5.0347809},
 title = {Hall effect in electrolyte solutions: Self-consistent Debye–Hückel–Onsager theory},
 author = {Budkov, Yury A. and Kalikin, Nikolai N.},
 journal = {The Journal of Chemical Physics},
 volume = {165},
 number = {6},
 year = {2026},
 month = {aug},
 issn = {1089-7690},
 publisher = {AIP Publishing},
 pages={064506}
}

@article{martin2021statistical,
  title={Statistical mechanics of active Ornstein-Uhlenbeck particles},
  author={Martin, David and O'Byrne, J{\'e}r{\'e}my and Cates, Michael E and Fodor, {\'E}tienne and Nardini, Cesare and Tailleur, Julien and Van Wijland, Fr{\'e}d{\'e}ric},
  journal={Physical Review E},
  volume={103},
  number={3},
  pages={032607},
  year={2021},
  publisher={APS},
   doi = {10.1103/physreve.103.032607},
}

@article{fodor2018statistical,
  title={The statistical physics of active matter: From self-catalytic colloids to living cells},
  author={Fodor, {\'E}tienne and Marchetti, M Cristina},
  journal={Physica A: Statistical Mechanics and its Applications},
  volume={504},
  pages={106--120},
  year={2018},
  publisher={Elsevier},
   doi = {10.1016/j.physa.2017.12.137},
}

@article{te2026colloquium,
  title={Colloquium: What do we mean by ‘active matter’?},
  author={te Vrugt, Michael and Liebchen, Benno and Cates, Michael E},
  journal={Reviews of Modern Physics},
  volume={98},
  number={3},
  pages={031001},
  year={2026},
  publisher={APS},
  doi = {10.1103/wd4f-q7kv}
}

@article{Martin1967,
 doi = {10.1103/physrev.161.143},
 title = {Sum Rules, Kramers-Kronig Relations, and Transport Coefficients in Charged Systems},
 author = {Martin, Paul C.},
 journal = {Physical Review},
 volume = {161},
 number = {1},
 pages = {143–155},
 year = {1967},
 month = {sept},
 issn = {0031-899X},
 publisher = {American Physical Society (APS)}
}

@article{Corkum1974,
 doi = {10.1139/p74-071},
 title = {Green–Kubo Formulae for a Plasma in an External Magnetic Field},
 author = {Corkum, P. B. and McLennan, J. A.},
 journal = {Canadian Journal of Physics},
 volume = {52},
 number = {6},
 pages = {516–522},
 year = {1974},
 month = {mar},
 issn = {1208-6045},
 publisher = {Canadian Science Publishing}
}

@article{Pavliotis2010,
 doi = {10.1093/imamat/hxq039},
 title = {Asymptotic analysis of the Green-Kubo formula},
 author = {Pavliotis, G. A.},
 journal = {IMA Journal of Applied Mathematics},
 volume = {75},
 number = {6},
 pages = {951–967},
 year = {2010},
 month = {june},
 issn = {1464-3634},
 publisher = {Oxford University Press (OUP)}
}

@article{Gaspard2013,
 doi = {10.1088/1367-2630/15/11/115014},
 title = {Multivariate fluctuation relations for currents},
 author = {Gaspard, Pierre},
 journal = {New Journal of Physics},
 volume = {15},
 number = {11},
 pages = {115014},
 year = {2013},
 month = {nov},
 issn = {1367-2630},
 publisher = {IOP Publishing}
}

@article{Shalchi2011,
 doi = {10.1103/physreve.83.046402},
 title = {Applicability of the Taylor-Green-Kubo formula in particle diffusion theory},
 author = {Shalchi, A.},
 journal = {Physical Review E},
 volume = {83},
 number = {4},
 pages = {046402},
 year = {2011},
 month = {apr},
 issn = {1550-2376},
 publisher = {American Physical Society (APS)}
}

@article{Lesnicki2021,
 doi = {10.1063/5.0052860},
 title = {On the molecular correlations that result in field-dependent conductivities in electrolyte solutions},
 author = {Lesnicki, Dominika and Gao, Chloe Y. and Limmer, David T. and Rotenberg, Benjamin},
 journal = {The Journal of Chemical Physics},
 volume = {155},
 number = {1},
 year = {2021},
 month = {july},
 issn = {1089-7690},
 publisher = {AIP Publishing},
 pages={014507}
}

@article{Nair2025,
 doi = {10.1063/5.0270981},
 title = {Ions at electrochemical interfaces: From explicit to implicit molecular solvent descriptions},
 author = {Nair, Swetha and Jeanmairet, Guillaume and Rotenberg, Benjamin},
 journal = {The Journal of Chemical Physics},
 volume = {163},
 number = {1},
 year = {2025},
 month = {july},
 issn = {1089-7690},
 publisher = {AIP Publishing},
 pages = {014107}
}

@article{Molina2009,
 doi = {10.1103/physreve.80.065103},
 title = {Models of electrolyte solutions from molecular descriptions: The example of NaCl solutions},
 author = {Molina, John Jairo and Dufrêche, Jean-François and Salanne, Mathieu and Bernard, Olivier and Jardat, Marie and Turq, Pierre},
 journal = {Physical Review E},
 volume = {80},
 number = {6},
 pages = {065103},
 year = {2009},
 month = {dec},
 issn = {1550-2376},
 publisher = {American Physical Society (APS)}
}

@article{Demery2026,
 doi = {10.1039/d5fd00149h},
 title = {Effect of solvent structure on the Wien effect and ionic correlations at the nanoscale},
 author = {Démery, Vincent and Toquer, Damien and Berthoumieux, Hélène},
 journal = {Faraday Discussions},
 year = {2026},
 issn = {1364-5498},
 publisher = {Royal Society of Chemistry (RSC)}
}

@article{Chun2015,
 doi = {10.1146/annurev-anchem-071114-040202},
 title = {Iontronics},
 author = {Chun, Honggu and Chung, Taek Dong},
 journal = {Annual Review of Analytical Chemistry},
 volume = {8},
 number = {1},
 pages = {441–462},
 year = {2015},
 month = {july},
 issn = {1936-1335},
 publisher = {Annual Reviews}
}

@article{Kavokine2021,
 doi = {10.1146/annurev-fluid-071320-095958},
 title = {Fluids at the Nanoscale: From Continuum to Subcontinuum Transport},
 author = {Kavokine, Nikita and Netz, Roland R. and Bocquet, Lydéric},
 journal = {Annual Review of Fluid Mechanics},
 volume = {53},
 number = {1},
 pages = {377–410},
 year = {2021},
 month = {jan},
 issn = {1545-4479},
 publisher = {Annual Reviews}
}

@article{Marbach2019,
 doi = {10.1039/c8cs00420j},
 title = {Osmosis, from molecular insights to large-scale applications},
 author = {Marbach, Sophie and Bocquet, Lydéric},
 journal = {Chemical Society Reviews},
 volume = {48},
 number = {11},
 pages = {3102–3144},
 year = {2019},
 issn = {1460-4744},
 publisher = {Royal Society of Chemistry (RSC)}
}

@article{Gatard2020,
 doi = {10.1016/j.coelec.2020.04.012},
 title = {Use of magnetic fields in electrochemistry: A selected review},
 author = {Gatard, Vivien and Deseure, Jonathan and Chatenet, Marian},
 journal = {Current Opinion in Electrochemistry},
 volume = {23},
 pages = {96–105},
 year = {2020},
 month = {oct},
 issn = {2451-9103},
 publisher = {Elsevier BV}
}

@article{Luo2022,
 doi = {10.1002/ange.202203564},
 title = {Electrochemistry in Magnetic Fields},
 author = {Luo, Songzhu and Elouarzaki, Kamal and Xu, Zhichuan J.},
 journal = {Angewandte Chemie},
 volume = {134},
 number = {27},
 year = {2022},
 month = {may},
 issn = {1521-3757},
 publisher = {Wiley},
 pages = {e202203564},
}

@article{Chen2024,
 doi = {10.1002/adfm.202415660},
 title = {Electrocatalysis under Magnetic Fields},
 author = {Chen, Jiang‐Bo and Ying, Jie and Tian, Yuan and Xiao, Yu‐Xuan and Yang, Xiao‐Yu},
 journal = {Advanced Functional Materials},
 volume = {35},
 number = {8},
 year = {2024},
 month = {sept},
 issn = {1616-3028},
 publisher = {Wiley},
 pages={2415660}
}

@article{Li2026,
 doi = {10.1073/pnas.2522696123},
 title = {Magnetic field–enhanced alkaline water electrolysis from laboratory to industry},
 author = {Li, Hao and Fang, Jiakun and Huang, Danji and Zhong, Zhiyao and Wen, Qunlei and Liu, Youwen and Hu, Kewei and Yang, Zhenhong and Lu, Ang and Ai, Xiaomeng and Zhai, Tianyou and Wen, Jinyu and Pan, Yuan},
 journal = {Proceedings of the National Academy of Sciences},
 volume = {123},
 number = {7},
 year = {2026},
 month = {feb},
 issn = {1091-6490},
 publisher = {National Academy of Sciences},
 pages={e2522696123}
}

@article{Feng2019,
 doi = {10.1103/physrevx.9.021024},
 title = {Free and Bound States of Ions in Ionic Liquids, Conductivity, and Underscreening Paradox},
 author = {Feng, Guang and Chen, Ming and Bi, Sheng and Goodwin, Zachary A. H. and Postnikov, Eugene B. and Brilliantov, Nikolai and Urbakh, Michael and Kornyshev, Alexei A.},
 journal = {Physical Review X},
 volume = {9},
 number = {2},
 pages = {021024},
 year = {2019},
 month = {may},
 issn = {2160-3308},
 publisher = {American Physical Society (APS)}
}

@article{Knotek1977,
 doi = {10.1016/0038-1098(77)90830-4},
 title = {The absence of a measurable hall effect in the superionic conductor RbAg4I5},
 author = {Knotek, M.L. and Seager, C.H.},
 journal = {Solid State Communications},
 volume = {21},
 number = {7},
 pages = {625–627},
 year = {1977},
 month = {feb},
 issn = {0038-1098},
 publisher = {Elsevier BV}
}

@article{Stuhrmann2002,
 doi = {10.1016/s0167-2738(02)00470-8},
 title = {Ionic Hall effect measured in rubidium silver iodide},
 author = {Stuhrmann, C},
 journal = {Solid State Ionics},
 volume = {154-155},
 pages = {109–112},
 year = {2002},
 month = {dec},
 issn = {0167-2738},
 publisher = {Elsevier BV}
}

@article{Gagliardi2016,
 doi = {10.1103/physrevb.94.134426},
 title = {Charge transport in superionic and melted AgI under a magnetic field studied via molecular dynamics},
 author = {Gagliardi, Luca and Bonella, Sara},
 journal = {Physical Review B},
 volume = {94},
 number = {13},
 pages = {134426},
 year = {2016},
 month = {oct},
 issn = {2469-9969},
 publisher = {American Physical Society (APS)}
}

@article{Mouhat2013,
 doi = {10.1080/00268976.2013.846486},
 title = {Charge transport simulations of NaCl in an external magnetic field: the quest for the Hall effect},
 author = {Mouhat, F. and Bonella, S. and Pierleoni, C.},
 journal = {Molecular Physics},
 volume = {111},
 number = {22-23},
 pages = {3651–3661},
 year = {2013},
 month = {dec},
 issn = {1362-3028},
 publisher = {Informa UK Limited}
}

@article{Coretti2020,
 doi = {10.1103/physreve.102.030101},
 title = {Fluctuation relations for systems in a constant magnetic field},
 author = {Coretti, Alessandro and Rondoni, Lamberto and Bonella, Sara},
 journal = {Physical Review E},
 volume = {102},
 number = {3},
 pages = {030101},
 year = {2020},
 month = {sept},
 issn = {2470-0053},
 publisher = {American Physical Society (APS)}
}

@article{Kanazawa2015,
 doi = {10.1103/physrevlett.114.090601},
 title = {Minimal Model of Stochastic Athermal Systems: Origin of Non-Gaussian Noise},
 author = {Kanazawa, Kiyoshi and Sano, Tomohiko G. and Sagawa, Takahiro and Hayakawa, Hisao},
 journal = {Physical Review Letters},
 volume = {114},
 number = {9},
 pages = {090601},
 year = {2015},
 month = {mar},
 issn = {1079-7114},
 publisher = {American Physical Society (APS)}
}

@article{Kanazawa2012,
 doi = {10.1103/physrevlett.108.210601},
 title = {Stochastic Energetics for Non-Gaussian Processes},
 author = {Kanazawa, Kiyoshi and Sagawa, Takahiro and Hayakawa, Hisao},
 journal = {Physical Review Letters},
 volume = {108},
 number = {21},
 pages = {210601},
 year = {2012},
 month = {may},
 issn = {1079-7114},
 publisher = {American Physical Society (APS)}
}

@article{Fodor2018,
 doi = {10.1103/physreve.98.062610},
 title = {Non-Gaussian noise without memory in active matter},
 author = {Fodor, \'Etienne and Hayakawa, Hisao and Tailleur, Julien and van Wijland, Fr\'ed\'eric},
 journal = {Physical Review E},
 volume = {98},
 number = {6},
 pages = {062610},
 year = {2018},
 month = {dec},
 issn = {2470-0053},
 publisher = {American Physical Society (APS)}
}

@article{Bonella2017,
 doi = {10.1103/physreve.96.012160},
 title = {Time-reversal symmetry for systems in a constant external magnetic field},
 author = {Bonella, Sara and Coretti, Alessandro and Rondoni, Lamberto and Ciccotti, Giovanni},
 journal = {Physical Review E},
 volume = {96},
 number = {1},
 pages = {012160},
 year = {2017},
 month = {july},
 issn = {2470-0053},
 publisher = {American Physical Society (APS)}
}

@article{Bonella2014,
 doi = {10.1209/0295-5075/108/60004},
 title = {Time reversal symmetry in time-dependent correlation functions for systems in a constant magnetic field},
 author = {Bonella, S. and Ciccotti, G. and Rondoni, L.},
 journal = {EPL (Europhysics Letters)},
 volume = {108},
 number = {6},
 pages = {60004},
 year = {2014},
 month = {dec},
 issn = {1286-4854},
 publisher = {IOP Publishing}
}

@book{Kubo1991,
 doi = {10.1007/978-3-642-58244-8},
 title = {Statistical Physics II},
 author = {Kubo, Ryogo and Toda, Morikazu and Hashitsume, Natsuki},
 journal = {Springer Series in Solid-State Sciences},
 year = {1991},
 isbn = {9783642582448},
 issn = {0171-1873},
 publisher = {Springer Berlin Heidelberg}
}

@article{Lucente2023,
 doi = {10.1088/1742-5468/ad063b},
 title = {Statistical features of systems driven by non-Gaussian processes: theory and practice},
 author = {Lucente, Dario and Puglisi, Andrea and Viale, Massimiliano and Vulpiani, Angelo},
 journal = {Journal of Statistical Mechanics: Theory and Experiment},
 volume = {2023},
 number = {11},
 pages = {113202},
 year = {2023},
 month = {nov},
 issn = {1742-5468},
 publisher = {IOP Publishing}
}

@article{vanKampen1961,
 doi = {10.1139/p61-056},
 title = {A POWER SERIES EXPANSION OF THE MASTER EQUATION},
 author = {van Kampen, N. G.},
 journal = {Canadian Journal of Physics},
 volume = {39},
 number = {4},
 pages = {551–567},
 year = {1961},
 month = {apr},
 issn = {1208-6045},
 publisher = {Canadian Science Publishing}
}

@misc{avni2026odd,
 eprint = {2609.32080},
 title = {Odd stochastic density functional theory},
 author = {Avni, Yael and Fruchart, Michel and Martin, David and Khain, Tali and Vitelli, Vincenzo},
 year = {2026},
 eprinttype = {arXiv},
 archivePrefix = {arXiv}
}

\end{document}